\documentclass[preprintnumbers,superscriptaddress,nofootinbib,aps,prd,floatfix]{revtex4-2}
\pdfoutput=1
\usepackage{amsmath,amssymb}
\usepackage{epsfig}
\usepackage{float}
\usepackage{graphicx}
\usepackage[usenames,dvipsnames]{color}
\usepackage{subfigure}
\usepackage{slashed}
\usepackage{hyperref,orcidlink}
\hypersetup{colorlinks=true, citecolor=blue, urlcolor=blue, linkcolor=blue}
\usepackage{tabularray}
\UseTblrLibrary{booktabs}

\begin{document}
	\title{Asymmetric long-lived dark matter and leptogenesis from type-III seesaw framework}
	\author{Satyabrata Mahapatra\orcidlink{https://orcid.org/0000-0002-4000-5071}}
\email{satyabrata@g.skku.edu}
\affiliation{Department of Physics and Institute of Basic Science, Sungkyunkwan University, Suwon 16419, Korea}

\author{Partha Kumar Paul\orcidlink{https://orcid.org/0000-0002-9107-5635}}
\email{ph22resch11012@iith.ac.in}
\affiliation{Department of Physics, Indian Institute of Technology Hyderabad, Kandi, Telangana-502285, India.}

\author{Narendra Sahu\orcidlink{https://orcid.org/0000-0002-9675-0484}}
\email{nsahu@phy.iith.ac.in}
\affiliation{Department of Physics, Indian Institute of Technology Hyderabad, Kandi, Telangana-502285, India.}

\author{Prashant Shukla\orcidlink{https://orcid.org/0000-0001-8118-5331}}
\email{pshukla@barc.gov.in}
\affiliation{Nuclear Physics Division, Bhabha Atomic Research Centre,
	Mumbai, 400085, India.}
\affiliation{Homi Bhabha National Institute, Anushakti Nagar, Mumbai,
	400094, India.}

	\begin{abstract}
		We propose a simple model in the type-III seesaw framework to explain the neutrino mass, asymmetric dark matter (ADM), and baryon asymmetry of the Universe. We extend 
		the standard model with a vector-like singlet lepton ($\chi$) and a hypercharge zero scalar triplet ($\Delta$) 
		in addition to three hypercharge zero triplet fermions($\Sigma_i~,i=1,2,3$). A $Z_2$ symmetry is imposed under which $\chi$ and 
		$\Delta$ are odd, while all other particles are even. As a result, the lightest $Z_2$ odd particle $\chi$ behaves 
		as a candidate of DM. In the early Universe, the $CP$-violating out-of-equilibrium decay of heavy triplet 
		fermions to the Standard Model lepton ($L$) and Higgs ($H$) generate a net lepton asymmetry, while that of triplet fermions to $\chi$ 
		and $\Delta$ generate a net asymmetric DM. The lepton asymmetry is converted to the required baryon 
		asymmetry of the Universe via the electroweak sphalerons, while the asymmetry in $\chi$ remains as a DM 
		relic that we observe today. We introduce a singlet scalar $\Phi$, with mass $M_\phi < M_\chi$, which not only 
		assists to deplete the symmetric component of $\chi$ through the annihilation process: $\bar{\chi} \chi \to \Phi \Phi$ 
		but also paves a path to detect DM $\chi$ at direct search experiments through $\Phi-H$ mixing. The electro-weak symmetry breaking induces a non-zero vacuum expectation value to $\Delta$, which leads to an unstable asymmetric DM ranging from a few MeV to hundreds of GeV. We then explore the displaced vertex signatures of the charged components of the scalar triplet $\Delta$ at colliders.
	\end{abstract}	
	\maketitle
	\noindent
	
	\setcounter{footnote}{0}
	\renewcommand*{\thefootnote}{\arabic{footnote}}
	
	\section{Introduction}\label{sec:intro}
	At present, the standard model (SM) of particle physics is the most comprehensive theoretical framework 
	for understanding elementary particles and their interactions. However, there exist observed 
	phenomena which can not be explained within the framework of the SM, implying that there exists physics 
	beyond the SM. 
	
	The most striking phenomenon that begs for physics beyond the SM is the present-day dark matter (DM) component 
	of the observed Universe. Irrefutable evidences from galaxy rotation curve, gravitational lensing, and large-scale 
	structure of the Universe combinely suggests that the present Universe is filled with a non-luminous, 
	non-baryonic form of matter, popularly called DM. The satellite-borne experiment 
	PLANCK, which precisely measures the temperature anisotropy of cosmic microwave background (CMB) sky, 
	allows us to estimate the DM component of the present Universe to be $26.8\%$. The relic density of DM is 
	conventionally expressed as: $\Omega_{\rm DM}h^2=0.120 \pm 0.001$ at 68\% confidence level~\cite{Aghanim:2018eyx}, where 
	$\Omega_{\rm DM}$ is the DM density parameter and $h$ is the reduced Hubble parameter defined as: 
	Hubble parameter/$(100 {\rm km} {\rm s}^{-1} {\rm Mpc}^{-1})$. However, none of the SM particles can mimic the properties of DM, implying that new fundamental particles must be introduced to explain DM. 
	
	Apart from the identity of DM, another appealing motivation for exploring physics beyond the SM is the sub-eV masses of light neutrinos and baryon asymmetry of the observed Universe. Despite compelling evidences 
	for existence of light neutrino masses from long baseline oscillation experiments~\cite{deSalas:2017kay, Aartsen:2015zva, Abe:2015awa} and cosmological data~\cite{Lesgourgues:2006nd, Wong:2011ip, Lattanzi:2017ubx}, 
	the origin of light neutrino masses is still unknown. While oscillation data is sensitive to the 
	mass squared differences of light neutrinos, the absolute mass scale is constrained by cosmological data to 
	be $\sum_i m_\nu < 0.12~ {\rm eV}$~\cite{Aghanim:2018eyx}. This implies that new physics beyond 
	the SM is required to account for the masses of light neutrinos, as the Higgs field, which is responsible for the origin of
	mass of all particles in the SM can not have any Yukawa coupling with left-handed neutrinos due to the 
	absence of right-handed neutrinos in the SM.

	The nature of neutrinos, either Dirac or Majorana, is hitherto unknown. Assuming that neutrinos are Majorana, the most 
	elegant way to introduce sub-eV masses of light neutrinos is to invoke seesaw mechanisms (type-I~\cite{Minkowski:1977sc, 
		Mohapatra:1979ia}, type-II~\cite{Ma:1998dx, Wetterich:1981bx}, type-III~\cite{Foot:1988aq}), which are ultraviolet 
	realizations of the dimension five Weinberg operator $\mathcal{O}_5 = L H L H/\Lambda$~\cite{Weinberg:1979sa, Ma:1998dn}. 
	Here, $L$ and $H$ represent the lepton and Higgs doublets of the SM, and $\Lambda$ is the scale of new physics. Note that the operator $\mathcal{O}_5$ breaks the lepton number by two units and paves a way to generate matter-antimatter asymmetry in the early Universe through 
	baryo-lepto-genesis route~\cite{Fukugita:1986hr, Luty:1992un,Mohapatra:1992pk, Flanz:1994yx,Buchmuller:2004nz,Ma:1998dx,Davidson:2008bu}. In the effective theory, the Majorana mass of the light neutrinos is given  by 
	$m_\nu = \langle H \rangle^2/\Lambda$. The seesaw mechanisms can further be extended to explain the origin 
	of asymmetric DM (ADM). See, for instance, ref.~\cite{Falkowski:2011xh,  Patel:2022xyv,Biswas:2018sib,Narendra:2018vfw,Nagata:2016knk} for origin of ADM in type-I seesaw and references~\cite{Arina:2011cu, Arina:2012fb, Arina:2012aj, Narendra:2019cyt} for the origin of ADM in type-II seesaw. For a review on ADM, see~\cite{Petraki:2013wwa,Zurek:2013wia}. 
	
	In this paper, we consider an extension of the SM within the type-III seesaw framework~\cite{Foot:1988aq}. Our motivation is to simultaneously explain the origin of neutrino mass, asymmetric DM, and baryon asymmetry. We extend the SM with a vector-like leptonic singlet fermion $\chi$ and a hypercharge zero triplet scalar $\Delta$ in addition to three hypercharge zero triplet 
	fermions $\Sigma_i$, $i=1,2,3$. We also impose a $Z_2$ symmetry under which $\chi$ and $\Delta$ are odd while all other 
	particles, including $\Sigma$, are even. As a result, the lightest $Z_2$ odd particle $\chi$ behaves as a candidate of DM. While the production mechanism of the DM relic is  unknown, here we consider the DM relic to be an asymmetric component 
	of $\chi$, generated through the $CP$-violating out-of-equilibrium decay of triplet fermions: 
	$\Sigma \to \chi \Delta$ in the early Universe. The simultaneous decay of $\Sigma \to LH$ also generates a net lepton 
	asymmetry is then converted to the required baryon asymmetry via electro-weak (EW) sphaleron transitions.  We introduce a singlet scalar $\Phi$, with mass $m_\Phi < m_\chi$, which not only helps in depleting the symmetric component of $\chi$ through 
	asymmetry is then converted to the required baryon asymmetry via electro-weak (EW) sphaleron transitions.  We introduce a singlet scalar $\Phi$, with mass $m_\Phi < m_\chi$, which not only helps in depleting the symmetric component of $\chi$ through 
	the annihilation process: $\bar{\chi}\chi \to \Phi \Phi$, but also paves a path for DM detection in direct search experiments 
	through $\Phi-H$ mixing. In other words, the null detection of DM can rule out large $\Phi-H$ mixing, which is though 
	favorable for depleting the symmetric component of DM. We scrutinize the parameter space using the constraints from CRESST-III \cite{CRESST:2019jnq}, LZ \cite{LZ:2022lsv}, XENONnT \cite{XENON:2023cxc} as well as showcase the projected sensitivities of DS-LM \cite{GlobalArgonDarkMatter:2022ppc} and DARWIN 
	\cite{DARWIN:2016hyl} experiments, which can probe some parts of the parameter space in the near future. The $Z_2$ symmetry breaks softly when $\Delta$ acquires an induced vacuum expectation value (vev) after electro-weak symmetry breaking (EWSB), thus resulting in an unstable DM in a wide mass range as well as eliminating the asymmetry in $\Delta$. 
	It is worth mentioning here that, by tuning the vev of $\Delta$, the DM can be made long-lived ({ \it, i.e.} $\tau_{\rm DM}>10^{27}$ s). Unlike the stable ADM scenarios, here, the long-lived DM paves the way for indirect detection.  Moreover, the model offers compelling signatures at colliders due to the participation of TeV-scale triplet scalar particles.
	
	In the original motivation of the ADM scenario~\cite{Petraki:2013wwa,Zurek:2013wia}, the $CP$ asymmetry parameters in the visible sector (arising from decay mode $\Sigma\rightarrow LH\equiv\epsilon_L$) and the dark sector (arising from the decay mode $\Sigma\rightarrow\chi\Delta\equiv\epsilon_\chi $) are assumed to be the same. This leads to $n_{B}\simeq n_{\rm DM}$.  However, we note that $CP$ asymmetry parameters  $\epsilon_L$ and $\epsilon_{\chi}$ may not be the same due to their dependence on different couplings. This mismatch can allow for a wide range of DM masses while still explaining the observed ratio of DM density to baryonic matter density, $\Omega_{\rm DM}/\Omega_B \sim 5$.

	The rest of the paper is organized as follows. In Section \ref{sec:model}, we outline the features of the model. Section \ref{sec:numass} discusses the generation of neutrino masses via the type-III seesaw mechanism. In Section \ref{sec:admlepto}, we examine darko-lepto-genesis from the out-of-equilibrium decay of heavy fermion triplets. Section \ref{sec:longliveddm} addresses the constraints on long-lived DM. Subsequently, we explore the detection prospects of the model in Section \ref{sec:detection} and conclude in Section \ref{sec:conclude}.

	\noindent
	\section{The Model}\label{sec:model}
	We extend the SM with three hypercharge-less triplet fermions $\Sigma_i$, $i=1,2,3$ with zero lepton number.  In addition, we introduce a 
	singlet scalar $\Phi$, a vector-like leptonic singlet fermion $\chi$ with lepton number one and a hypercharge zero triplet 
	scalar $\Delta$. On top of the SM gauge symmetry, we impose an additional discrete symmetry $Z_2$ under which $\chi$ and $\Delta$ are odd, while all other particles are even. The charge assignment of these fields 
	under $SU(3)_c \otimes SU(2)_L \otimes U(1)_Y \otimes Z_2$ is given in Table \ref{tab:tab1}. The lightest $Z_2$ odd particle $\chi$ serves as a candidate of DM in our model.

	\begin{table}[h!]
		\begin{center}
				\begin{tabular}{|c|c|c|c|}
					\hline \multicolumn{2}{|c}{Fields}&  \multicolumn{1}{|c|}{ $ SU(3)_C \otimes SU(2)_L \otimes U(1)_Y $  $\otimes  Z_2 $  } \\ \hline
					{Fermions} &  $\Sigma_i (i=1,2,3)$&  ~~1 ~~~~~~~~~~~~3~~~~~~~~~~0~~~~~~~~~~ + \\ [0.5em] 
					& $\chi$  & ~1 ~~~~~~~~~~~1~~~~~~~~~~0~~~~~~~~~~ - \\
					[0.5em] \cline{1-3}
					{Scalar} & 
					$\Delta$ & ~1 ~~~~~~~~~~3~~~~~~~~~~~0~~~~~~~~~~ - \\
					& $\Phi$ & ~~1~~~~~~~~~~~~~1~~~~~~~~~~~0~~~~~~~~~ +\\
					
					\hline
				\end{tabular}
				\caption{Charge assignment of BSM fields under the gauge group $G \equiv G_{\rm SM} \otimes Z_2$,  where $G_{\rm SM}\equiv SU(3)_c \otimes SU(2)_L \otimes U(1)_Y$ .}
				\label{tab:tab1}
			\end{center}
		\end{table}

		\noindent The relevant Lagrangian in our model is given by\footnote{We suppress the generation indices unless stated explicitly in any other part of the paper.}:
		\begin{eqnarray}
			\mathcal{L}&=&{\rm Tr}[\overline{\Sigma} i \gamma_\mu D^\mu \Sigma] + {\rm Tr}[(D_\mu \Delta)^\dagger (D_\mu \Delta)] + (D_\mu H)^\dagger (D_\mu H) +\frac{1}{2}~\partial_\mu\Phi~\partial^{\mu}\Phi -\big[ \frac{1}{2}{\rm Tr}[m_\Sigma \overline{\Sigma^c}\Sigma] \nonumber\\
			&+& \sqrt{2}~ y_{_\Sigma} \overline{L}  \Tilde{H} \Sigma  + 
			y_\chi {\rm Tr}[\overline{\Sigma}~\Delta~\chi]  + {\rm h.c.} \big]  +~\overline{\chi}i\gamma_\mu\partial^\mu\chi-M_\chi \overline{\chi}\chi  -\lambda_{\rm DM} \Phi \bar{\chi}\chi - V(H,\Delta, \Phi),\, 
			\label{lagrangian} 
		\end{eqnarray}
		where $\Tilde{H}=i \sigma_2 H^*$ and the scalar potential $V(H,\Delta,\Phi)$ is given by:
		\begin{eqnarray}
			V(H,\Delta, \Phi)&=&\frac{1}{2} \mu^2_\Phi \Phi^2 -\mu_H^2 (H^\dagger H) +\mu_\Delta^2 {\rm Tr}[\Delta^\dagger\Delta] +\lambda_H(H^\dagger H)^2 + \lambda_\Delta{\rm Tr}[(\Delta^\dagger\Delta)^2] +\lambda_{H\Delta} (H^\dagger H) {\rm Tr}[\Delta^\dagger\Delta] \nonumber\\
			& +& \sqrt{2} \mu H^\dagger (\vec{\sigma} \cdot \vec{\Delta}) H + \frac{\lambda_{\Phi}}{4}\Phi^4+\frac{\rho_1}{\sqrt{2}}\Phi(H^\dagger H)+\frac{\rho_2}{\sqrt{2}}\Phi{\rm Tr}[(\Delta^\dagger\Delta)]+\frac{\lambda_{H\Phi}}{2}(H^\dagger H)\Phi^2,
			\label{Eq:scalarpot}
		\end{eqnarray}
		where $\sigma$ in the tri-linear term represents the Pauli-spin matrix. Here it is worth mentioning that, $\Delta$ being odd under the $Z_2$ symmetry, the tri-linear term: $\sqrt{2} \mu H^\dagger (\Vec{\sigma} \cdot \Vec{\Delta}) H$  in Eq. \ref{Eq:scalarpot}, breaks the $Z_2$ symmetry softly. The EW phase transition induces a nonzero vev of $\Delta$ and $\Phi$. The small vev of $\Delta$ facilitates mixing between $\chi$ and $\Sigma$, implying that $\chi$ is a decaying DM in this model. Similarly the small vev of $\Phi$ enables a small mixing between $H-\Phi$, thus creating a portal for detection of DM via direct search experiments.
		
		The adjoint representation of the hypercharge-less triplet fermion field ($\Sigma$) and the triplet scalar field ($\Delta$) is given by:  
		\begin{equation}
			\Sigma=\begin{pmatrix}
				\frac{\Sigma^0}{\sqrt{2}} & \Sigma^+\\
				\Sigma^- & -\frac{\Sigma^0}{\sqrt{2}}
			\end{pmatrix} \,,~~
			\Delta=\begin{pmatrix}
				\frac{\delta^0}{\sqrt{2}} & \delta^+\\
				\delta^- & -\frac{\delta^0}{\sqrt{2}}
			\end{pmatrix}.\
			\label{fieldrep}
		\end{equation}
		The quantum fluctuations around the minima of scalar fields $H$, $\Delta$ and $\Phi$ are given as\footnote{After the EWSB, $\Phi$ obtains an induced vev, which gives a small correction to the DM mass. The DM mass after the EWSB becomes, $M^\prime_{\chi}=M_\chi+\lambda_{{\rm DM}}u \simeq M_\chi$ where the correction term $\lambda_{{\rm DM}}u$ is negligible. As this small correction does not affect our phenomenological results, we use $M_{\rm DM}=M_{\chi}$ throughout our manuscript.}:
		\begin{equation}
			H=\frac{1}{\sqrt{2}}\begin{pmatrix}
				0\\
				v_0+h
			\end{pmatrix}\,,~~
			\Delta=\begin{pmatrix}
				\frac{v_1+s}{\sqrt{2}} & 0\\
				0 & -\frac{v_1+s}{\sqrt{2}}
			\end{pmatrix}
			~~{\rm and}~~ \Phi=\phi+u.\
		\end{equation}
		By minimizing the scalar potential w.r.t $v_0,~v_1$, and $u$, we get
		\begin{eqnarray}
			\mu^2_H&=&\lambda_Hv^2_0+\lambda_{H\Delta}v^2_1+\frac{1}{2}\lambda_{H\Phi}u^2-\mu v_1+\frac{1}{\sqrt{2}}\rho_1u,\nonumber\\
			\mu^2_\Delta&=&-\lambda_\Delta v^2_1-\frac{1}{2}\lambda_{H\Delta}v_0^2-\frac{1}{4}\frac{\mu v^2_0}{v_1}-\frac{1}{\sqrt{2}}\rho_2u,\nonumber\\
			\mu^2_\Phi&=&-\lambda_\Phi u^2-\frac{1}{2}\lambda_{H\Phi}v_0^2-\frac{1}{2\sqrt{2}}\frac{\rho_1v^2_0}{u}-\frac{1}{\sqrt{2}}\frac{\rho_2v^2_1}{u}.
		\end{eqnarray}
		
		The neutral CP even mass squared matrix in the basis ($h~s~\phi$), can be expressed as
		\begin{equation}
			\mathcal{M}^2=\begin{pmatrix}
				2\lambda_{H}v^2_0 & 2\lambda_{H\Delta}v_0v_1-\mu v_0 & \frac{\rho_1}{\sqrt{2}} v_0+\lambda_{H\Phi}uv_0\\
				2\lambda_{H\Delta}v_0v_1-\mu v_0 & 4\lambda_{\Delta}v^2_1+\frac{\mu v^2_0}{2v_1} & \sqrt{2}\rho_2 v_1\\
				\frac{\rho_1}{\sqrt{2}} v_0+\lambda_{H\Phi}uv_0 & \sqrt{2}\rho_2 v_1 & 2\lambda_{\Phi}u^2-\frac{1}{2\sqrt{2}}\frac{\rho_{1}v^2_0+2\rho_2v^2_1}{u}
			\end{pmatrix}.
		\end{equation}
		This matrix can be diagonalized by considering an orthogonal matrix  $\mathcal{O}=\mathcal{O}_{12}(\theta_{12})\cdot\mathcal{O}_{23}(\theta_{23})\cdot\mathcal{O}_{13}(\theta_{13})$.
		The mixing angle $\theta_{12}$ is then given by
		\begin{equation}
			\tan(2\theta_{12})=\frac{2(2\lambda_{H\Delta}v_0v_1-\mu v_0)}{4\lambda_{\Delta}v^2_1+\frac{\mu v^2_0}{2v_1}-2\lambda_{H}v_0^2}.
		\end{equation}
		By taking $\lambda_H\sim0.13$, $\lambda_{\Delta}=\mathcal{O}(1)$, $\lambda_{H\Delta}=\mathcal{O}(10^{-2})$,  $v_1=\mathcal{O}(1)$ eV\footnote{In section \ref{sec:longliveddm}, we will show that the required vev of the scalar triplet $\Delta$ to 
			be $v_1<\mathcal{O}(1)$ keV to keep lifetime of DM, $\tau_{\rm DM}>10^{27}$ s.}, and $\mu=\mathcal{O}(10^{-10})$ GeV, it leads to the $H-\Delta$ mixing: $\theta_{12}\equiv\theta_{H\Delta} \approx \mathcal{O}(10^{-12})$. Therefore, for all practical purposes $\theta_{12}$ is effectively set to zero {\it. i.e.} $\theta_{12}=0$. This also leads to the $\Delta-\Phi$ mixing angle, $\theta_{23}$ to be
		\begin{equation}			\tan(2\theta_{23})\simeq\frac{2\sqrt{2}\rho_2v_1}{2\lambda_{\Phi}u^2-\frac{1}{2\sqrt{2}}\frac{\rho_{1}v^2_0+2\rho_2v^2_1}{u}-4\lambda_{\Delta}v^2_1-\frac{\mu v^2_0}{2v_1}}.
		\end{equation}
		By taking $\rho_1=-\mathcal{O}(1)$ MeV, $\rho_2=\mathcal{O}(1)$ MeV, and $\lambda_{\Phi}u^2\sim \mathcal{O}(1)$ $\rm MeV^2$, we get $\theta_{23}\approx\mathcal{O}(10^{-16})$. Thus the mixing between $\Delta$ and $\Phi$ can be safely neglected. The angle $\theta_{13}$ which characterizes the mixing between SM Higgs and $\Phi$ is found to be
		\begin{equation}			\tan(2\theta_{13})\simeq\frac{2(\frac{\rho_1}{\sqrt{2}} v_0+\lambda_{H\Phi}uv_0)}{2\lambda_{\Phi}u^2-\frac{1}{2\sqrt{2}}\frac{\rho_{1}v^2_0+2\rho_2v^2_1}{u}-2\lambda_H v_0^2}\equiv\tan 2\gamma.
		\end{equation}
		Assuming $\lambda_{H\Phi}\sim \mathcal{O}(1)$ and the other parameters as defined above, we get $\theta_{13}\approx\mathcal{O}(10^{-5})$. Larger values of $\theta_{13}$ are also possible by adjusting the other parameters. This results in two mass eigenstates $h_1$ and $h_2$ with masses $M_{h_1}$ and $M_{h_2}$ given as
		\begin{eqnarray}
			M^2_{h_1}&\simeq& \lambda_{H}v_0^2+\lambda_\phi u^2+\sqrt{\lambda_{H}^2v_0^4+\lambda_\phi^2 u^4+v_0^2u^2(\lambda_{H\Phi}-2\lambda_H\lambda_\Phi)},\nonumber\\
			M^2_{h_2}&\simeq& \lambda_{H}v_0^2+\lambda_\phi u^2-\sqrt{\lambda_{H}^2v_0^4+\lambda_\phi^2 u^4+v_0^2u^2(\lambda_{H\Phi}-2\lambda_H\lambda_\Phi)}\equiv M^2_\phi,
		\end{eqnarray}
		where we set $\rho_1,\rho_2\rightarrow0$ for simplicity. We identify $h_1$ to be the SM Higgs with mass $M_{h_1}$ = 125 GeV, and $h_2$ be dominantly $\Phi$ with mass $M_{h_2}=M_\phi$. In section \ref{sec:depele}, we redefine this mixing angle $\theta_{13}$ to be $\gamma$. 
		Since $\theta_{12}$ and $\theta_{23}$ are negligible compared to $\theta_{13}$, $\Delta$ effectively decouples from $H$ and $\Phi$. Therefore, for all practical purposes, the mass eigenstate of $\Delta$ can be taken as the flavor eigenstate with the mass given by
		\begin{eqnarray}
			M_{\Delta}=\sqrt{\frac{\mu v^2_0}{2v_1}+4\lambda_{\Delta}v^2_1}.
		\end{eqnarray}
		
		From Eq. \ref{lagrangian}, the Lagrangian for triplet scalar $\Delta$ can be expanded as: 
		\begin{eqnarray}
			\mathcal{L}_{\Delta}&=&(\partial^\mu \delta^{+^*})(\partial_\mu \delta^{+})+(\partial^\mu \delta^{-^*})(\partial_\mu \delta^{-})+(\partial^\mu \delta^{0^*})(\partial_\mu \delta^{0})+ \mathcal{L}_{int},\  
		\end{eqnarray}
		where
		\begin{eqnarray}
			\mathcal{L}_{int}&=&[i4g(\partial^\mu\delta^0)W^+_\mu\delta^-+i4g(\partial_\mu\delta^+)W_\mu^-\delta^0+i4g(\partial^\mu\delta^-)(c_w Z_\mu+s_w A_\mu)\delta^+ +h.c.]+8g^2W^{+\mu}W_\mu^-[\delta^-\delta^+  \nonumber\\
			&+& \delta^0\delta^0] -4g^2W^{-\mu}W^-_\mu\delta^+\delta^+-[8g^2W^+_\mu(c_w Z_\mu+s_w A_\mu)\delta^-\delta^0+h.c.]+8g^2(c_w Z_\mu+s_w A_\mu)(c_w Z_\mu+\nonumber\\
			& &s_w A_\mu)\delta^+\delta^-+ \frac{1}{2}\lambda_{H\Delta}\big[2v_0h^2(\delta_0^2+2\delta^+\delta^-)+(h^2\delta_0^2+2h^2\delta^+\delta^-)\big],\
		\end{eqnarray}
		where $s_w(c_w)=\sin(\theta_w)(\cos(\theta_w))$, $\theta_w$ is the Weinberg angle.\\
		\begin{figure}[h]
			\centering
			\includegraphics[width=7.5cm,height=7.5cm]{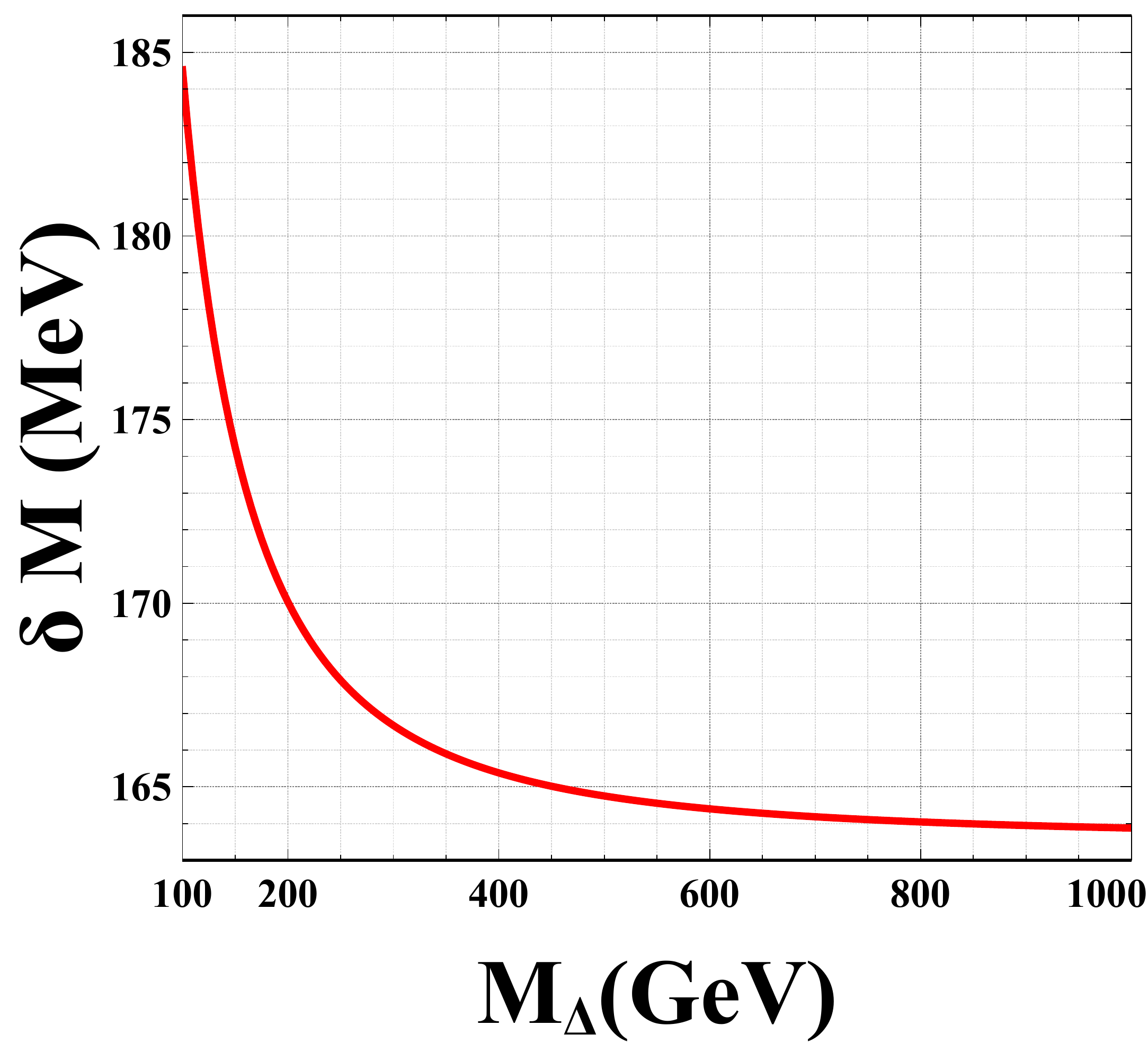}
			\caption{Variation of scalar triplet mass splitting $\delta M$ with respect to $M_\Delta$.}
			\label{fig:dmvsm}
		\end{figure}
		
		The mass splitting between $\delta^+$ and $\delta^0$ due to EW corrections is given as,~\cite{Cirelli:2005uq}
		\begin{equation}
			\delta M= \frac{\alpha_2 M_{\Delta}}{4\pi}\left[s_w^2 f\left(\frac{M_Z}{M_\Delta}\right)+{f\left(\frac{M_W}{M_\Delta}\right)-f\left(\frac{M_Z}{M_\Delta}\right)}\right],\
		\end{equation}
		where
		\begin{equation}
			f(r)=-\frac{r[2r^3 \ln r +(r^2-4)^{\frac{3}{2}}\ln A]}{4}
		\end{equation}
		\begin{equation}
			A=\frac{(r^2-2-r\sqrt{r^2-4})}{2}
		\end{equation}
		and $\alpha_2=\frac{g^2}{4\pi}$. In Fig. \ref{fig:dmvsm}, We have shown $\delta M$ as a function of scalar triplet mass $M_\Delta$. 
		We see that the maximum mass splitting that can be achieved for $M_\Delta = 100$ GeV is about 185 MeV. This implies that the decay of 
		$\delta^\pm$ is phase space suppressed and may give displaced vertex signature. We will discuss more about it in section \ref{collider_signature}.

		\section{Neutrino Mass}\label{sec:numass}
		The sub-eV masses of light neutrinos can be obtained through the type-III seesaw 
		mechanism \cite{Foot:1988aq,Ma:1998dn}. The Lagrangian responsible for neutrino mass generation is
		\begin{eqnarray}
			-\mathcal{L}_{\nu- mass} & \supset  \frac{1}{2}M_\Sigma Tr[\overline{\Sigma^c}\Sigma]+ \sqrt{2} y_{_\Sigma} \overline{L}  \Tilde{H} \Sigma + h.c.
		\end{eqnarray}
		In the effective theory, the masses of light neutrinos can be obtained as:
		\begin{equation}
			-\mathcal{L}_{\nu - mass}=\begin{pmatrix}
				\overline{\nu_{L}} & \overline{(\Sigma^{0})^c}
			\end{pmatrix} \begin{pmatrix}
				0 & \frac{y_{_\Sigma} v_0}{\sqrt{2}}\\
				\frac{y_{_\Sigma} v_0}{\sqrt{2}} & M_\Sigma
			\end{pmatrix} \begin{pmatrix}
				(\nu_{L})^c\\
				\Sigma^0
			\end{pmatrix}.\
		\end{equation}
		Diagonalizing the above mass matrix, we get the Majorana mass of the light neutrino and the heavy fermion triplet as:
		\begin{equation}
			m_\nu \simeq - \frac{y_{_\Sigma}^2 v_0^2}{2 M_\Sigma},\
		\end{equation}
		and 
		\begin{equation}
			M^\prime_\Sigma \simeq M_\Sigma + \frac{y_{_\Sigma}^2 v_0^2}{2 M_\Sigma}.\,
		\end{equation}
		
		Considering 3 generations of the heavy fermion triplets, the light neutrino mass matrix is given by
		\begin{equation}
			(m_\nu)_{\alpha \beta} = (m_D)_{\alpha i} \left( M^{-1}\right)_{ij} (m_D^T)_{j\beta},\
			\label{Eq:massmatrix}
		\end{equation}		
		where $m_D=\frac{y_{_\Sigma}~v_0}{\sqrt{2}}$ is the $3\times 3$ Dirac mass matrix and $M$ is the $3\times 3$ triplet fermion mass matrix.
		Using the Casas-Ibarra parametrization \cite{Casas:2001sr}, we can calculate the Yukawa coupling matrix as
		\begin{equation}
			y_{_\Sigma}=\frac{\sqrt{2}}{v_0}\left(U_{\rm PMNS}.\sqrt{\hat{m}_\nu}.R.\sqrt{\hat{M_\Sigma}}\right),\
			\label{Eq:casasibarra}
		\end{equation}
		where $U_{\rm PMNS}$ is the lepton mixing matrix, $\hat{m}_\nu$ is $3\times3$ diagonal light neutrino mass matrix with eigenvalues $m_1$, $m_2$, and $m_3$; $\hat{M}_\Sigma$ is $3\times3$ diagonal triplet fermion mass matrix with eigenvalues $M_{\Sigma_1}$, $M_{\Sigma_2}$, and $M_{\Sigma_3}$, and R is an arbitrary rotation matrix with complex rotation angle, $\theta$.

		\section{Leptogenesis and Asymmetric Dark Matter}\label{sec:admlepto}
		\subsection{Co-genesis scenario}
\begin{figure}[h]
			\centering
			\includegraphics[scale=0.1]{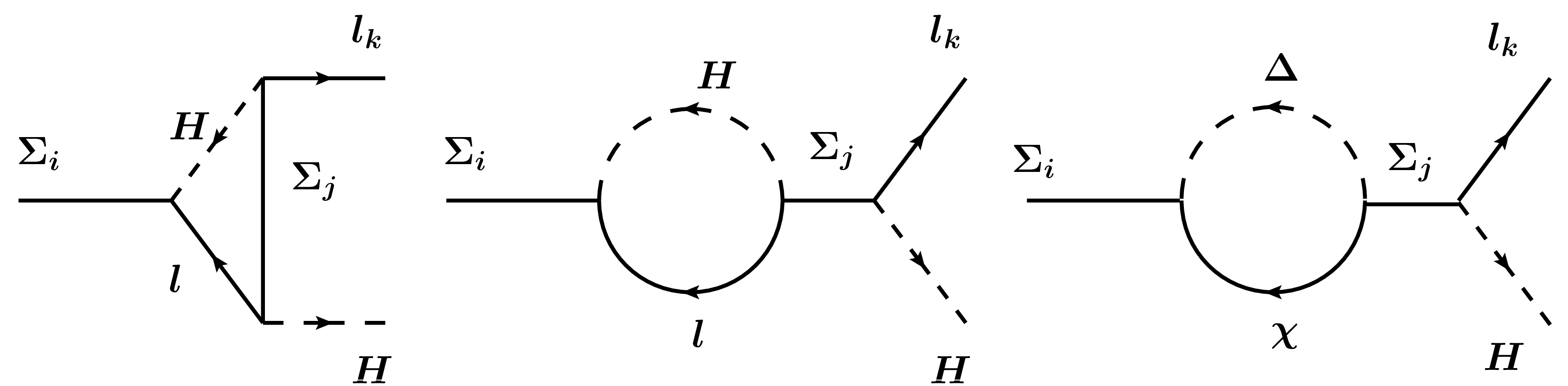}
			\includegraphics[scale=0.1]{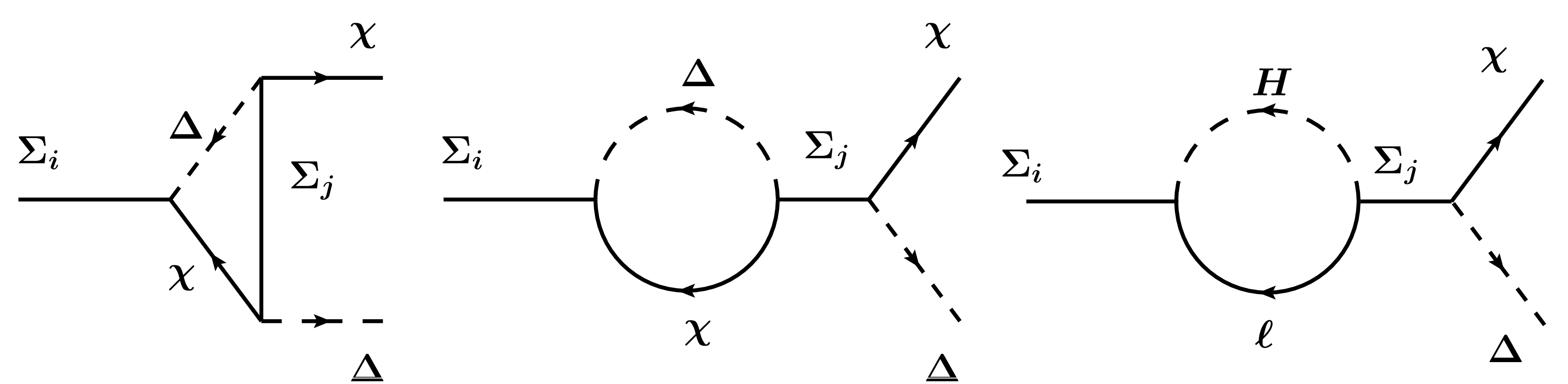}
			\caption{Loop level diagrams for generating $CP$ asymmetries in both visible and dark sectors through the decay of fermion triplets.}
			\label{fig:cpasym}
		\end{figure}

		We assume the hypercharge zero fermion triplets to be much heavier than the SM fields so that in the effective theory, 
		they can naturally give rise to Majorana masses of light neutrinos through the type-III seesaw~\cite{Foot:1988aq}. Unlike the 
		case of type-I seesaw, here, the triplet fermions have additional gauge interactions which keep them in thermal equilibrium at a temperature above their mass scales. In an expanding Universe, as the temperature falls below the mass scale of a triplet 
		fermion, the latter goes out of equilibrium and decays to SM fields (lepton and Higgs). If the corresponding decay mode 
		violates $C$ and $CP$, then a net lepton asymmetry can be generated~\cite{Hambye:2012fh,Vatsyayan:2022rth,Fischler:2008xm}. 
		Here, we consider that the fermion triplets decay simultaneously to visible and dark sectors, generating both matter and DM ($\chi$) asymmetries. We consider a hierarchical mass spectrum for the triplet fermions: $M_{\Sigma_1}<M_{\Sigma_2}<M_{\Sigma_3}$. Any asymmetry produced by $\Sigma_2$ and $\Sigma_3$ decays is washed out by lepton number violating interactions mediated by $\Sigma_1$. Consequently, the final asymmetry stems from $\Sigma_1$ decay via $\Sigma_1 \to LH$ (visible sector) and $\Sigma_1 \to \chi \Delta$ (dark sector). The $CP$ asymmetry parameters for the visible sector ($\epsilon_L$) and dark sector ($\epsilon_\chi$) arising from the interference between tree-level and one-loop diagrams in $\Sigma_1$ decay, as shown in Fig. \ref{fig:cpasym}, are given as \cite{Hambye:2003rt,Falkowski:2011xh,Borah:2024wos}
		\begin{eqnarray}
			\epsilon_L&=&\frac{\Gamma(\Sigma_1\rightarrow L H)-\Gamma(\Sigma_1\rightarrow \overline{L} H^\dagger)}{\Gamma_{\Sigma_1}}\nonumber\\&=&-\frac{1}{16\pi}\frac{M_1}{((y^\dagger y)_{11}+y_{\chi_1} y^*_{\chi_1})}\sum_j\frac{1}{M_j}\big\{{\rm Im}[(y^\dagger y)^2_{1j}]+2{\rm Im}[(y^\dagger y)_{1j}y_{\chi_1}y^*_{\chi_j}]\big\} \\
			\epsilon_\chi&=&\frac{\Gamma(\Sigma_1\rightarrow \chi \Delta)-\Gamma(\Sigma_1\rightarrow \overline{\chi} \Delta^\dagger)}{\Gamma_{\Sigma_1}}\nonumber\\&=&-\frac{1}{16\pi}\frac{M_1}{((y^\dagger y)_{11}+y_{\chi_1} y^*_{\chi_1})}\sum_j\frac{1}{M_j}\big\{{\rm Im}[(y_{\chi_1}y^*_{\chi_j})^2]+2{\rm Im}[(y^\dagger y)_{1j}y_{\chi_1}y^*_{\chi_j}]\big\}.\
		\end{eqnarray}
		The total decay rate $\Gamma_{\Sigma_1}$ is the sum of partial decay rates:
		\begin{equation}
			\Gamma_{\Sigma_1}=\Gamma(\Sigma_1\rightarrow L H)+\Gamma(\Sigma_1\rightarrow \chi \Delta),\
			\label{Eq:Decayrate}
		\end{equation}
		where 
		\begin{equation}
			\Gamma(\Sigma_1\rightarrow L H)=\frac{\tilde{m}_1 M_{\Sigma_1}^2}{8\pi v_0^2} ~ ; ~ \Gamma(\Sigma_1\rightarrow \chi \Delta)=\frac{\tilde{m}_{dm} M_{\Sigma_1}^2}{8\pi v_0^2}.
		\end{equation}
		Here $\tilde{m}_1$ is the effective neutrino mass, defined as
		\begin{equation}
			\tilde{m}_1=\frac{(m^\dagger_D m_D)_{11}}{M_{\Sigma_1}},\,
		\end{equation}
		and the effective mass parameter for the dark sector is
		\begin{equation}
			\tilde{m}_{dm}= \frac{y_{\chi}^2 v_0^2}{4M_{\Sigma_1}}.\,
		\end{equation}
		The branching ratios for the two sectors are defined as,
		\begin{equation}
			Br_l = \frac{\Gamma(\Sigma_1\rightarrow L H)}{\Gamma_{\Sigma_1}} ~ ; ~
			Br_\chi = \frac{\Gamma(\Sigma_1\rightarrow \chi \Delta)}{\Gamma_{\Sigma_1}}.\
		\end{equation}

\begin{figure}[h]
			\includegraphics[scale=0.43]{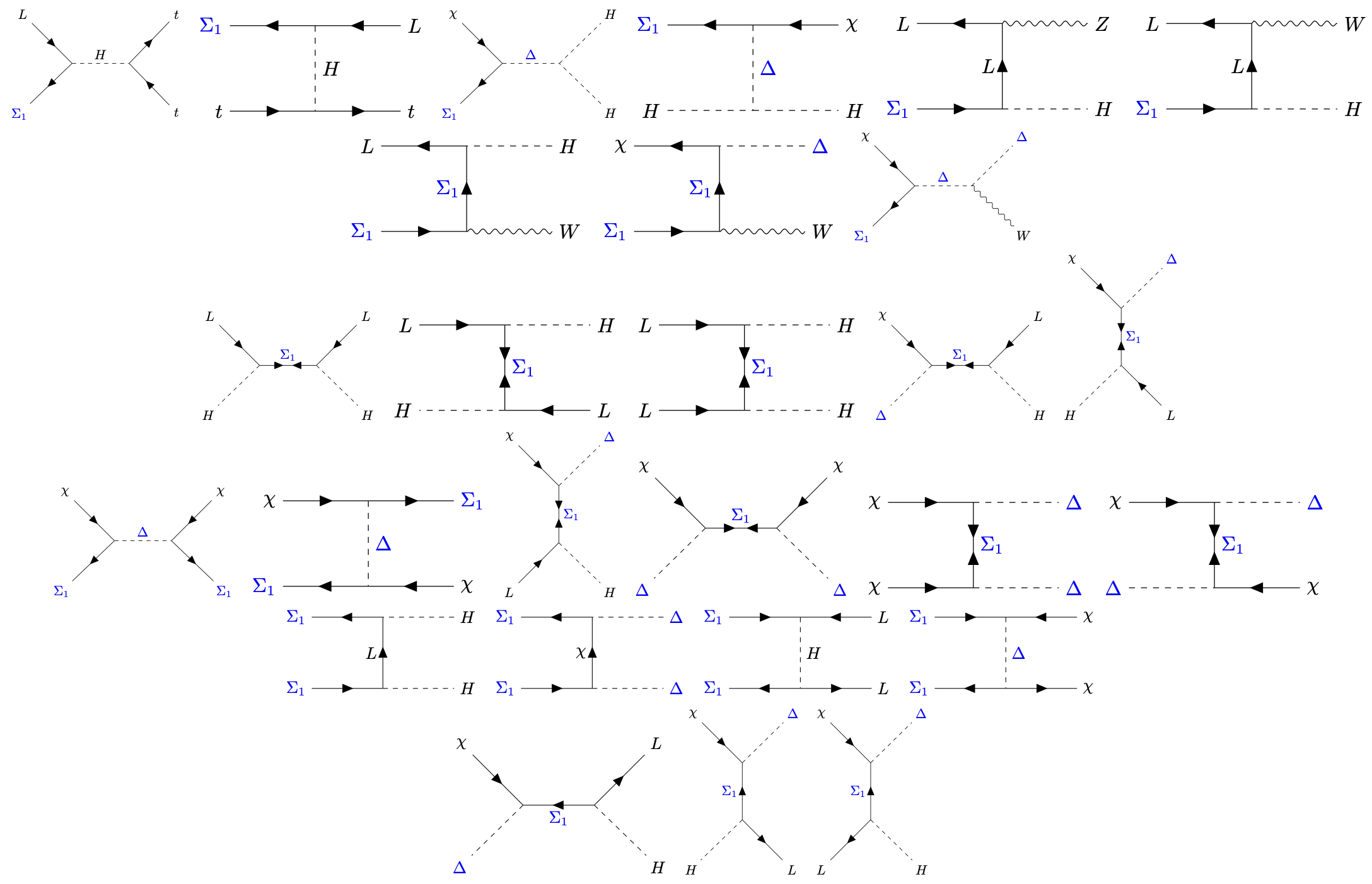}
			\caption{Feynman diagrams for $\Delta$L=1 scatterings are given in 1st row and 2nd row. Similarly, $\Delta$L=2 scatterings are given in the 3rd row and 4th row. The $\Delta$L=0 scatterings are shown in the 5th row, which contributes to the evolution of the triplet. $\Delta L=0$ transfer processes are shown in the 6th row, which transfers the asymmetries between the visible and dark sectors.}
			\label{fig:newdeltal2}
		\end{figure}

		The evolution of the number density $\Sigma_1$ and the asymmetries generated via its decay can be obtained by solving 
		the relevant Boltzmann equations \cite{Buchmuller:2004nz}. We define a dimensionless variable $Y_x=\frac{n_x}{n_\gamma}$, 
		where $n_x$ is the number density of any species $x$ and $n_\gamma$ is the photon number density which is given as $n_\gamma=\frac{2\zeta(3)T^3}{\pi^2}$. The relevant processes that dictate the evolution of the number density $\Sigma_1$ and the asymmetries in both the sectors are the decays, inverse 
		decays, the gauge scatterings, lepton number conserving ($\Delta$L = 0)  transfer processes, and the lepton number violating scatterings ($\Delta$L=1,2). We have shown all processes in Fig. \ref{fig:newdeltal2}.
		
		The Boltzmann equations for tracking the evolution of $\Sigma_1$ as well as the asymmetries can be written as:
		\begin{eqnarray}
			\frac{d Y_{\Sigma_1}}{ d z}&=&- \frac{\Gamma_D}{{\rm H} z} (Y_{\Sigma_1} - Y_{\Sigma_1}^{eq})- \frac{\Gamma_1}{{\rm H}z} (Y_{\Sigma_1} - Y_{\Sigma_1}^{eq}) -\frac{(\Gamma_0+\Gamma_A)}{{\rm H}z} \frac{\big(Y^2_{\Sigma_1} - (Y_{\Sigma_1}^{eq})^2\big)}{Y_{\Sigma_1}^{eq}},\label{eq:BEY}
		\end{eqnarray}
		\begin{eqnarray}
			\frac{d Y_{\Delta L}}{d  z}&=&\epsilon_L~ \frac{\Gamma_D}{{\rm H}z} ~(Y_{\Sigma_1} - Y_{\Sigma_1}^{eq}) -\left( \frac{1}{2} \frac{\Gamma_D}{{\rm H}z} ~\frac{Y_{\Sigma_1}^{eq}}{Y_l^{eq}}~ Br_l + \frac{(\Gamma^W_{1L}+\Gamma^W_{2L}+\Gamma_{2W})}{{\rm H}z} \right) Y_{\Delta L}\nonumber\\&-&\frac{\Gamma_{\Sigma_1}}{\rm H_1} Br_l Br_\chi \left(I_{T_+}(z)(Y_{\Delta L}+Y_{\Delta \chi})+ I_{T_-}(z)(Y_{\Delta L}-Y_{\Delta \chi})\right),\label{eq:BEYl}
		\end{eqnarray}
		\begin{eqnarray}
			\frac{d Y_{\Delta \chi}}{d z}&=&\epsilon_\chi~ \frac{\Gamma_D}{{\rm H}z} ~(Y_{\Sigma_1} - Y_{\Sigma_1}^{eq}) -\left( \frac{1}{2} \frac{\Gamma_D}{{\rm H}z} ~\frac{Y_{\Sigma_1}^{eq}}{Y_\chi^{eq}}~ Br_\chi + \frac{\Gamma^W_{1\chi}+\Gamma^W_{2\chi}+\Gamma_{2W}}{{\rm H}z} \right) ~Y_{\Delta \chi}\nonumber\\&-&\frac{\Gamma_{\Sigma_1}}{\rm H_1} Br_l Br_\chi \left(I_{T_+}(z)(Y_{\Delta \chi}+Y_{\Delta L})+ I_{T_-}(z)(Y_{\Delta \chi}-Y_{\Delta L})\right),\
			\label{eq:BEYx}
		\end{eqnarray}
		where $z=\frac{M_{\Sigma_1}}{T}$ , $T$ is the temperature of thermal bath, ${\rm H} = 1.66\sqrt{g_*}\frac{T^2}{M_{\rm pl}}$ is the Hubble parameter, ${\rm H_1}$ denotes the ${\rm H}(z=1)$, $g_*$ is the effective number of degrees of freedom, $M_{\rm pl}$ is the Planck mass and ``$eq$'' denotes equilibrium value. In the Boltzmann equation, $\Gamma_0$ and $\Gamma_1$ represent the scatterings with $\Delta L=0$ and $\Delta L=1$, respectively, contributing to the evolution of fermion triplet abundances. $\Gamma^{W}_{1L}$ and $\Gamma^W_{2L}$ denote the washout processes with $\Delta L=1$ and $\Delta L=2$, respectively, involving only visible sector particles in the asymmetry erasure. Conversely, $\Gamma^{W}_{1\chi}$ and $\Gamma^W_{2\chi}$ represent the washout processes with $\Delta L=1$ and $\Delta L=2$, respectively, incorporating only dark sector particles. $\Gamma_{2W}$ signifies the washout process with $\Delta L=2$ involving both visible and dark sectors. $\Gamma_D$ stands for the total decay rate of $\Sigma_1$, as mentioned in Eq. \ref{Eq:Decayrate}, while $\Gamma_A$ denotes the rate of gauge interactions. The last terms in Eq \ref{eq:BEYl} and \ref{eq:BEYx} represent the transfer terms which are due to the $\Delta L$=0 processes. Analytical expressions for all the scattering cross-sections including $I_{T_+}$ and $I_{T_-}$ are given in Appendix \ref{appen1}. $Y_{\Sigma}^{\rm eq}=\frac{3}{8}z^2 K_2(z)$, $K_2(z)$ is the modified Bessel function of second kind, $Y_{\chi}^{\rm eq}=\frac{3}{8}*(\frac{M_\chi}{M_{\Sigma_1}})^2* z^2 K_2(\frac{M_\chi}{M_{\Sigma_1}}z)$  and $Y_l^{\rm eq}=\frac{3}{4}$. 
		\begin{figure}[h]
			\centering
			\includegraphics[width=7.5cm,height=7.5cm]{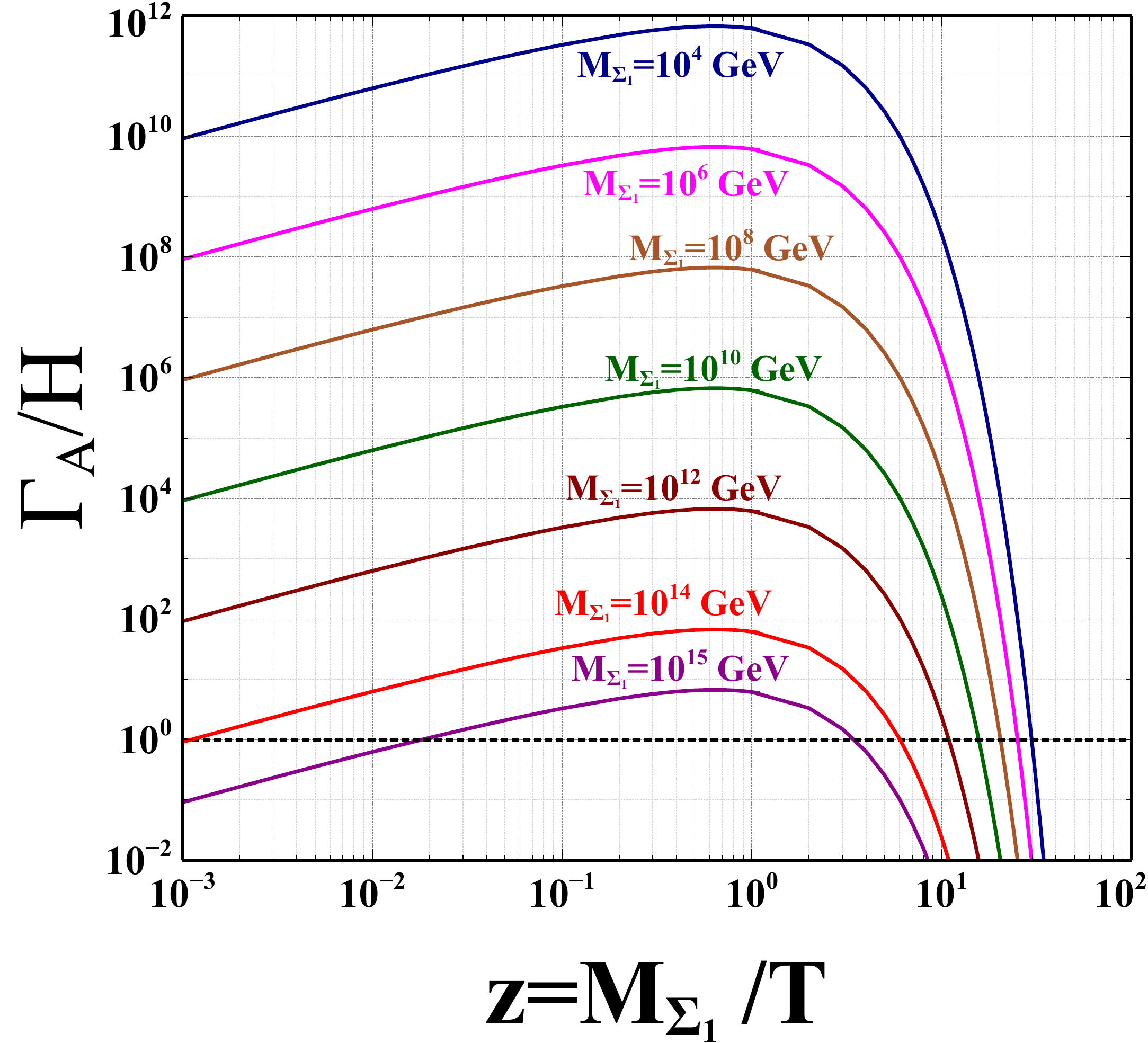}
			\caption{Evolution of interaction rate of gauge scatterings as compared to Hubble for different masses of ${\Sigma_1}$. The horizontal dashed black line represents $\Gamma_A/{\rm H}=1$.}
			\label{fig:gaugeint}
		\end{figure}
		
		Let us first focus on the gauge interactions. The processes: $~\Sigma_1 \overline{\Sigma_1^\prime} \leftrightarrow G G^\prime, ~ f \bar{f}$, contribute to the gauge scattering, where $G$,$G^\prime$ are SM gauge bosons. As the gauge interactions do not involve any lepton number violating processes, they will not contribute to the asymmetry generation in either sector. Rather, we will see that these gauge interactions will thermalize the triplet fermions quickly in the early Universe. We compare the total rate of gauge scattering processes with the Hubble parameter in Fig. \ref{fig:gaugeint}\footnote{The gauge scattering cross-section used in this work is adopted from~\cite{Hambye:2003rt,AristizabalSierra:2010mv} which is roughly 7 times larger than the one used in~\cite{Hambye:2012fh}. 
			Following the convention for thermal averaged cross-section given in \cite{Gondolo:1990dk}, we found that the $\Gamma_A/H$  in this work is roughly one order of magnitude larger compared to that in~\cite{Hambye:2012fh} for $0<z<10$. 
		} . 
		We see from Fig \ref{fig:gaugeint} that for 
		$z \le  1$, gauge interaction rates for $M_{\Sigma_1}<10^{14}$ GeV are larger than the Hubble rate, and hence the corresponding processes are in thermal equilibrium. Therefore, even if we start with zero initial abundance of $\Sigma_1$ with $M_{\Sigma_1} < 10^{14}$ GeV, 
		due to the gauge interactions, $\Sigma_1$ will reach thermal equilibrium very quickly. Thus 
		, for $M_{\Sigma_1}$ below $10^{14}$ GeV, the produced asymmetry is independent of the initial condition on the triplet number density. 
		In other words, irrespective of the values of Yukawa couplings $y_{_\Sigma}$ and $y_\chi$, the initial number density of fermion triplets 
		can be set to their equilibrium value. 
		
		Now we turn to analyze the evolution of the asymmetries produced by the decay of $\Sigma_1$ to $LH$ and $\chi \Delta$. 
		Note that the first terms on right-hand side (RHS) of Eq. \ref{eq:BEYl} and Eq. \ref{eq:BEYx} are the source terms for the 
		asymmetry which come from the $CP$ violating decay of $\Sigma_1$. The other terms on the RHS of Eq. \ref{eq:BEYl} and Eq. \ref{eq:BEYx} contribute to the washout. Depending on the strength of the interaction rates, washout regimes can be classified into two categories, namely, weak washout (WW) and strong washout (SW). The parameter which can distinguish between the two washout regimes can be defined as
		\begin{equation}
			K=\frac{\Gamma( z=\infty)}{{\rm H}(z=1)}=\frac{\tilde{m}_1}{m_*},\
		\end{equation}
		where $m_*$ is the equilibrium neutrino mass, defined as \cite{Buchmuller:2004nz},
		\begin{equation}
			m_*=\frac{16\pi^{5/2}\sqrt{g_*}v_0^2}{3\sqrt{5}M_{pl}}\simeq 1.08\times10^{-3}~ \rm eV.\
		\end{equation}
		When $K\gg1$, {\it i.e.} $\Gamma\gg\rm H$, the processes fall into the strong washout regime, while for $K\ll1$, {\it i.e.} $\Gamma\ll\rm H$, the processes fall into weak washout regime.
		\begin{figure}[h]
			\centering
			\includegraphics[width=9cm,height=7.5cm]{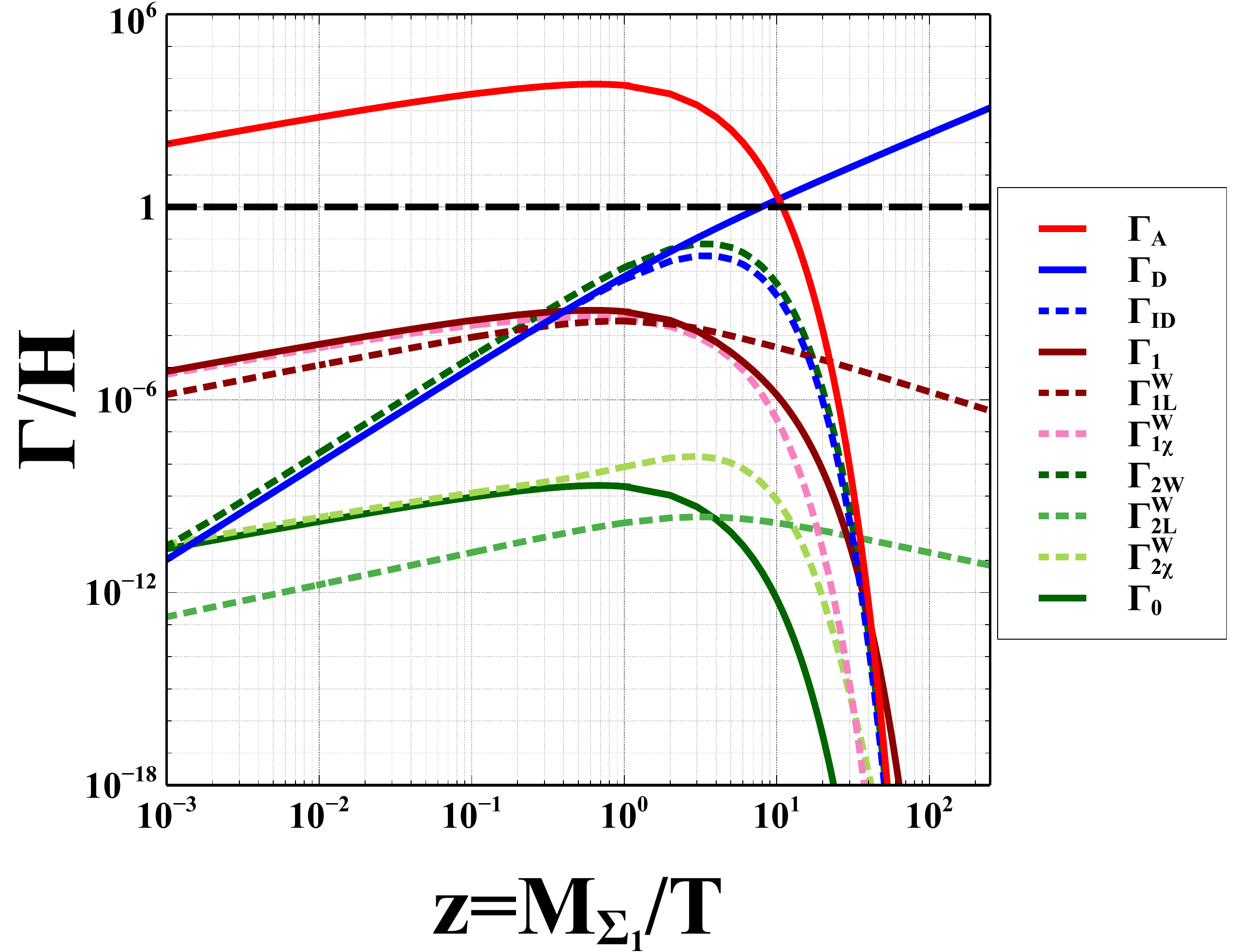}
			\caption{Comparison of various interaction rates with the Hubble parameter H.  The corresponding interaction rates are given in Appendix \ref{appen1}}.
			\label{fig:intrate}
		\end{figure}
		In Fig \ref{fig:intrate} , we have shown interaction rates of the different processes involved in leptogenesis as a function of $z$. For this comparison, we have taken $M_{\Sigma_1}=10^{12}$ GeV, $M_\Delta=10^3$ GeV, $M_\chi=1$ GeV, $\tilde{m}_1=2.1\times10^{-14}$ GeV, $\tilde{m}_{dm}=2.1\times10^{-14}$ GeV, $y_\chi=1.177\times10^{-3}$ and $y_{_\Sigma}=8.331\times10^{-4}$. Note that the value of 
		$y_{_\Sigma}=8.331\times10^{-4}$ is calculated using Eq \ref{Eq:casasibarra} so that it satisfy the neutrino mass. We see that in 
		the very early universe {\it, i.e.} $z\ll1$, all the lepton number violating interaction rates are much smaller than the gauge interactions. 
		Moreover, in the early Universe, the decay and inverse decay rates are evolving at the same rates, and after $z\sim3$, due to the Boltzmann suppression, the inverse decay rate falls exponentially.
		\begin{figure}[h]
			\centering
			\includegraphics[width=7.5cm,height=7.5cm]{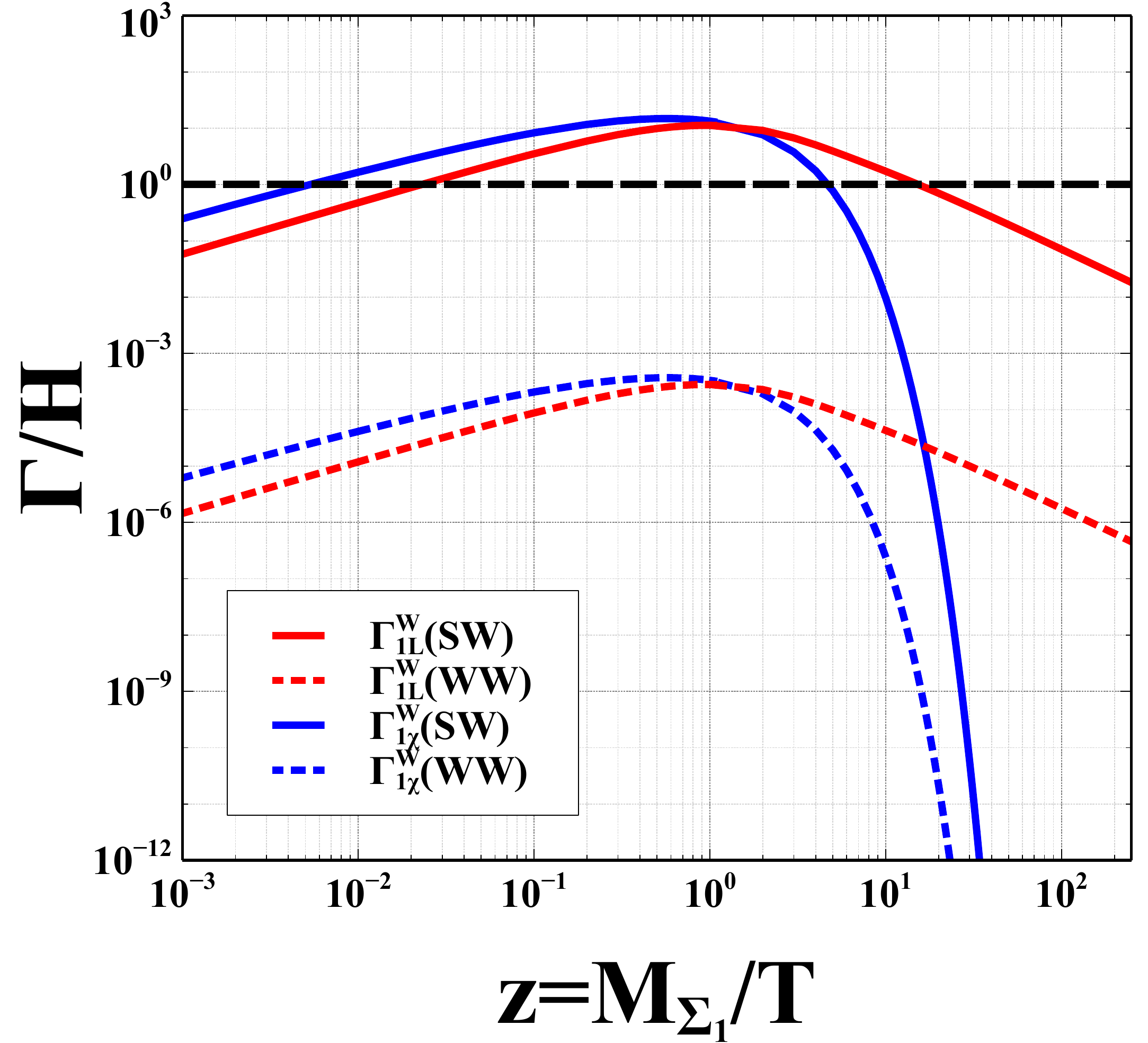}~~
			\includegraphics[width=7.5cm,height=7.5cm]{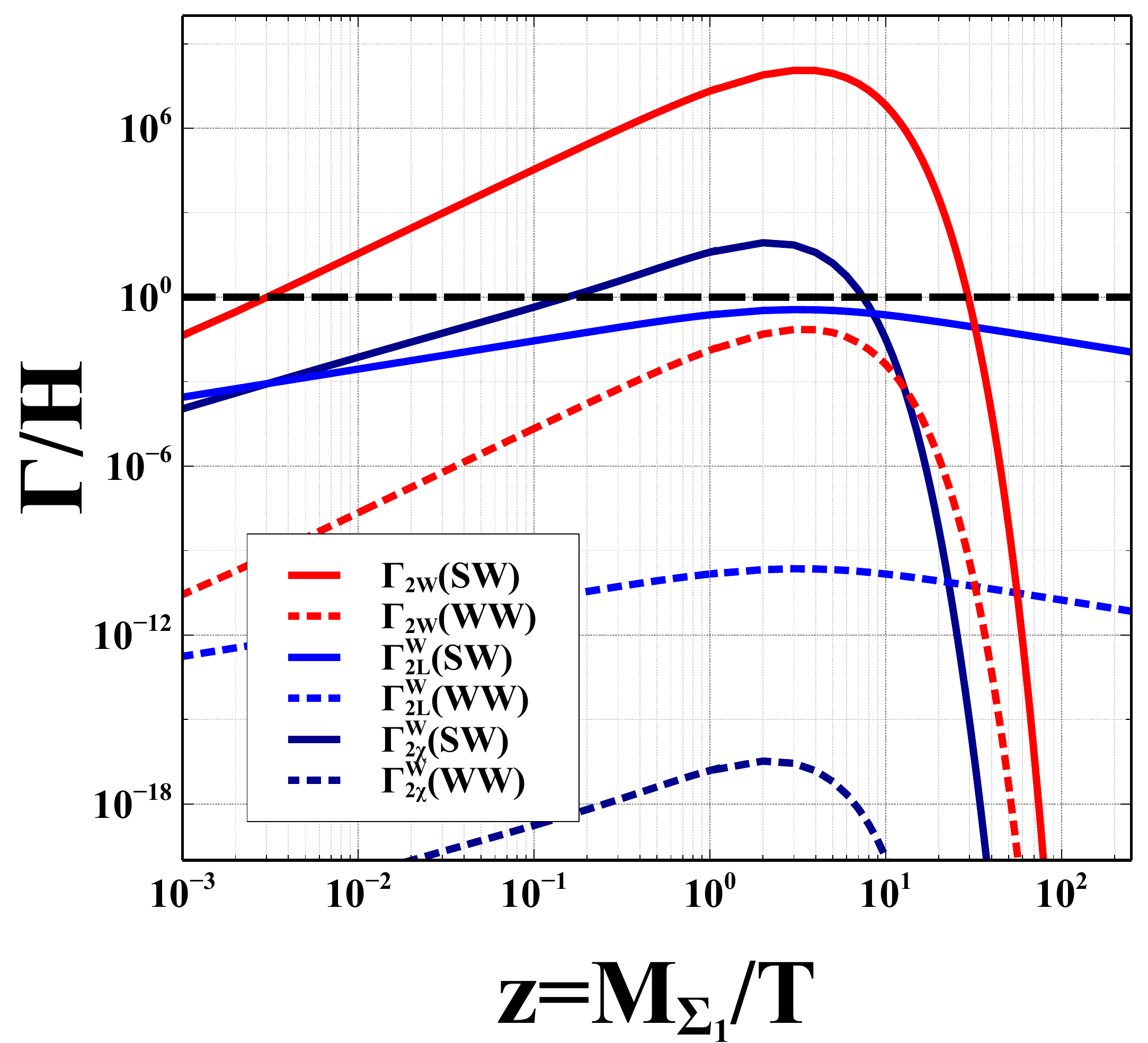}
			\caption{Comparison of $\Delta$L= 1 [left] and $\Delta$L=2 [right] washout interactions in strong and weak washout regimes.}
			\label{fig:deltaL2}
		\end{figure}
		We compare the interaction rates of $\Delta$L=1 and $\Delta$L=2 washout processes in the left and right panel of Fig~\ref{fig:deltaL2} respectively, considering both strong and weak washout regimes. We have fixed the masses as $M_{\Sigma_1}=10^{12}$ GeV, $M_\Delta=10^3$ GeV, $M_\chi=1$ GeV and the Yukawa couplings are fixed to be $y_{_\Sigma}=0.1667$, $y_\chi=0.2358$ for strong washout and  $y_{_\Sigma}=8.331\times10^{-4}$, $y_\chi=1.177\times10^{-3}$ for weak washout.
		
		\subsection{Gauge vs Yukawa regime}
		In Fig. \ref{fig:gaugeint}, it is evident that gauge interactions keep $\Sigma_1$ in thermal equilibrium for a prolonged duration as $M_{\Sigma_1}$ decreases. The processes contributing to the establishment of thermal equilibrium for triplets include gauge interactions, inverse decay, and scatterings. When $\tilde{m}_1$ (or $\tilde{m}_{dm}$) is small, triplets remain in equilibrium primarily due to gauge interactions alone, constituting what is commonly termed the "Gauge regime." Within this regime, triplets adhere to the equilibrium number density until the gauge interactions go out-of-equilibrium, which occurs relatively later for lighter triplets in comparison to the heavier ones. This can be easily read from Fig \ref{fig:gaugeint}. Since the lighter triplets remain in equilibrium for a longer period, their number density suffers larger exponential suppression in comparison to the heavier ones.  As a result, the lighter triplets need larger $CP$ asymmetries in comparison to heavier ones to account for the observed baryon asymmetry.
		
		Even if the gauge interactions prove insufficient to maintain thermal equilibrium for relatively heavier triplets, they can still be in equilibrium through sufficient inverse decay and Yukawa-mediated scatterings by choosing large $\tilde{m}_1$ (or $\tilde{m}_{dm}$). This is called the "Yukawa regime," in contrast to the "Gauge regime."
		In the typical type-III leptogenesis case, the Gauge and Yukawa regimes are defined by only two free parameters: $M_{\Sigma_1}$ and $\tilde{m}_1$. However, in our case, triplets can decay into both the visible and dark sectors, characterized by $\tilde{m}_1$ and $\tilde{m}_{dm}$ respectively. Both $\tilde{m}_1$ and $\tilde{m}_{dm}$ contribute to maintaining triplets in equilibrium through inverse decay, resulting in a total of three free parameters to define the Gauge and Yukawa regimes. 
		
\begin{figure}[h]
			\centering
			\includegraphics[width=7.5cm,height=7.5cm]{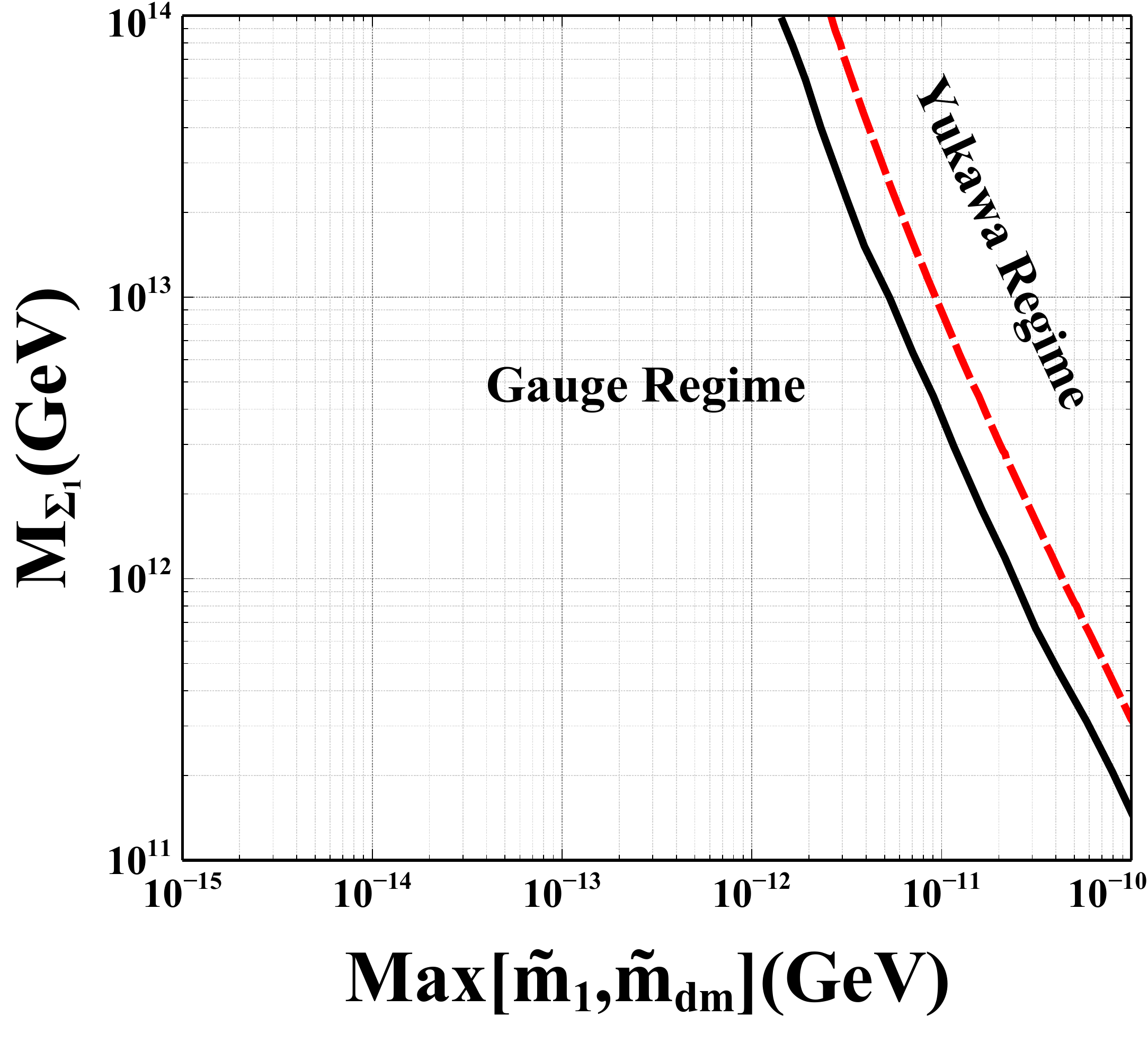}
			\caption{Delineation of Gauge versus Yukawa regime in the plane of $M_{\Sigma_1}$ and ${\rm Max}[\tilde{m_1},\tilde{m}_{ dm}]$.}
			\label{fig:GY}
		\end{figure}

		To delineate the Gauge and Yukawa regimes, we proceed as follows: first, we calculate the value of $z=z_A$ at which gauge interactions decouple for a fixed $M_{\Sigma_1}$. Then, we determine the minimum value of $Max\{\tilde{m}_1,\tilde{m}_{dm}\}$ for which Yukawa interactions remain active at $z=z_A$, ensuring $(\Gamma_D n^{eq}_{\Sigma_1})/({\rm H}n^{eq}_l)|_{z=z_A}\gtrsim1$. In Fig. \ref{fig:GY}, we illustrate the two regimes in the $M_{\Sigma_1}-Max\{\tilde{m}_1,\tilde{m}_{dm}\}$ plane. The black line represents our case, while the red dashed line corresponds to the typical type-III seesaw leptogenesis scenario ({\it, i.e.} $\tilde{m}_{ dm}\to0$).\footnote{
			As mentioned earlier, the gauge interaction rate in our calculation is approximately one order of magnitude larger than that discussed in \cite{Hambye:2012fh}. Consequently, for a fixed mass of $\Sigma_1$, the departure from equilibrium occurs later compared to the scenario outlined in \cite{Hambye:2012fh}, resulting in a larger value for the parameter $z_A$. This, in turn, necessitates a larger $\tilde{m}_{1}$ value in comparison to the one given in \cite{Hambye:2012fh} to ensure the equilibrium of inverse decays in the case of type-III seesaw leptogenesis. As a result, the boundary between the Gauge and Yukawa regimes shifts upward compared to that in \cite{Hambye:2012fh}, as indicated by the red dashed line.} As we have observed in the co-genesis scenario, the triplets can remain in thermal equilibrium due to both $\tilde{m}_1$ and $\tilde{m}_{dm}$, as opposed to only $\tilde{m}_1$ in the conventional type-III leptogenesis case. Since both $\tilde{m}_1$ and $\tilde{m}_{dm}$ contribute to the inverse decay, a smaller value of $Max\{\tilde{m}_1,\tilde{m}_{dm}\}$ is required to maintain the triplets in thermal equilibrium compared to the $\tilde{m_1}$ value in the type-III leptogenesis case (as $\tilde{m}_{ dm}\to0$). Consequently, we note that the boundary separating the "Gauge" and "Yukawa" regimes shifts downward as compared to the red dashed line.
		
		Now, the asymmetries in both the sectors can be written in terms of dimensionless quantities as
		\begin{equation}
			Y_{\Delta L}\equiv \frac{n_{\Delta L}}{n_\gamma}=\epsilon_L \kappa_L \frac{n_{\Sigma_1}}{n_\gamma} ~ ~ ; ~ ~ 
			Y_{\Delta \chi}\equiv \frac{n_{\Delta \chi}}{n_\gamma}=\epsilon_\chi \kappa_\chi \frac{n_{\Sigma_1}}{n_\gamma},\label{Eq:asymm}
		\end{equation}
		where $\epsilon_L$, $\epsilon_\chi$ are the $CP$ asymmetries and $\kappa_L$, $\kappa_\chi$ are the efficiencies in visible and dark sectors respectively.
		In Eq \ref{Eq:asymm} the efficiency $\kappa_i$ is defined as
		\begin{equation}
			\kappa_i=\frac{Y_{\Delta i}}{\epsilon_i~ Y_{\Sigma_1}^{eq}|_{T\gg M_{\Sigma_1}}} ~~;~~~~ i=L,~\chi.\
		\end{equation}
		The efficiency is defined in such a way that its value is $\approx$ 1 when there is no washout, and it has a value $<$ 1 when there is washout. Here, it is important to note that the $CP$ violating out-of-equilibrium decay of $\Sigma_1$ not only generates asymmetries in $L$ and $\chi$ but also in $H$ and $\Delta$. However, the later fields acquire vevs at the EW scale and thereby decay to SM fermions, effectively erasing the corresponding asymmetries. See, for instance, refs~\cite{Arina:2011cu, Arina:2012aj, Falkowski:2011xh}.  Additionally, it is important to note that while the decay mode $\Delta \to \chi \nu H$ via $\Sigma$ mediation could potentially replenish symmetric DM at later times, this process is significantly suppressed compared to the dominant two-body decays of $\Delta$ into SM fermions through mixing with the SM Higgs. This suppression is primarily due to the large mass of the triplet fermion. Consequently, we safely omit tracking the asymmetry in $\Delta$ from our analysis.

		Before EWSB, $\chi$ doesn't have any Majorana mass. Therefore, DM anti-DM oscillation \cite{Arina:2011cu,Cirelli:2011ac,Tulin:2012re} is absent before EWSB.   Following EWSB, the triplet scalar $\Delta$ acquires a small induced vev, and $\chi$ gains a minor Majorana mass, as detailed in Appendix \ref{app:pseduodirac}. As discussed in Appendix \ref{app:dmantidmosc}, this small Majorana mass induces a negligibly small oscillation probability between DM and anti-DM. In particular, for BP2 in Table \ref{tab:tab3}, the oscillation probability is calculated to be $2.6\times10^{-55}$ for $M_\chi/T=0.1$. This implies that DM anti-DM oscillation doesn't contribute to the washout effect.

		Once the lepton asymmetry is created, it then gets converted to baryon asymmetry via $(B+L)$ violating sphaleron processes at a temperature above the EW phase transition. As given in Appendix  \ref{App:LtoB} baryon asymmetry is estimated as,
		\begin{eqnarray}
			\eta_{_B}&=&-3\times\frac{4}{3 \left(3+\frac{1}{N}\right)-\frac{4 (N+1)}{m+2 N}} \frac{Y_{\Delta L}}{f}=3\frac{S}{f} Y_{\Delta L}.\
		\end{eqnarray}
		Here, with the number of Higgs doublets $m=1$ and the number of fermion generations $N=3$, we have $S=-0.518519$. The calculation details are provided in Appendix \ref{App:LtoB}. Additionally, the dilution factor $f=30.5627$\footnote{$g^*_s=106.75+\frac{7}{8}\times6+\frac{7}{8}\times4+3+1=119.5$, is the relativistic degrees of freedom (d.o.f) at the onset of leptogenesis in the type-III seesaw scenario, $g^*_0=3.91$ is the present day relativistic d.o.f.} is computed assuming standard photon production from the onset of leptogenesis until recombination, as given in \cite{Buchmuller:2004nz}.

		The ratio of the DM relic density ($\Omega_{\chi}h^2$) to baryon relic density ($\Omega_{B}h^2$) is given as \cite{Arina:2011cu}
		\begin{equation}
			R\equiv\frac{\Omega_{\chi}h^2}{\Omega_B h^2}=\frac{f}{3S}\frac{M_{\chi}}{M_p}\frac{\epsilon_{\chi}}{\epsilon_L}\frac{\kappa_\chi}{\kappa_L}=\frac{f}{3S}\frac{M_{\chi}}{M_p}\frac{Y_{\Delta \chi}}{Y_{\Delta L}},\
			\label{eq:omegaratio}
		\end{equation}
		where $M_p$ is the mass of the proton.
		
		It has to be noted that $Y_{\Delta \chi}$ in Eq \ref{eq:omegaratio} is not observed by PLANCK experiment \cite{Aghanim:2018eyx}, rather $\Omega_{\chi}h^2$ $\propto$ $M_{\chi}Y_{\Delta \chi}$ is the observed quantity which can be parameterized as
		\begin{eqnarray}
			\Omega_{\chi}h^2\simeq 0.12 \times \bigg(\frac{M_{\chi}}{1 \rm GeV}\bigg)\bigg(\frac{Y_{\Delta \chi}}{4.26\times 10^{-10}}\bigg).\label{Eq:DMrelic}
		\end{eqnarray}
		In the above Equation, $Y_{\Delta \chi}\equiv f(\epsilon_\chi, \tilde{m}_{dm}, M_{\Sigma_1})$ can be obtained by solving the relevant Boltzmann equations \ref{eq:BEY}, \ref{eq:BEYl}, and \ref{eq:BEYx}, which are independent of $M_{\chi}$. Therefore, for a typical solution of $Y_{\Delta\chi}$, an appropriate value of $M_{\chi}$ has to be supplemented to get the observed value of $\Omega_{\chi}h^2$.
		
		Now we turn to comment on $Y_{\Delta L}\equiv f(\epsilon_L, \tilde{m}_{1},M_{\Sigma_1})$. Recognizing that the permissible values of $\epsilon_L$ for lepton asymmetry is intrinsically linked to neutrino masses, which are tightly constrained by both oscillation experiments and cosmological data, we can derive constraints on the $\epsilon_L$ values. The upper bound on $\epsilon_L$ was first shown in~\cite{Davidson:2002qv} for the conventional leptogenesis scenarios, which also implies a lower bound on the lightest RHN mass in the type-I seesaw leptogenesis. For the type-III seesaw scenario it is given by $|\epsilon_L|\lesssim\frac{1}{8\pi}\frac{M_{\Sigma_1} (m_3-m_1)}{v_0^2}$~\cite{Hambye:2012fh,Hambye:2005tk}. Assuming the normal ordering of neutrino masses ($m_3>m_2$$\gg$ $m_1$), there exists a lower bound on the lightest triplet fermion in the usual type-III leptogenesis to be $M_{\Sigma_1}$$\gtrsim 3\times10^{10}$ GeV \cite{Hambye:2012fh,Hambye:2005tk}. However, this upper bound on $\epsilon_L$ gets modified by a factor involving the dark sector parameters in the cogenesis scenario, and in the hierarchical case, it is given by \cite{Falkowski:2011xh,Davidson:2002qv,Hambye:2003rt} 
		\begin{eqnarray}
			|\epsilon_L|\lesssim\frac{1}{8\pi}\frac{M_{\Sigma_1} (m_3-m_1)}{v_0^2}\mathcal{C}, \label{eq:DIbound}
		\end{eqnarray}
		and the modification factor $C$ is given by:
		\begin{equation}
			\mathcal{C}\simeq\left\{
			\begin{array}{l}
				1 ~~~~~~~~~~~~~~~~~,~~Br_{L}\gg Br_{\chi}\\
				\sqrt{\frac{y^2_{\chi_2}M_{\Sigma_1}}{y^2_{\chi_1}M_{\Sigma_2}}} ~~~~,~~Br_{L}\ll Br_{\chi}\\
			\end{array}
			\right.
		\end{equation}
		When the $\Sigma$ is dominantly decaying to the visible sector ($\textit{i.e.}~~ Br_L\gg Br_\chi$), the case resembles the usual type-III leptogenesis. However, when decays to the dark sector dominate, the bound is modified by $\mathcal{C}$, consequently altering the lower bound on $M_{\Sigma_1}$. This modification will be evident in our subsequent analysis and results. 
		On the other hand, the dark sector lacks such information on the $CP$ asymmetry parameter $\epsilon_\chi$ as $\tilde{m}_{\rm dm}$ is not related to any of the low energy parameters. Therefore, we do not have any prior information on $\epsilon_\chi$.      
		We conduct a comprehensive numerical scan of our independent parameters
		\begin{center}
			\{$M_{\Sigma},M_{\chi},y_{\chi_{1,2,3}},\theta$\}
		\end{center}
		varying them randomly in a range as in Table ~\ref{tab:tab2}. Here, $\theta$ represents the complex mixing angle that parametrizes the orthogonal matrix $R$ in the Casas-Ibarra parametrization, used to derive $y_{\Sigma_i}$ values consistent with neutrino oscillation data. In our scans, we also maintain a hierarchical structure among the triplet fermion masses, following the ratio $M_{\Sigma_1}:M_{\Sigma_2}:M_{\Sigma_3} = 1:10:50$.  
		\begin{table}[h]
			\centering
			\resizebox{5cm}{!}{			
				\begin{tabular}{|c|c|}
					\hline\hline
					Parameter & Scanned range\\ 
					\hline 
					$M_\chi$ (GeV) & $10^{-3}-10^{3}$\\ 
					\hline
					$M_{\Sigma_1}$ (GeV) & $10^{7} -10^{14}$\\
					\hline 
					$Re(y_{\chi 1}),~Im(y_{\chi 1})$ & $10^{-5}-\sqrt{4\pi}$\\
					\hline 
					$Re(y_{\chi 2}),~Im(y_{\chi 2})$ & $10^{-5}-\sqrt{4\pi}$\\
					\hline 
					$Re(y_{\chi 3}),~Im(y_{\chi 3})$ & $10^{-5}-\sqrt{4\pi}$\\
					\hline 
					$Re(\theta),~Im(\theta)$ & $10^{-5}-2\pi$ \\
					\hline\hline
			\end{tabular}}
			
			\caption{The range in which the free parameters are randomly varied for the numerical analysis.}
			\label{tab:tab2}
		\end{table}	
		
		\begin{figure}[h]
			\centering	\includegraphics[width=7.5cm,height=7.5cm]{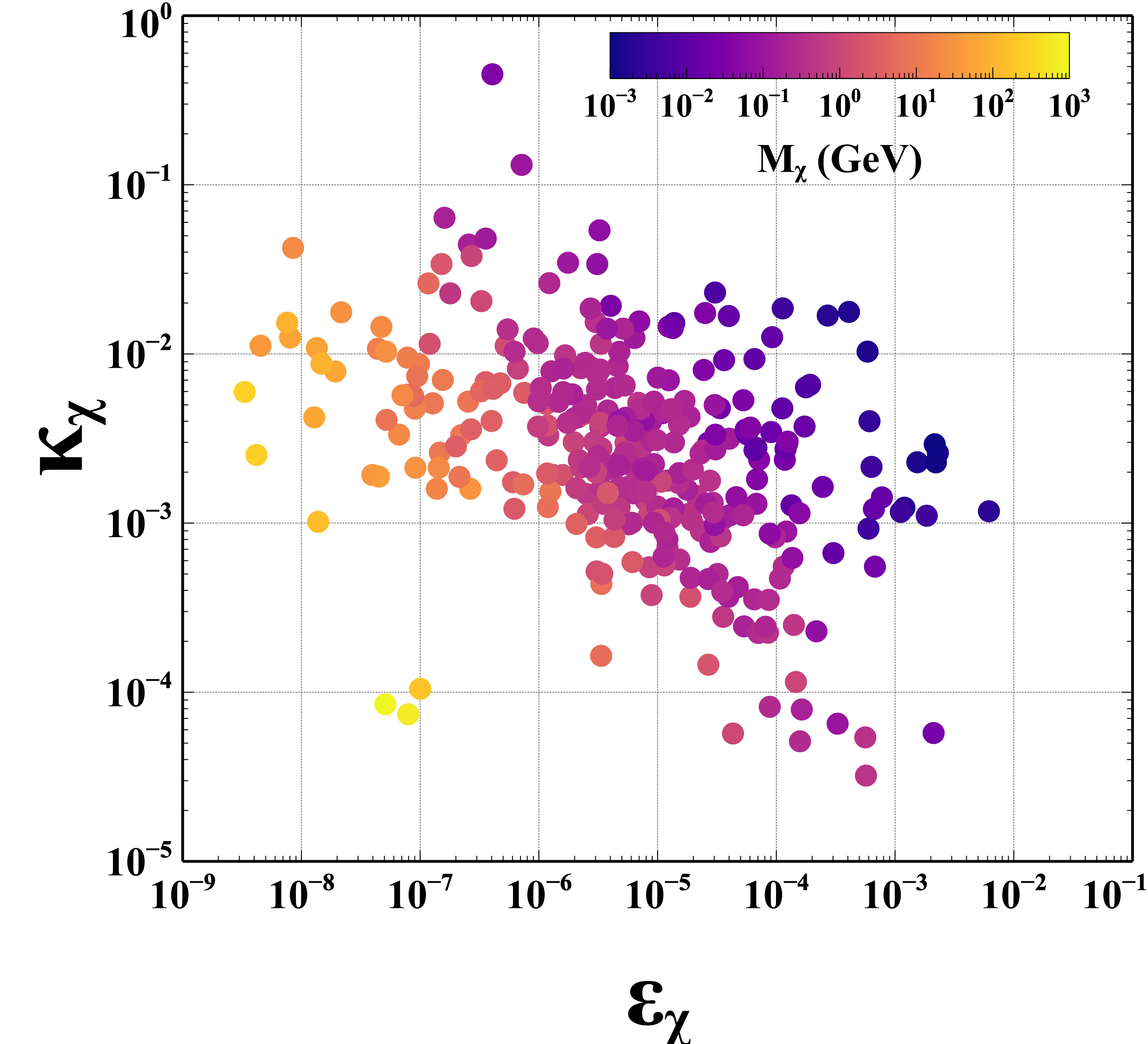}\includegraphics[width=7.5cm,height=7.5cm]{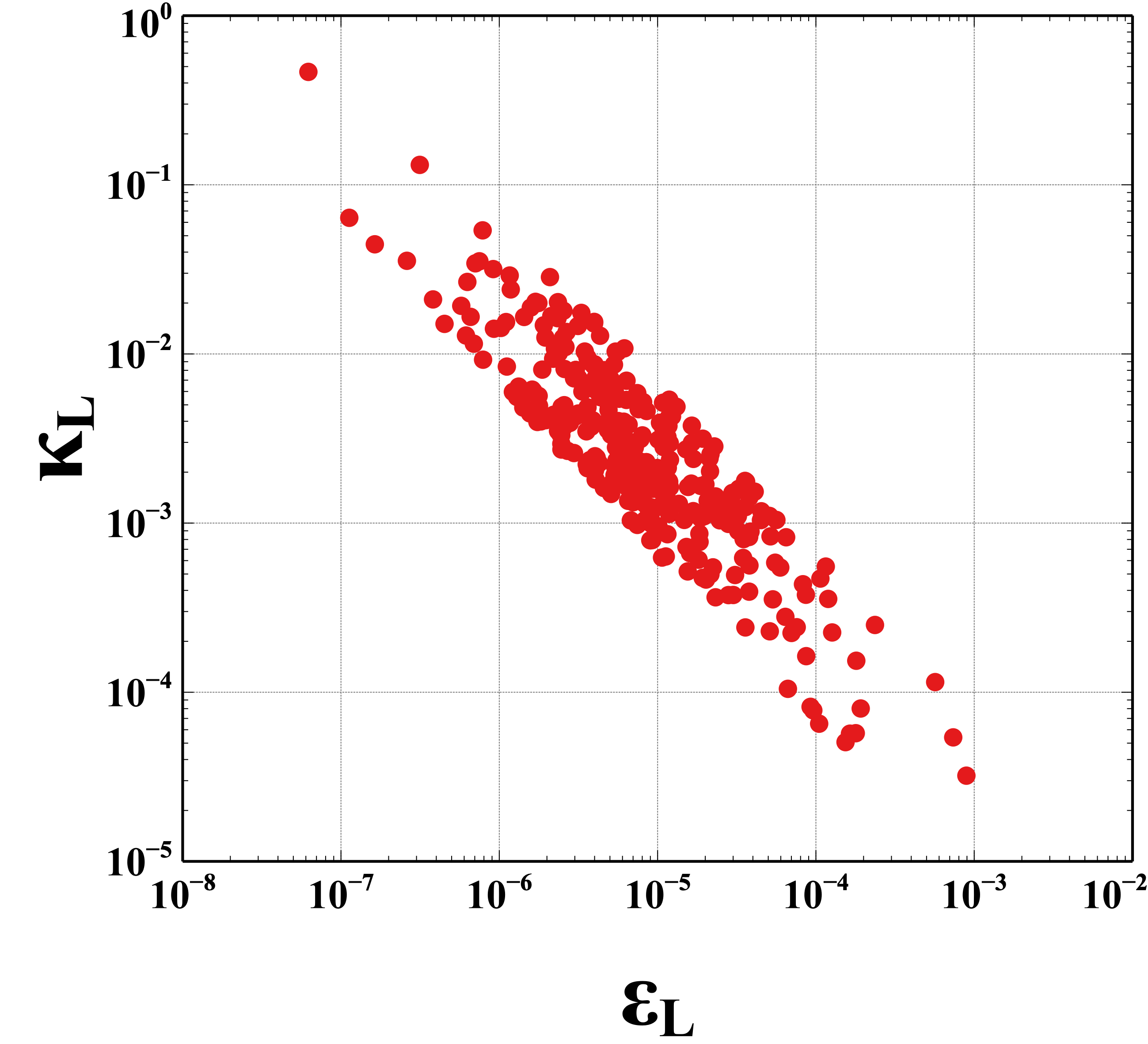}  \\      \includegraphics[width=7.5cm,height=7.5cm]{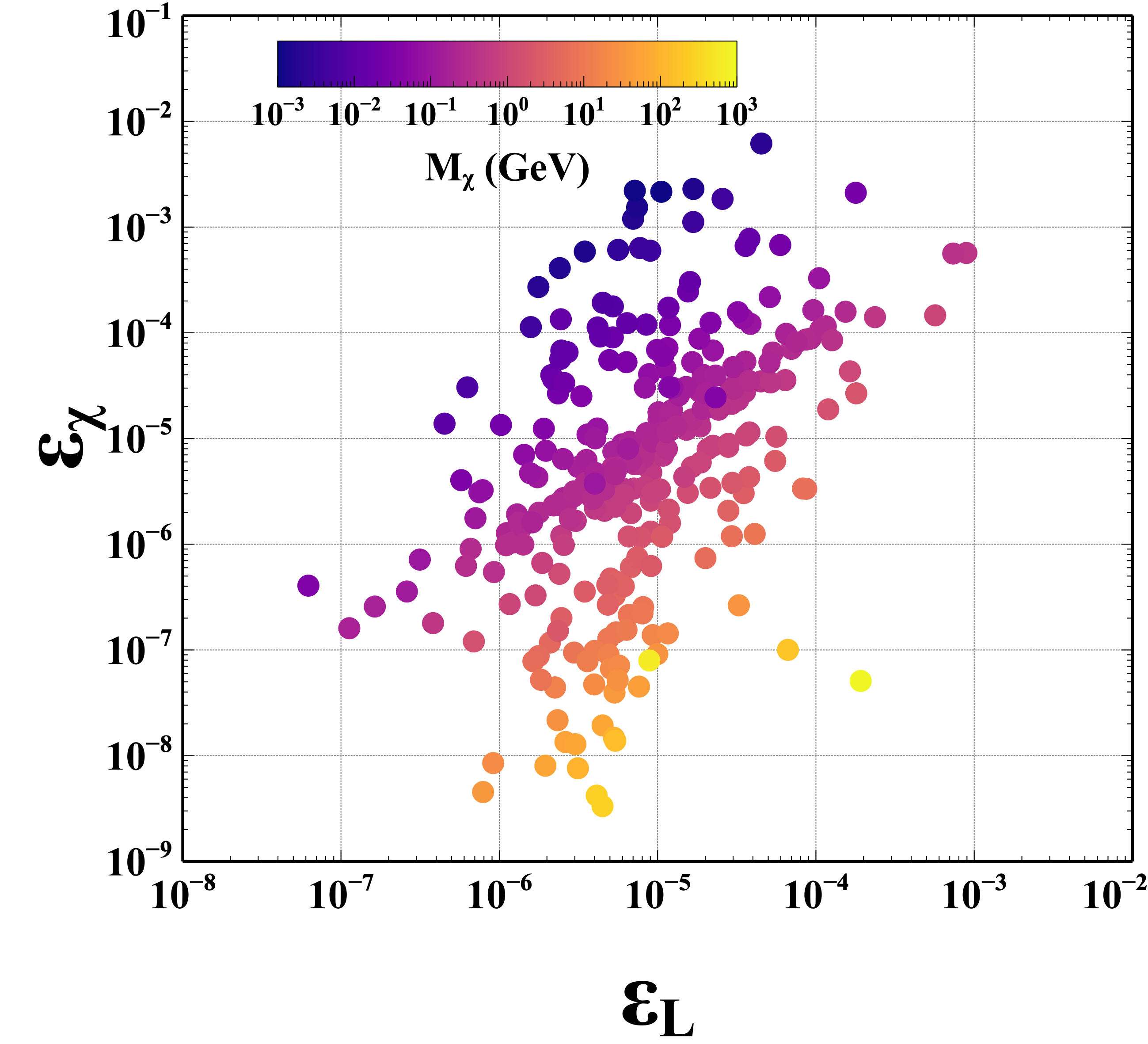}       \includegraphics[width=7.5cm,height=7.5cm]{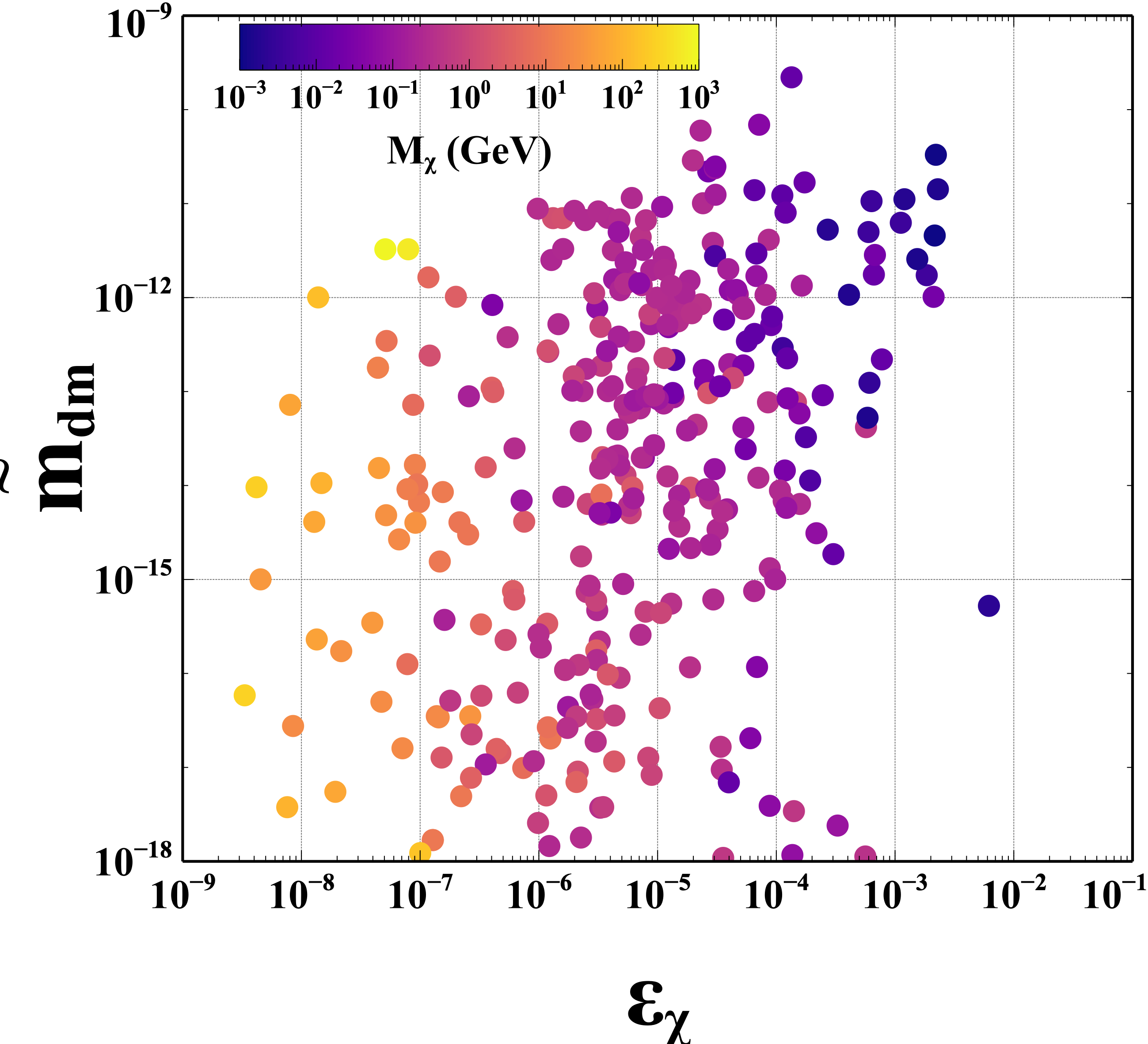}   \caption{The parameter space in $\kappa_\chi$-$\epsilon_{\chi}$  plane  as a function of $M_\chi$ [top panel left].  The dependency of $\kappa_L$ is shown as a function of  $\epsilon_L$ in the top panel right. Parameter space in $\epsilon_{\chi}$ vs $\epsilon_L$ [bottom panel left] and $\tilde{m}_{dm}$ dependence on $\epsilon_\chi$ [bottom panel right].}
			\label{fig:parscan}
		\end{figure}
		
		For simplicity, we have fixed the triplet scalar mass ($M_{\Delta}$) at 1 TeV and explored the parameter space satisfying the correct lepton asymmetry and DM relic density, as shown in Fig. \ref{fig:parscan}. In the top-left panel of Fig. \ref{fig:parscan}, we plot $\kappa_\chi$ as a function of $\epsilon_\chi$. The plot illustrates that as $\epsilon_\chi$ increases, $M_\chi$ decreases to maintain the correct relic density of DM, consistent with Eq. \ref{eq:omegaratio}, where $\Omega_{\chi}h^2 \propto \epsilon_\chi \kappa_\chi M_\chi$. Additionally, for a fixed $M_\chi$, $\kappa_\chi$ decreases as $\epsilon_\chi$ increases, and vice versa, which is evident from the same figure.
		The top-right panel of Fig. \ref{fig:parscan} depicts the visible sector efficiency factor as a function of the visible sector $CP$ asymmetry. As we move towards the large washout regime, the efficiency factor decreases, necessitating an increase in the $CP$ asymmetry to achieve the correct lepton asymmetry. This trend is clearly visible in the same plot. A key observation here is that, unlike leptogenesis in the type-I scenario, where the efficiency factor approaches 1 in the small washout regime (for smaller couplings), in our scenario, the efficiency factor never reaches unity. This is because, in the present scenario, the dynamics of $\Sigma_1$ are governed by gauge as well as Yukawa interactions.
		In the bottom-right panel of Fig. \ref{fig:parscan}, we observe that for large DM masses, $\epsilon_\chi$ is small, while for small DM masses, $\epsilon_\chi$ is larger, regardless of whether the washout regime is small or large. Furthermore, the bottom-left panel highlights that the correct lepton asymmetry and DM relic density can be achieved across a wide range of DM masses with $CP$ asymmetries spanning the ranges $10^{-9} < \epsilon_\chi < 6 \times 10^{-3}$ and $4 \times 10^{-8} < \epsilon_L < 10^{-3}$.

		\begin{table}[H]
			\centering
			\resizebox{17cm}{!}{
				\begin{tblr}{
						colspec={|l|l|l|l|l|l|l|l|l|l|},
						row{1}={font=\bfseries},
						column{1}={font=\itshape}
					}
					\toprule BPs&$M_1\rm(GeV)$&$M_\chi\rm(GeV)$& $y_{\chi_1}$ &$y_{\chi_2}$ &$y_{\chi_3}$ & $\theta$ & $\epsilon_L$ & $\epsilon_\chi$  \\
					\toprule
					BP1& $6.04\times10^{13}$&0.85 & $3.01\times10^{-2}$ & $(2.39+i0.19)\times10^{-2}$ & $(2.65+i0.96)\times10^{-1}$ & $6.27+i5.49\times10^{-3}$ & $3.61\times10^{-5}$ & $1.07\times10^{-5}$\\ \hline
					
					BP2&$1.98\times10^{12}$&17 & $6.78\times10^{-5}$ & $(1.92+i0.39)$ & $(0.56+i3.63)\times10^{-1}$& $(0.36+i1.19)\times10^{-2}$ &$9.32\times10^{-6}$&$1.38\times10^{-7}$\\\hline
					
					BP3& $3.92\times10^{9}$&1.48&$1.35\times10^{-3}$&$(0.036+i2.06)\times10^{-1}$&$1.11\times10^{-2}+i1.54$&$3.57+i2.86\times10^{-2}$&$9.05\times10^{-6}$&$1.32\times10^{-6}$\\\hline
					
					BP4&$4.44\times10^{11}$&920&$9.78\times10^{-3}$&$0.19+i2.82\times10^{-4}$&$2.52+i0.54$&$1.40\times10^{-4}+i0.52$&$1.92\times10^{-4}$&$5.09\times10^{-8}$\\
					\bottomrule
			\end{tblr}}
			\caption{Benchmark points chosen for showcasing the evolution of visible sector and dark sector asymmetries.
			}
			\label{tab:tab3}
		\end{table}
		
		From the parameter scan, we have selected four benchmark points, as listed in Table \ref{tab:tab3}, to illustrate the evolution of the dark sector and visible sector asymmetries, along with the comoving abundance of $\Sigma_1$. 
		A noteworthy observation in this scenario is that even when starting with a zero initial abundance of ${\Sigma_1}$, gauge interactions rapidly bring it into thermal equilibrium. Consequently, the evolution closely resembles the case where the initial abundance is assumed to be thermal.
		
		\begin{figure}[h]
			\centering        \includegraphics[width=7.5cm,height=7.5cm]{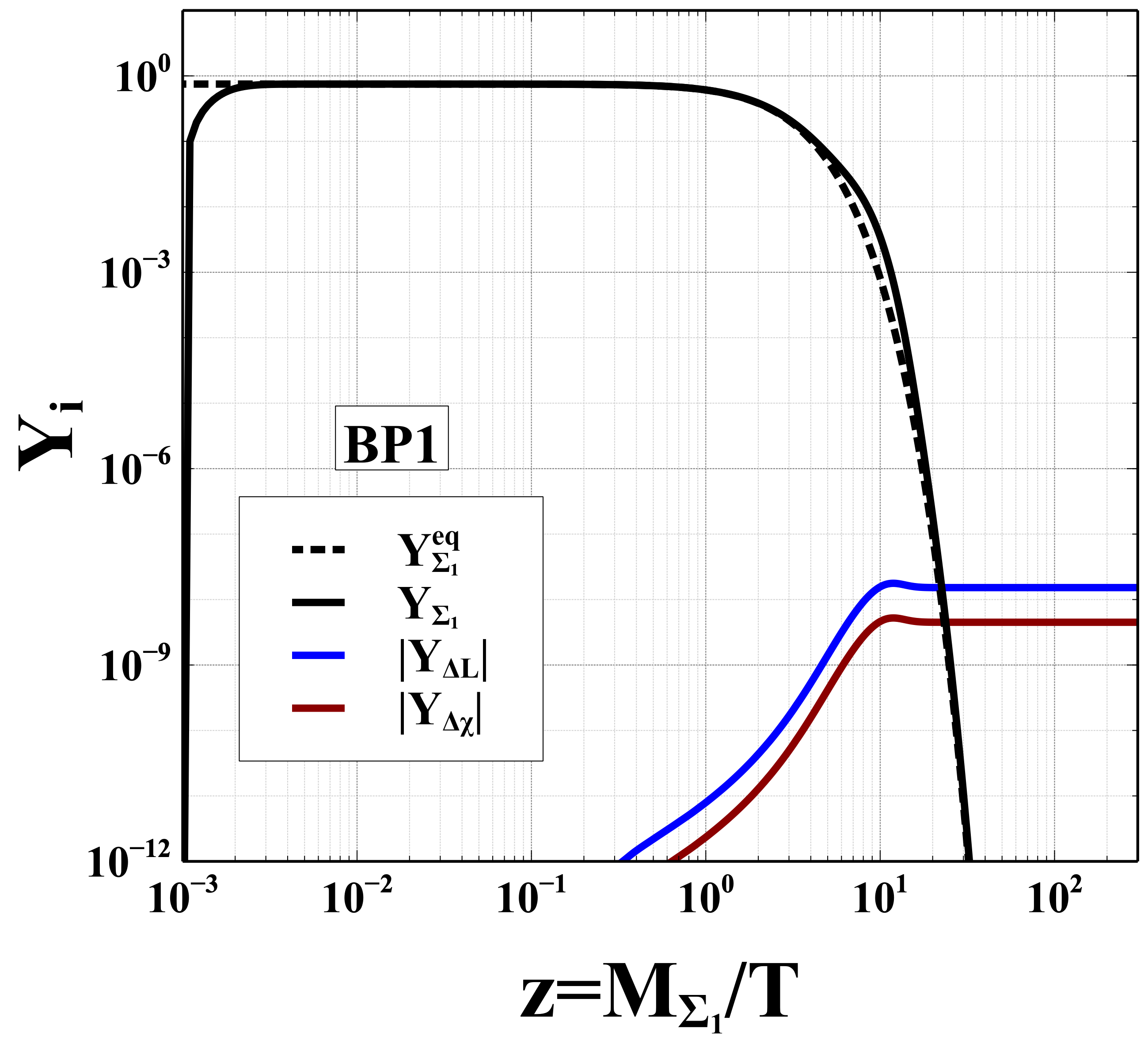}\includegraphics[width=7.5cm,height=7.5cm]{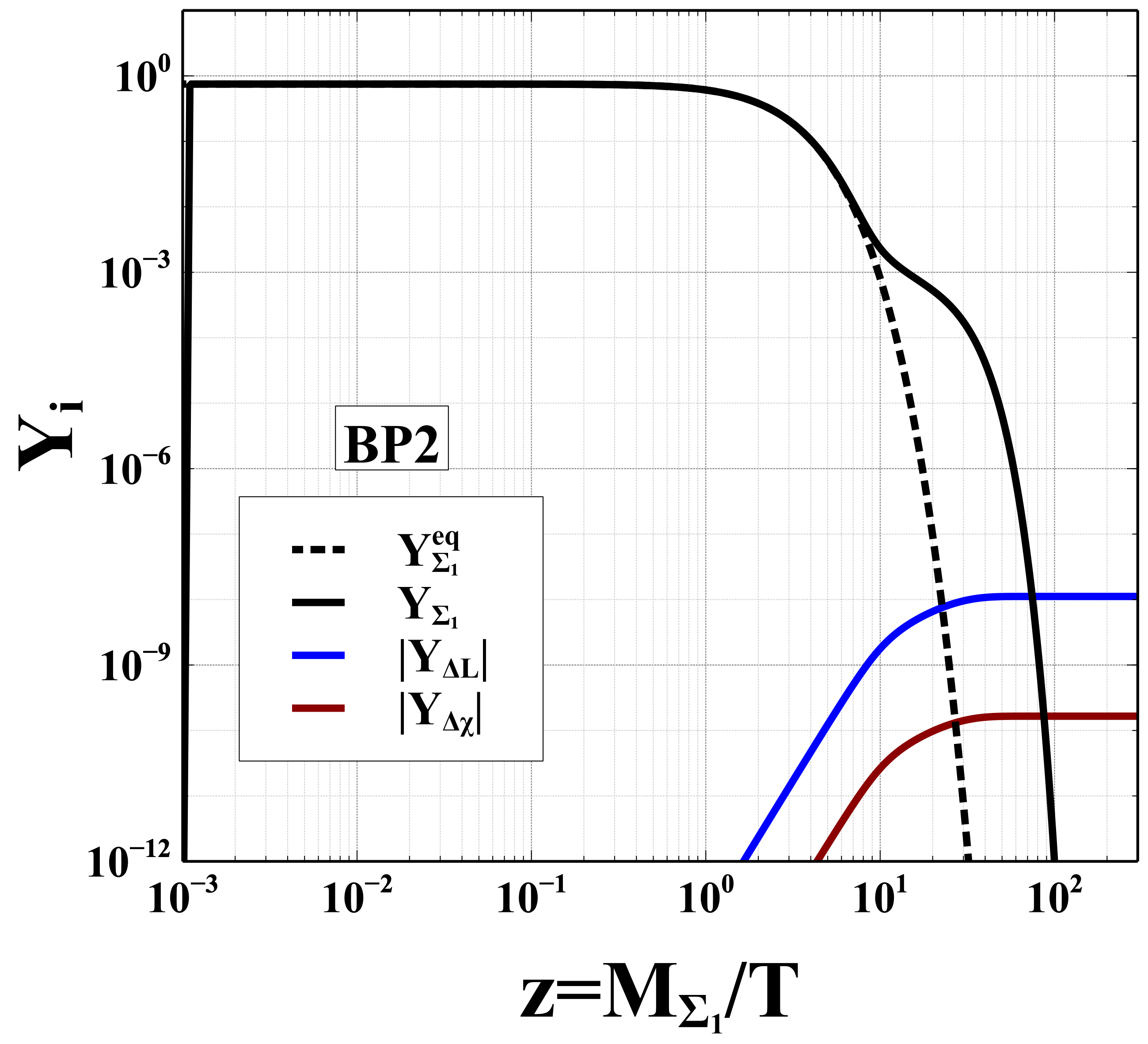}
			\caption{Cosmological evolutions of both sector asymmetries for  BP1 (left), and BP2 (right).}
			\label{fig:bp12}
		\end{figure}
		
		We present the evolution of the triplet abundance alongside the visible and dark sector asymmetries for BP1 in Fig.~\ref{fig:bp12} (left). The asymmetries are generated as the triplet decays into both sectors. For the chosen parameters, the branching ratio to the visible sector is $3.59\%$, and for the dark sector, it is $96.41\%$ with $\tilde{m}_1=8.46\times10^{-15}$ GeV, $\tilde{m}_{dm}=2.27\times10^{-13}$ GeV. The relatively larger Yukawa couplings result in a suppression of the dark sector asymmetry due to its dominant branching ratio.  However, we also see a corresponding suppression in the visible sector asymmetry caused by the transfer terms that couple the two sectors, as in this co-genesis framework, the asymmetries in the two sectors are interconnected, with the dynamics of one influencing the other. Both asymmetries experience suppression due to washout effects and eventually stabilize to give the required lepton asymmetry and DM relic density.
		In Fig.~\ref{fig:bp12} (right), we show the evolution for BP2. Here, the triplets reach thermal equilibrium due to gauge interactions, remaining in equilibrium as long as these interactions dominate over the Hubble expansion rate. However, since the Yukawa coupling in BP2 is not strong enough to maintain equilibrium,  once the gauge interactions fall below the Hubble rate, the triplets decouple, which is around $z \sim 10$. Their subsequent decay generates asymmetries in both the visible and dark sectors. For BP2, the branching ratios to the visible and dark sectors are $99.61\%$ and $0.39\%$, respectively with $\tilde{m}_1=9.11\times10^{-15}$ GeV, $\tilde{m}_{dm}=3.51\times10^{-17}$ GeV. However, since the Yukawa couplings are small and washout effects are negligible, we observe no suppression in the final asymmetry.
		\begin{figure}[h]
			\centering        \includegraphics[width=7.5cm,height=7.5cm]{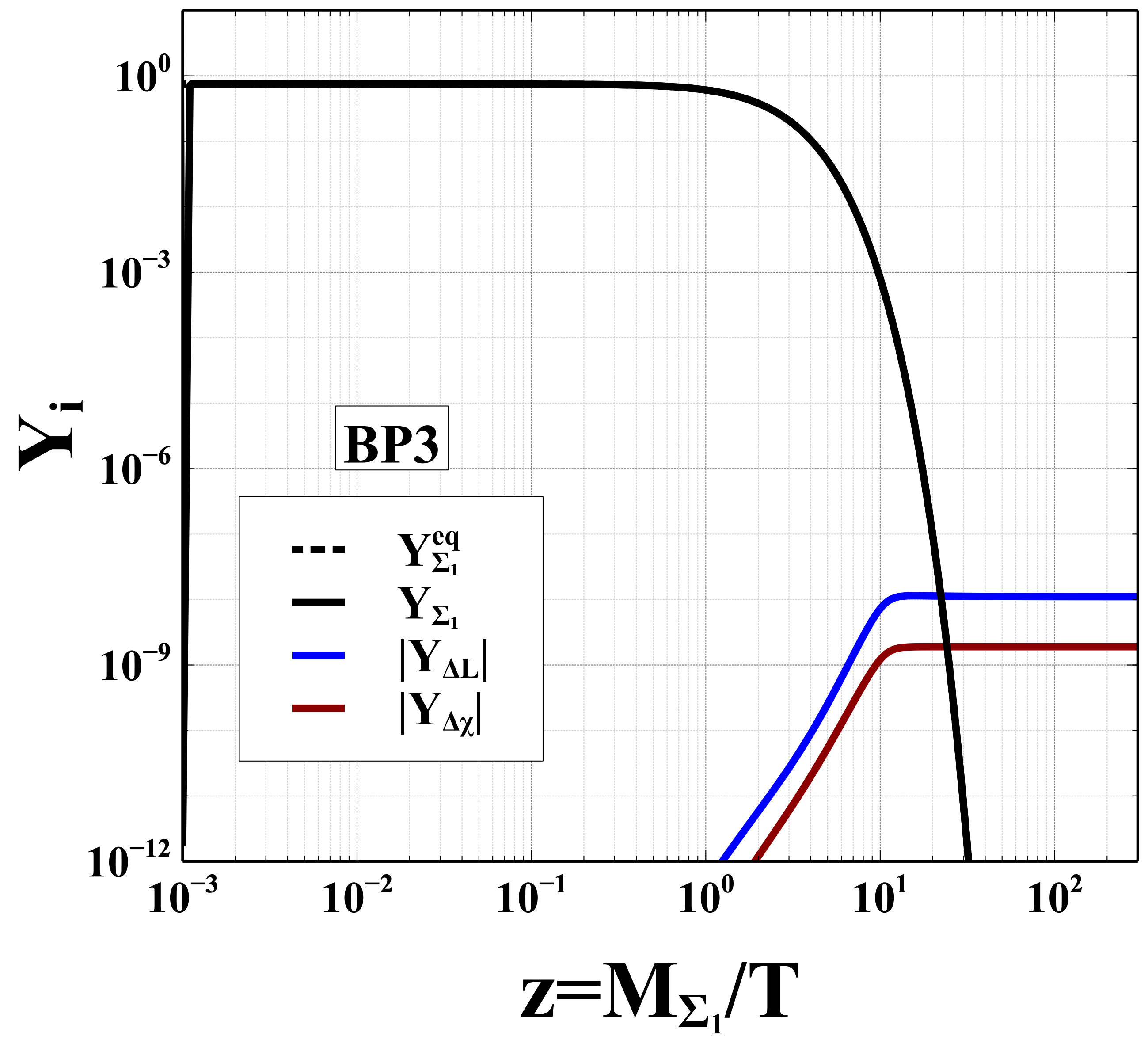}
			\includegraphics[width=7.5cm,height=7.5cm]{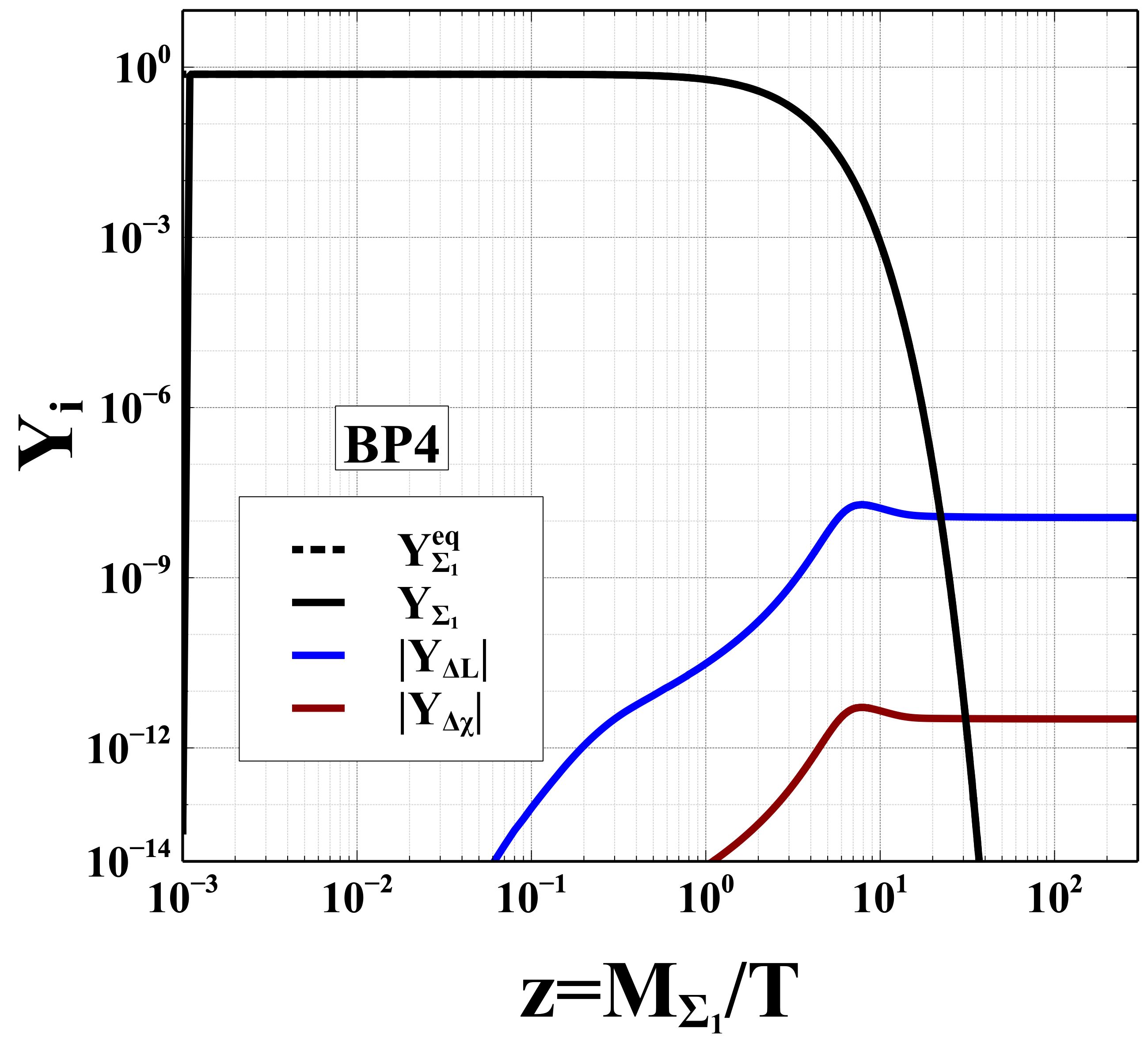}
			\caption{Cosmological evolutions of both sector asymmetries for  BP3 (left), and BP4 (right).}
			\label{fig:bp34}
		\end{figure}
		
		In Fig.~\ref{fig:bp34} (left), we present the evolution for BP3, where the mass of the fermion triplet is an order of magnitude smaller than the lower bound typically required for type-III seesaw leptogenesis. Notably, in the co-genesis framework, the triplet mass can be reduced by several orders of magnitude, depending on the choice of dark sector Yukawa couplings and the mass hierarchy among the triplet generations, as discussed in Eq.~\ref{eq:DIbound}. In the standard type-III seesaw scenario, the $CP$ asymmetry parameter in the visible sector is primarily determined by the Yukawa couplings, which are tied to the light neutrino mass, thereby imposing a strict lower bound on the triplet mass. However, this bound can be relaxed in the co-genesis scenario due to contributions to the $CP$ asymmetry from the dark sector couplings via self-energy correction diagrams involving dark sector particles in the loop. 
		This benchmark demonstrates how the triplet mass bound can be lowered by an order of magnitude while still generating the correct lepton asymmetry. Further relaxation of this bound is possible by fine-tuning couplings and triplet mass hierarchies. The branching ratio to the visible sector is $69.68\%$, while the branching ratio to the dark sector is $30.32\%$ with $\tilde{m}_1=1.61\times10^{-11}$ GeV, $\tilde{m}_{dm}=7.00\times10^{-12}$ GeV. For BP3, we observe no suppression in the final asymmetry. This is attributed to the smaller triplet mass, which keeps the triplets in equilibrium for a longer duration, while the couplings are not strong enough to induce significant washout of the asymmetries.
		Fig. \ref{fig:bp34} (right) illustrates the evolution for BP4, featuring a relatively heavy DM candidate with $M_{\chi}=920$ GeV.  For this benchmark, the branching ratio to the visible sector is $99.35\%$, while the branching ratio to the dark sector is $0.65\%$ with $\tilde{m}_1= 2.89\times10^{-11}$ GeV, $\tilde{m}_{dm}=3.26\times10^{-12}$ GeV. Due to the large Yukawa couplings, $\Sigma_1$ remains in thermal equilibrium for an extended period. This results in a significant washout of the asymmetry in both sectors before eventually stabilizing to the observed values. 
		
		\subsection{Depletion of symmetric component of $\chi$}\label{sec:depele}
\begin{figure}[h]
			\centering
			\includegraphics[scale=1.2]{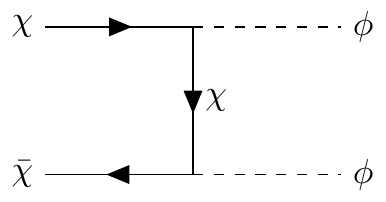}
			\caption{Annihilation of symmetric component of $\chi$.}
			\label{fig:xxtopp}
		\end{figure}

		As has already been discussed before that in this scenario, the asymmetric component of $\chi$ is accountable for the observed DM relic; this 
		requires the symmetric component of $\chi$ to be annihilated away completely. This can be achieved through the annihilation process: $\bar{\chi}\chi\rightarrow\phi \phi$ 
		as shown in Fig \ref{fig:xxtopp}. 	
		\noindent The corresponding cross-section is given in Appendix \ref{appen2}. In Fig \ref{fig:cxxtophiphi}, we show the annihilation cross-section ($\bar{\chi}\chi\rightarrow\phi \phi$) as a function of $M_\chi$ for different values $\lambda_{\rm DM}$. Here, it should be noted that for the complete annihilation of the symmetric part of the DM $\chi$, its freeze-out cross-section should be greater than $2.6\times10^{-9}$ $\rm GeV^{-2}$.
		\begin{figure}[h]
			\centering
			\includegraphics[width=7.5cm,height=7.5cm]{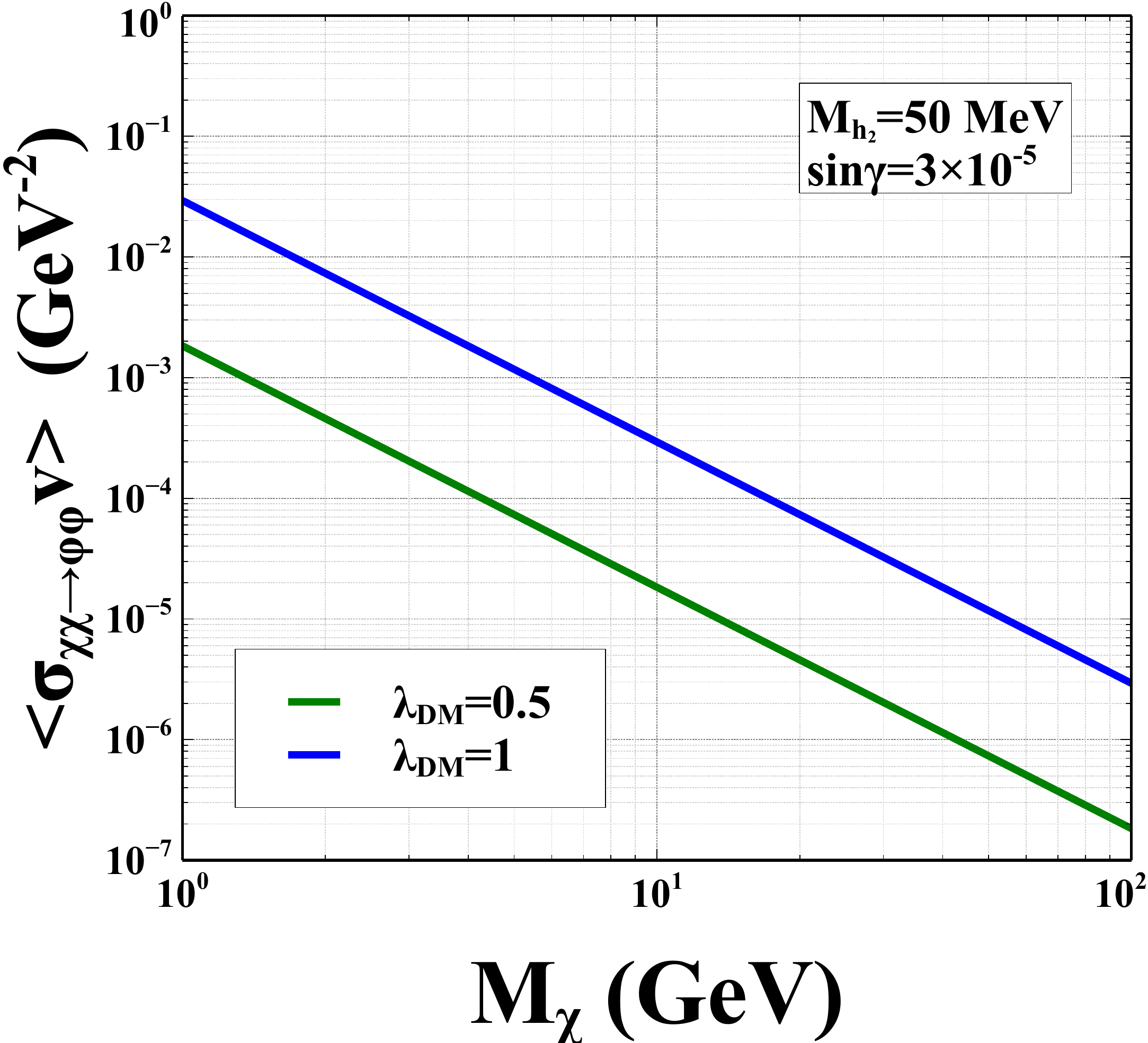}
			\caption{ Annihilation cross-section for $\chi$ as a function of $M_\chi$ for 2 different values of $\lambda_{\rm DM}$.}
			\label{fig:cxxtophiphi}
		\end{figure}
		
		Here it is worth mentioning that, once $\Phi$ is produced from annihilation of $\chi$, it decays to SM fermions through $\Phi-H$ mixing. For a typical benchmark value of $M_{h_2}$=50 MeV ($h_2\equiv\phi$) and $\Phi-H$ mixing, $\sin(\gamma)$ = $3\times10^{-5}$ used in Fig~\ref{fig:cxxtophiphi}, the decay rate of $\phi$ is found to be $\Gamma_\phi= 7.691\times10^{-24}$ GeV ($\tau_\phi=0.0856$ s) considering the accessible decay mode ($e^+e^-$), which is safe from the BBN constraint. Throughout our analysis, we ensure that $M_{h_2}$ and $\sin(\gamma)$ is such that the lifetime of $\phi$,  $\tau_\phi<\tau_{_{BBN}}= 1~s$,  keeping the BBN predictions intact.

		\section{Long lived Dark Matter}\label{sec:longliveddm}
\begin{figure}[h]	
			\includegraphics[scale=0.45]{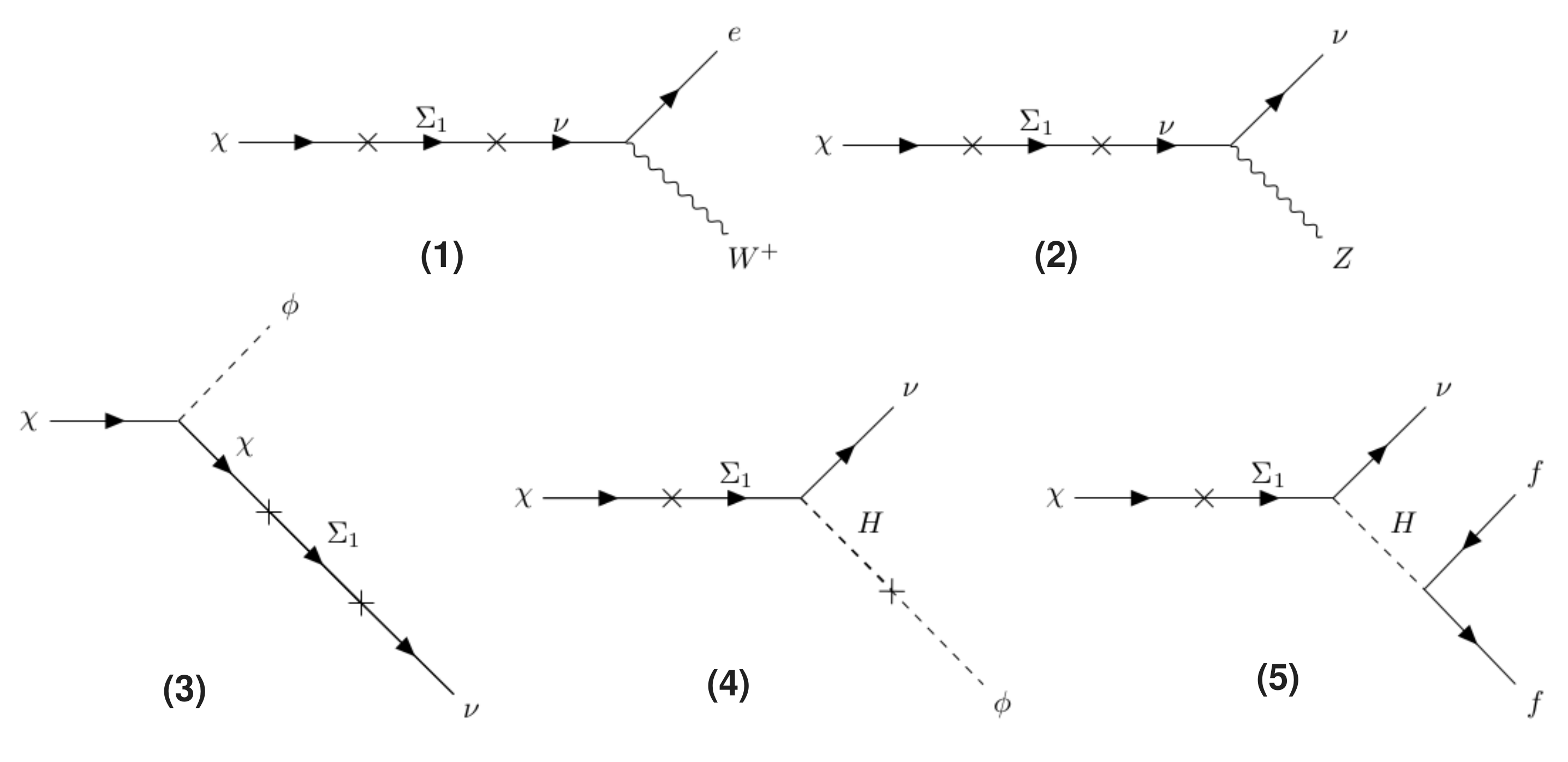}
			\centering
			\caption{Different decay modes of $\chi$.}
			\label{fig:chidecay}
		\end{figure}

		As the $Z_2$ symmetry, which guarantees the stability of DM (under which $\chi$ and $\Delta$ are odd) gets broken softly when $\Delta$ acquires an induced vev after EWSB, the DM $\chi$ becomes unstable and decays to the SM particles, and $\phi\equiv h_2$. For $\chi$ to qualify as a viable DM candidate, its lifetime ($\tau_{\rm DM}$) should satisfy $\tau_{\rm DM}>10^{27}$ sec\cite{Baring:2015sza,Mahapatra:2023zhi}.
		In Fig \ref{fig:chidecay}, the Feynman diagrams of the different possible decay channels are shown.
		The corresponding decay rates are given as
		\begin{eqnarray}	
			\Gamma_{{\chi}}^{(1)} (\chi\rightarrow e W)=\frac{{G_{{\chi}}^{(1)}}^2 M^3_\chi}{16\pi M_W^2}\bigg(1+\frac{m^4_e}{M^4_\chi}-\frac{2M^4_W}{M^4_\chi}-\frac{2m^2_e}{M^2_\chi}+\frac{M^2_W}{M^2_\chi}+\frac{m^2_{e}M^2_W}{M^4_\chi}\bigg)\bigg( 1+\frac{m_e^4}{M_\chi^4} +\frac{M_W^4}{M_\chi^4}-\frac{2m_e^2}{M_\chi^2}-\frac{2M_W^2}{M_\chi^2}-\frac{2m_e^2M_W^2}{M_\chi^4}\bigg)^{\frac{1}{2}},\nonumber\\ \label{eq:2body1}
		\end{eqnarray}	
		\begin{eqnarray}
			\Gamma_{{\chi}}^{(2)}(\chi\rightarrow \nu Z)=\frac{{G_{{\chi}}^{(2)}}^2 M^3_\chi}{16\pi M_Z^2}\bigg(1+\frac{m^4_\nu}{M^4_\chi}-\frac{2M^4_Z}{M^4_\chi}-\frac{2m^2_\nu}{M^2_\chi}+\frac{M^2_Z}{M^2_\chi}+\frac{m^2_{\nu}M^2_Z}{M^4_\chi}\bigg)\bigg( 1+\frac{m_\nu^4}{M_\chi^4} +\frac{M_Z^4}{M_\chi^4}-\frac{2m_\nu^2}{M_\chi^2}-\frac{2M_Z^2}{M_\chi^2}-\frac{2m_\nu^2M_Z^2}{M_\chi^4}\bigg)^{\frac{1}{2}},\nonumber\\\label{eq:2body2}
		\end{eqnarray}
		
		\begin{eqnarray}
			\Gamma_{{\chi}}^{(3)}(\chi\rightarrow \phi \nu)=&\frac{{G_{{\chi}}^{(3)}}^2 M_\chi}{16\pi} \bigg( 1+\frac{m_\nu^2}{M_\chi^2} -\frac{M_\phi^2}{M_\chi^2}+\frac{2m_\nu}{M_\chi}\bigg)\bigg( 1+\frac{m_\nu^4}{M_\chi^4} +\frac{M_\phi^4}{M_\chi^4}-\frac{2m_\nu^2}{M_\chi^2}-\frac{2M_\phi^2}{M_\chi^2}-\frac{2m_\nu^2M_\phi^2}{M_\chi^4}\bigg)^{\frac{1}{2}},\nonumber\\ \label{eq:2body3}
		\end{eqnarray}
		
		\begin{eqnarray}
			\Gamma_{{\chi}}^{(4)}(\chi\rightarrow \phi\nu)&=&\frac{{G_{{\chi}}^{(4)}}^2 M_\chi}{16\pi} \bigg( 1+\frac{m_\nu^2}{M_\chi^2} -\frac{M_\phi^2}{M_\chi^2}+\frac{2m_\nu}{M_\chi}\bigg)\bigg( 1+\frac{m_\nu^4}{M_\chi^4} +\frac{M_\phi^4}{M_\chi^4}-\frac{2m_\nu^2}{M_\chi^2}-\frac{2M_\phi^2}{M_\chi^2}-\frac{2m_\nu^2M_\phi^2}{M_\chi^4}\bigg)^{\frac{1}{2}},\nonumber\\ \label{eq:2body4}
		\end{eqnarray}	
		
		\begin{eqnarray}
			\Gamma_{{\chi}}^{(5)}(\chi\rightarrow \nu f \bar{f})&=&\frac{{G_{{\chi}}^{(5)}}^2 M_\chi^5}{192\pi^3} \bigg( \frac{1}{4}(1-4s_w^2+8s_w^4)((1-14\frac{m_f^2}{M_\chi^2}-2\frac{m_f^4}{M_\chi^4}-12\frac{m_f^6}{M_\chi^6})\sqrt{1-4\frac{m_f^2}{M_\chi^2}}\nonumber\\&+&12\frac{m_f^4}{M_\chi^4}(\frac{m_f^4}{M_\chi^4}-1)\log\bigg[\frac{1-3\frac{m_f^2}{M_\chi^2}-(1-\frac{m_f^2}{M_\chi^2})\sqrt{1-4\frac{m_f^2}{M_\chi^2}}}{\frac{m_f^2}{M_\chi^2}(1+\sqrt{1-4\frac{m_f^2}{M_\chi^2}})}\bigg])\nonumber\\&+&2s_w^2(2s_w^2-1)(\frac{m_f^2}{M_\chi^2}(2+10\frac{m_f^2}{M_\chi^2}-12\frac{m_f^4}{M_\chi^4})\sqrt{1-4\frac{m_f^2}{M_\chi^2}}\nonumber\\&+&6\frac{m_f^4}{M_\chi^4}(1-2\frac{m_f^2}{M_\chi^2}+2\frac{m_f^4}{M_\chi^4})\log\bigg[\frac{1-3\frac{m_f^2}{M_\chi^2}-(1-\frac{m_f^2}{M_\chi^2})\sqrt{1-4\frac{m_f^2}{M_\chi^2}}}{\frac{m_f^2}{M_\chi^2}(1+\sqrt{1-4\frac{m_f^2}{M_\chi^2}})}\bigg])   \bigg) ,\\\nonumber \label{eq:3body}
		\end{eqnarray}
		where the
		effective couplings $G_{{\chi}}^{(1)}$, $G_{{\chi}}^{(2)}$, $G_{{\chi}}^{(3)}$, $G_{{\chi}}^{(4)}$, and $G_{{\chi}}^{(5)}$ can be given as	
		\begin{equation}
			G_{{\chi}}^{(1)} =\big(y_\chi v_1\big)\bigg(\frac{1}{M_{\Sigma_1}}\bigg)\big(y_{_\Sigma}v_0\big)\bigg(\frac{1}{M_{\chi}}\bigg)\bigg(\frac{g}{2~ \cos\theta_w}\bigg),\ \label{eq:gx1}
		\end{equation}
		\begin{equation}
			G_{{\chi}}^{(2)} =\big(y_\chi v_1\big)\bigg(\frac{1}{M_{\Sigma_1}}\bigg)\big(y_{_\Sigma}v_0\big)\bigg(\frac{1}{M_{\chi}}\bigg)\bigg(\frac{g}{\sqrt{2}}\bigg),\ \label{eq:gx2}
		\end{equation}	
		\begin{equation}
			G_{{\chi}}^{(3)} =\lambda_{\rm DM}\bigg(\frac{1}{M_{\chi}}\bigg)\big(y_{\chi}v_1\big)\bigg(\frac{1}{M_{\Sigma_1}}\bigg)\big(y_{_\Sigma}v_0\big),\ \label{eq:gx3}
		\end{equation}	
		\begin{equation}
			G_{{\chi}}^{(4)} =\big(y_{\chi}v_1\big)\bigg(\frac{1}{M_{\Sigma_1}}\bigg)(y_{_{\Sigma}}) (\sin(\gamma)),\ \label{eq:gx4}
		\end{equation}	
		\begin{equation}
			G_{{\chi}}^{(5)} =\big(y_\chi v_1\big)\bigg(\frac{1}{ M_{\Sigma_1}}\bigg)(y_{_\Sigma})\bigg(\frac{1}{M_{h_1}^2}\bigg)\bigg(\frac{m_f}{v_0}\bigg),\ \label{eq:gx5}
		\end{equation}
		where $y_{_\Sigma}=\frac{\sqrt{2m_{\nu}M_{\Sigma_1}}}{v_0}$, and $\sin(\gamma)\sim\frac{\rho_1 v_0}{M_{h_1}^2}$ is the mixing between $H$ and $\phi$.

		\begin{figure}[h]
			\centering	\includegraphics[width=7.5cm,height=7.5cm]{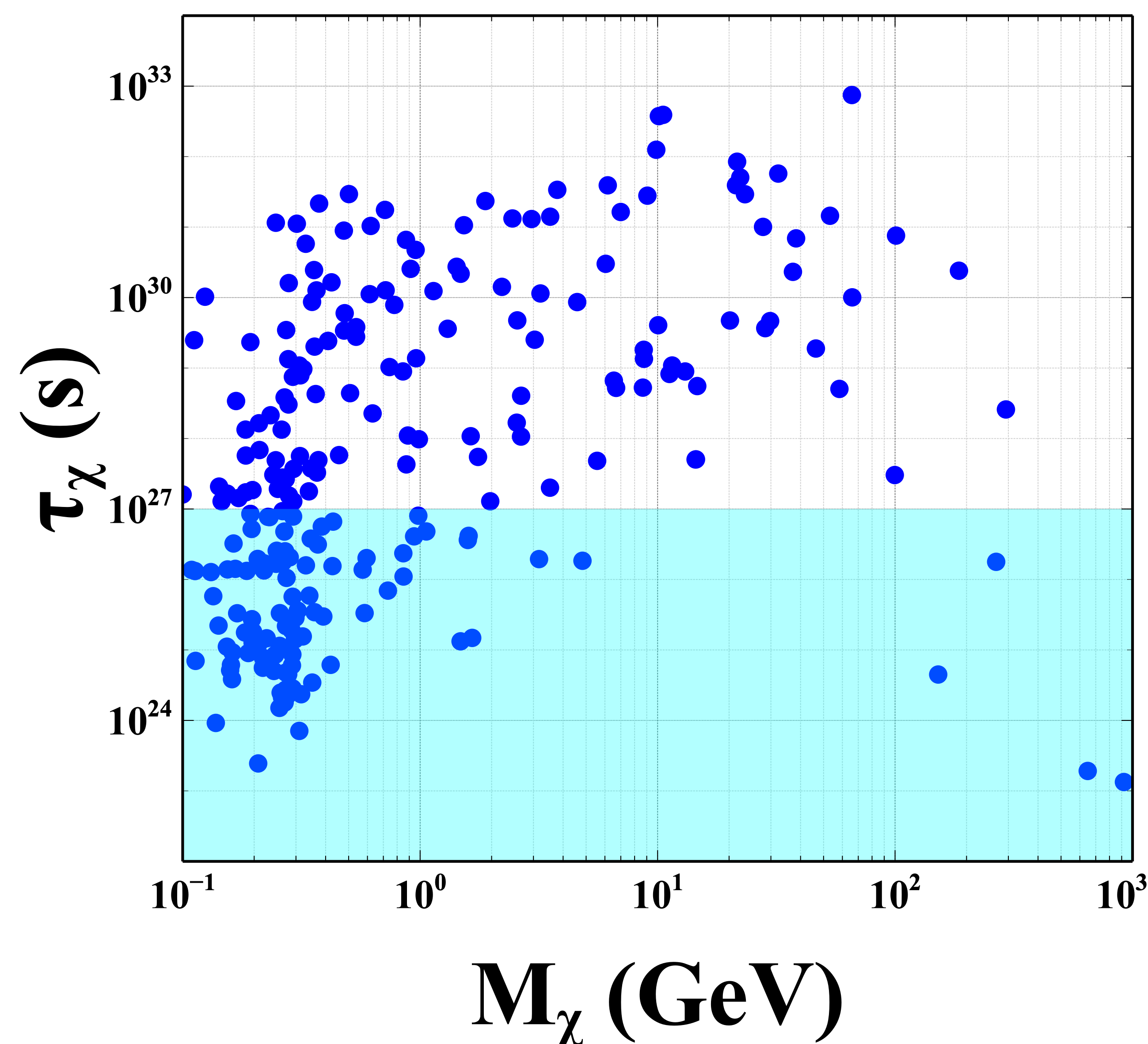}
			\includegraphics[width=7.5cm,height=7.5cm]{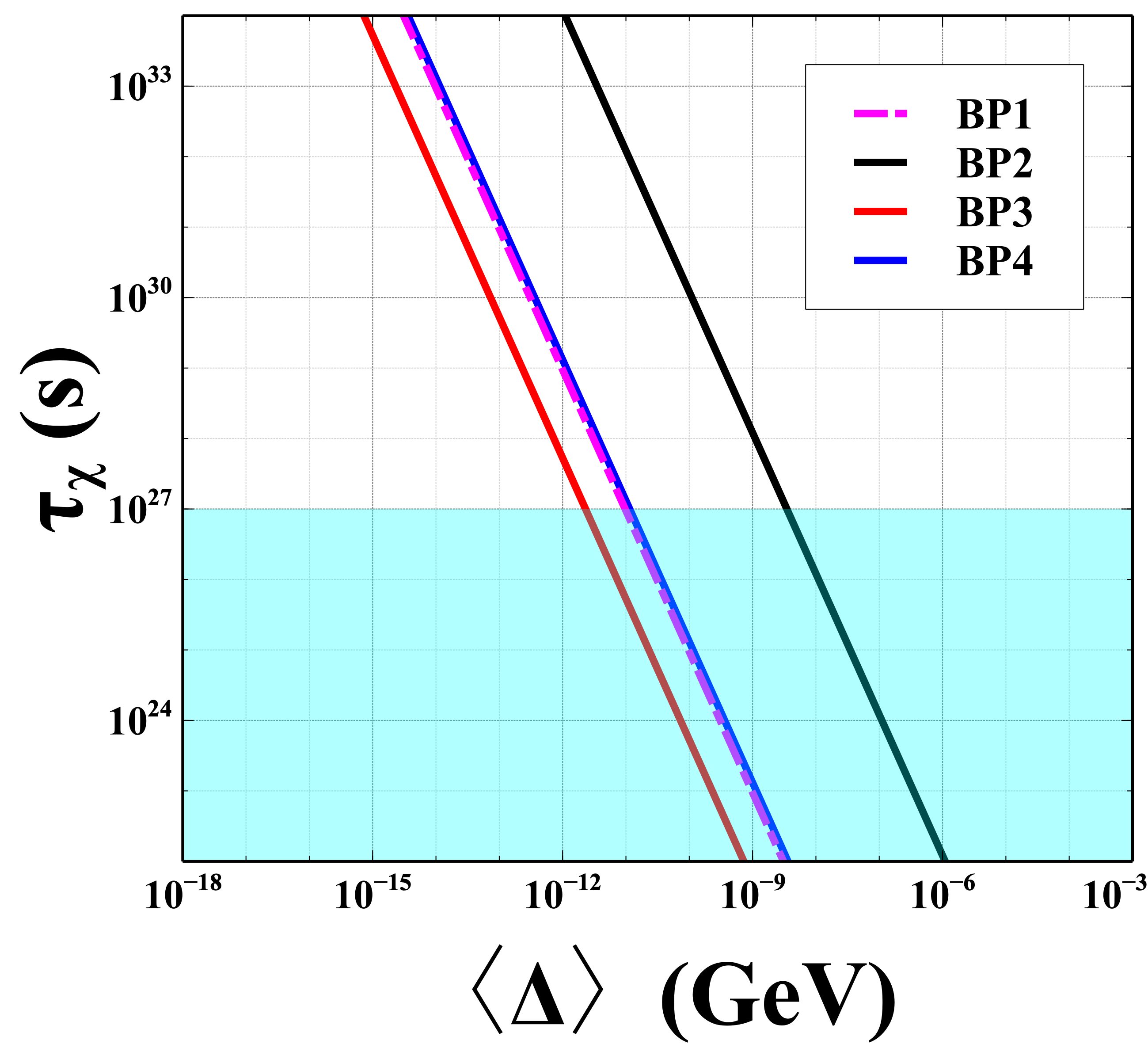}
			\caption{DM lifetime as a function of: (i) DM mass $M_\chi$ (left), and (ii) vev of the triplet scalar $\Delta$ (right). We fixed $\lambda_{\rm DM}=1$, $\sin(\gamma)=10^{-5}$, and $M_{h_2}=50$ MeV in right panel.}
			\label{fig:dmlifetime}
		\end{figure}

		When the DM mass falls within 100 MeV $< M_\chi < M_W$, the dominant decay channels are $\chi\rightarrow \nu \phi$ and $\chi\rightarrow \nu f\bar{f}$. In this range, the decay widths follow the hierarchy $\Gamma_{\chi}^{(3)}\gg\Gamma_{\chi}^{(4)},\Gamma_{\chi}^{(5)}$. As $\Gamma_{\chi}^{(3)}\propto\frac{1}{M_\chi}$, increasing DM mass leads to a longer lifetime.
		For $M_W < M_\chi < M_Z$, the $\chi\rightarrow W^+e^-$ channel becomes accessible, resulting in $\Gamma_{\chi}^{(1)}>\Gamma_{\chi}^{(3)}\gg\Gamma_{\chi}^{(5)},\Gamma_{\chi}^{(4)}$. Since $\Gamma_{\chi}^{(1)}\propto M_\chi$, the DM lifetime decreases with increasing mass.
		In the range $M_Z < M_\chi < M_{h_1}$, the $\chi\rightarrow Z\nu$ channel opens, yielding $\Gamma_{\chi}^{(1)}\sim\Gamma_{\chi}^{(2)}>\Gamma_{\chi}^{(3)}\gg\Gamma_{\chi}^{(5)},\Gamma_{\chi}^{(4)}$. Both $\Gamma_{\chi}^{(1)}$ and $\Gamma_{\chi}^{(2)}$ are proportional to $M_\chi$, further reducing the DM lifetime.
		For $M_\chi > M_{h_1}$, the $\chi\rightarrow H\nu$ channel becomes available, further increasing the decay width and decreasing the lifetime.
		The cyan shaded region in Fig \ref{fig:dmlifetime} is excluded by the constraint $\tau_{\rm DM} < 10^{27}$ sec. As we can see from the left panel of Fig~\ref{fig:dmlifetime}, this analysis shows that asymmetric DM is viable across a wide mass range from MeV to TeV.
		In the above discussion, we have fixed the values of $v_1= 1~{\rm eV},~\lambda_{\rm DM}=1$, $M_{h_2}=50$ MeV and $\sin(\gamma)=10^{-5}$ which are consistent with the direct detection constraint that we will discuss in Sec \ref{sec:directdetection}. In Sec \ref{sec:directdetection}, we will see that product of $\lambda_{\rm DM}$ and $\sin(\gamma)$ can not be less than $\mathcal{O}(10^{-7})$. This implies that there exists an upper bound on $v_1$ from the constraint $\tau_{\rm DM}>10^{27}$ s. Since $\tau_{\rm DM}$$\propto \frac{1}{v_1^2}$, from the left panel pf Fig \ref{fig:dmlifetime} it is apparent that the largest value of $v_1$ can not be larger than $\mathcal{O}(1)$ keV in order to be compatible with all other phenomenologies.
		The right panel of Fig \ref{fig:dmlifetime} illustrates the DM lifetime for benchmark points that satisfy both the correct relic density and lepton asymmetry, as presented in Table \ref{tab:tab3}.

		\section{Detection Prospects}\label{sec:detection}
		\subsection{Constraints from Direct Detection of Dark Matter}\label{sec:directdetection}
\begin{figure}[h]
			\centering
			\includegraphics[scale=1]{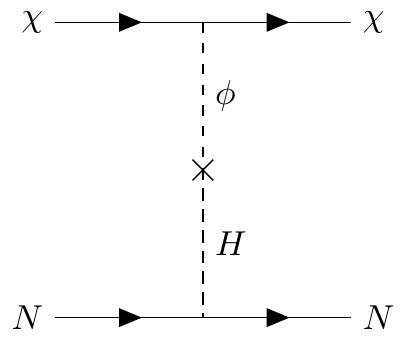}
			\caption{The spin-independent scattering of DM-nucleon (N) via Higgs portal.}
			\label{fig:DDdiag}
		\end{figure}

		The spin-independent (SI) elastic scattering of DM is possible through $H-\Phi$ mixing, where DM particles can scatter off the target nuclei at terrestrial laboratories. The Feynman diagram for this process is shown in Fig \ref{fig:DDdiag}.
		The spin-independent elastic scattering cross-section of DM per nucleon can be expressed as \cite{Ellis:2008hf}
		\begin{equation}
			\sigma^{SI}_{DM-N}=\frac{\mu_r^2}{\pi A^2}[Z f_p + (A-Z) f_n]^2,\
		\end{equation}
		where $\mu_r=\frac{M_\chi m_n}{M_\chi+m_n}$ is the reduced mass, $m_n$ is the nucleon (proton or
		neutron) mass, $A$ is the mass number of target nucleus, $Z$ is the atomic number of target nucleus. The $f_p$ and $f_n$ are the interaction strengths of proton and neutron with DM, respectively, and are given as
		\begin{equation}
			f_{p,n}=\sum_{q=u,d,s} f_{T_q}^{p,n} \alpha_q \frac{m_{p,n}}{m_q}+\frac{2}{27} f_{T_G}^{p,n} \sum_{q=c,t,b}\alpha_q \frac{m_{p,n}}{m_q},\
		\end{equation}
		where
		\begin{equation}
			\alpha_q=\lambda_{\rm DM}*\frac{m_q}{v_0}*\sin(\gamma)\cos(\gamma)(\frac{1}{M^2_{h_2}}-\frac{1}{M^2_{h_1}}),\
		\end{equation}
		with $\gamma$ being the mixing angle between $H$ and $\Phi$. The values of $f_{T_q}^{p,n}$, $f_{T_G}^{p,n}$ can be found in \cite{Ellis:2000ds}. 
		\begin{figure}[h]
			\centering
			\includegraphics[width=7.5cm,height=7.5cm]{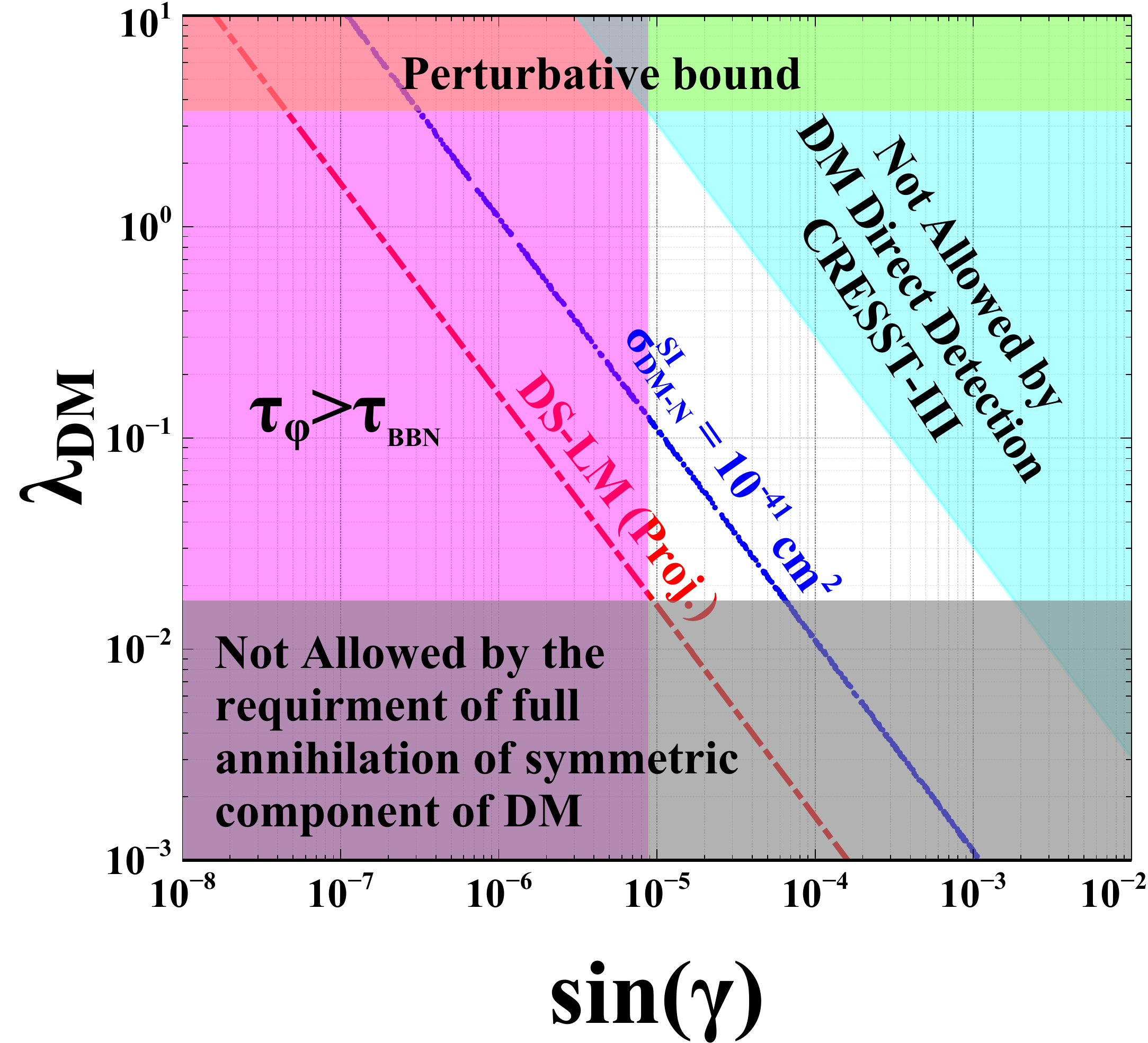}
			\includegraphics[width=7.5cm,height=7.5cm]{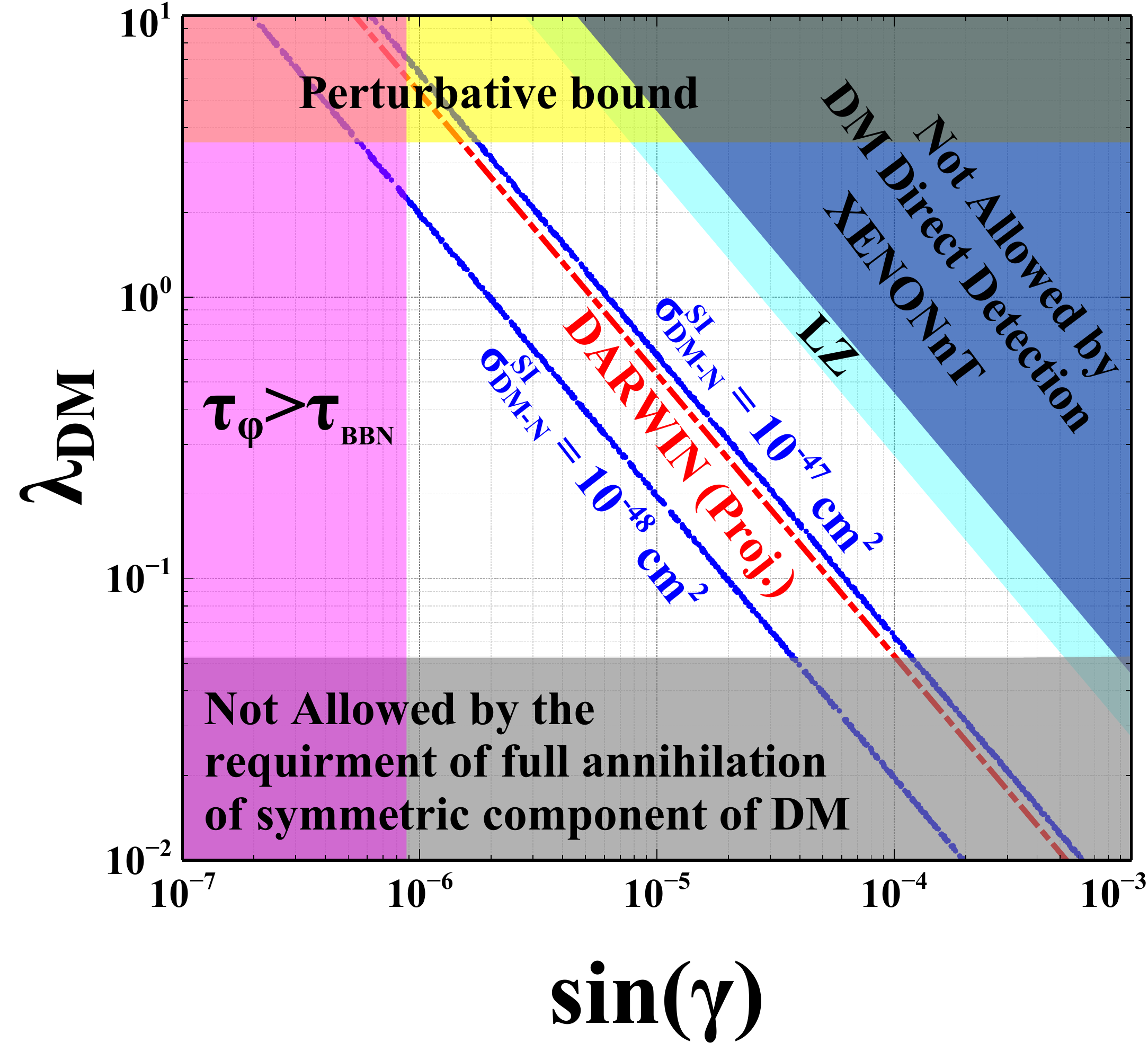}
			\caption{The spin-independent scattering of DM-nucleon (N) via Higgs portal. [left]: $M_\chi=1$ GeV, $M_{h_2}=50$ MeV, [right]: $M_\chi=10$ GeV, $M_{h_2}=5$ GeV.}
			\label{fig:lvsgamma}
		\end{figure}
		In Fig \ref{fig:lvsgamma}, we showcase the constraints from DM direct search, full annihilation of symmetric component of DM and perturbative bound on the Yukawa coupling $\lambda_{\rm DM}$ in the plane of $\lambda_{\rm DM}$ vs $\sin(\gamma)$ for two different set of $M_{\chi}$ and $M_{h_2}$. $\sin(\gamma)$ can not be extremely small. For a given value of $M_{h_2}$, $\sin(\gamma)$ is lower bounded by BBN. For a conservative value of $\tau_{_{BBN}}$=1 sec, the magenta-shaded region, shown in the left and right panels of Fig \ref{fig:lvsgamma}, is disfavored. The cyan shaded region is disallowed by the direct detection constraints from the CRESST-III (LZ) experiment \cite{CRESST:2019jnq,LZ:2022lsv} which corresponds to spin-independent cross-section $7.96\times10^{-39}$ $\rm cm^2$ ($1.98\times10^{-46}$ $\rm cm^2$) for $1$ GeV ($10$ GeV) DM. We have also shown the XENONnT \cite{XENON:2023cxc} bound in blue color for 10 GeV DM. The projected sensitivities of the DS-LM\cite{GlobalArgonDarkMatter:2022ppc} and DARWIN \cite{DARWIN:2016hyl} experiments have also been shown for respective DM masses $1$ GeV and $10$ GeV by red dot-dashed lines. Here, it should be mentioned that for the complete annihilation of the symmetric part of the DM $\chi$, its freeze-out cross-section should be~ $>2.6\times10^{-9}$ $\rm GeV^{-2}$. The gray shaded region gets ruled out from the requirement of complete annihilation of symmetric component for DM as in this region $\lambda_{\rm DM}$ is small and hence $\big<\sigma v\big>_{\chi\chi\to\phi\phi}$~$ <2.6\times10^{-9}~ \rm GeV^{-2}$. The yellow shaded region is ruled out from the perturbativity constraints on $\lambda_{\rm DM}$ {\it i.e.} $\lambda_{\rm DM} <\sqrt{4\pi}$. We have also shown SI DM-nucleon scattering cross-section in the same plane: $\{10^{-41}~\rm cm^2$\} for 1 GeV DM and \{$10^{-47}~\rm cm^2$ and $10^{-48}~\rm cm^2$\} for 10 GeV DM. Clearly, future direct search experiments with enhanced sensitivities will be able to probe this parameter space\cite{SuperCDMS:2016wui,Agnes:2018oej,GlobalArgonDarkMatter:2022ppc,DARWIN:2016hyl}.
		
		\subsection{Collider Signature}\label{collider_signature}
\begin{figure}[h]
			\centering
			\includegraphics[scale=0.65]{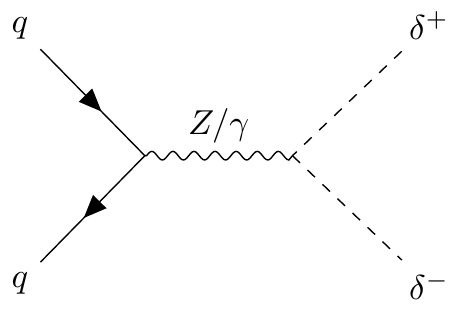} ~
			\includegraphics[scale=0.5]{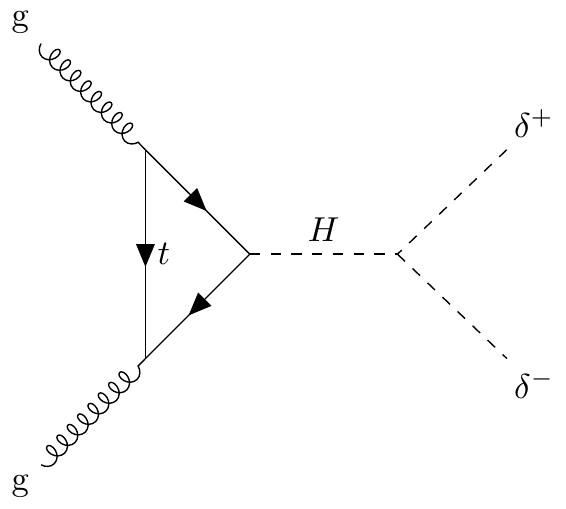}~
			\includegraphics[scale=0.5]{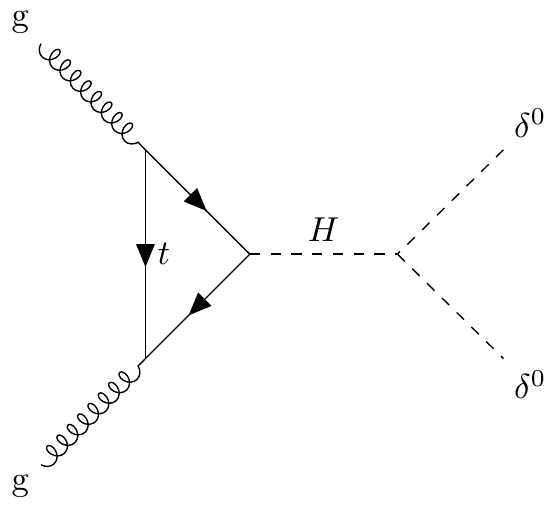}\vspace{8mm}\\
			\includegraphics[scale=0.65]{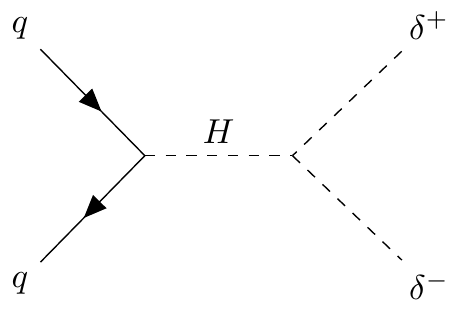} ~
			\includegraphics[scale=0.65]{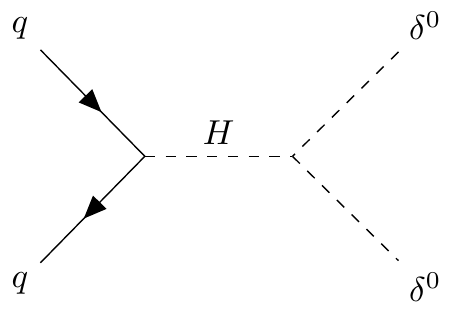}~~
			\includegraphics[scale=0.65]{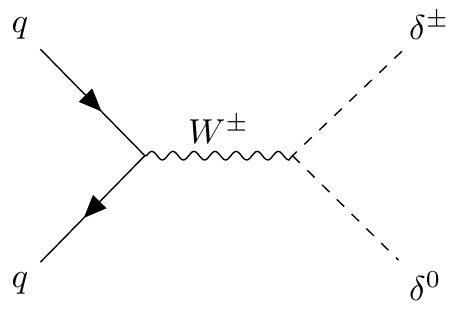}
			\caption{Feynman diagrams for the production of $\delta^\pm$, $\delta^0$ at hadron collider.}
			\label{fig:deltaprod}
		\end{figure}

		In this section, we briefly mention the potential detection prospects of this scenario at collider experiments. 
		In Fig \ref{fig:deltaprod}, the production channels of the scalar triplet through proton-proton collision is shown.  Once the charged scalar ($\delta^\pm$) is produced, because of phase space suppression, it can travel a certain distance before decaying to neutral scalar ($\delta^0$) and $\pi^\pm$, leaving displaced vertex signature at the colliders as already mentioned at the end of section~\ref{sec:model}. 

\begin{figure}[htb]
			\centering
			\includegraphics[width=7.5cm,height=7.5cm]{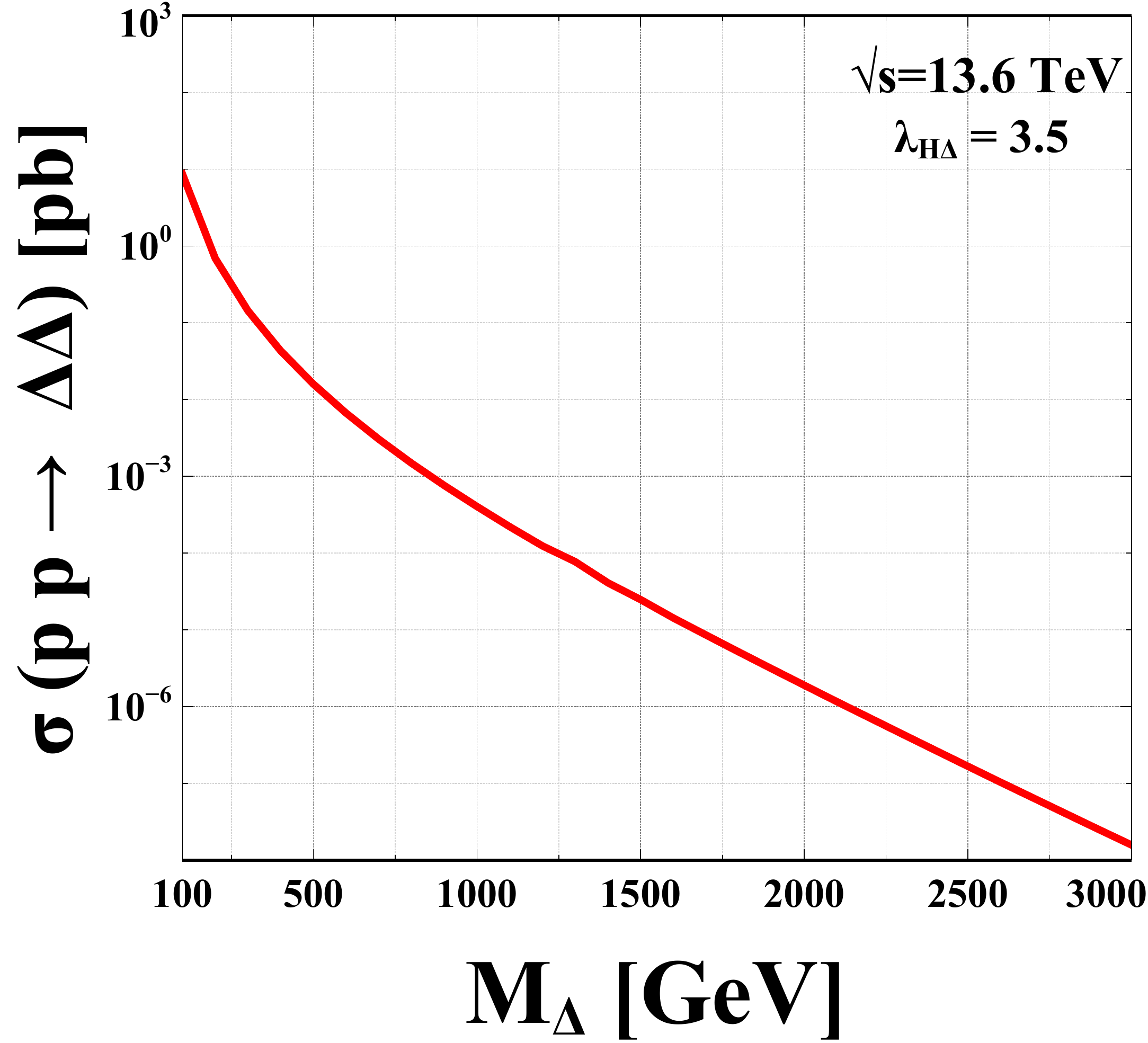}
			\includegraphics[width=7.5cm,height=7.5cm]{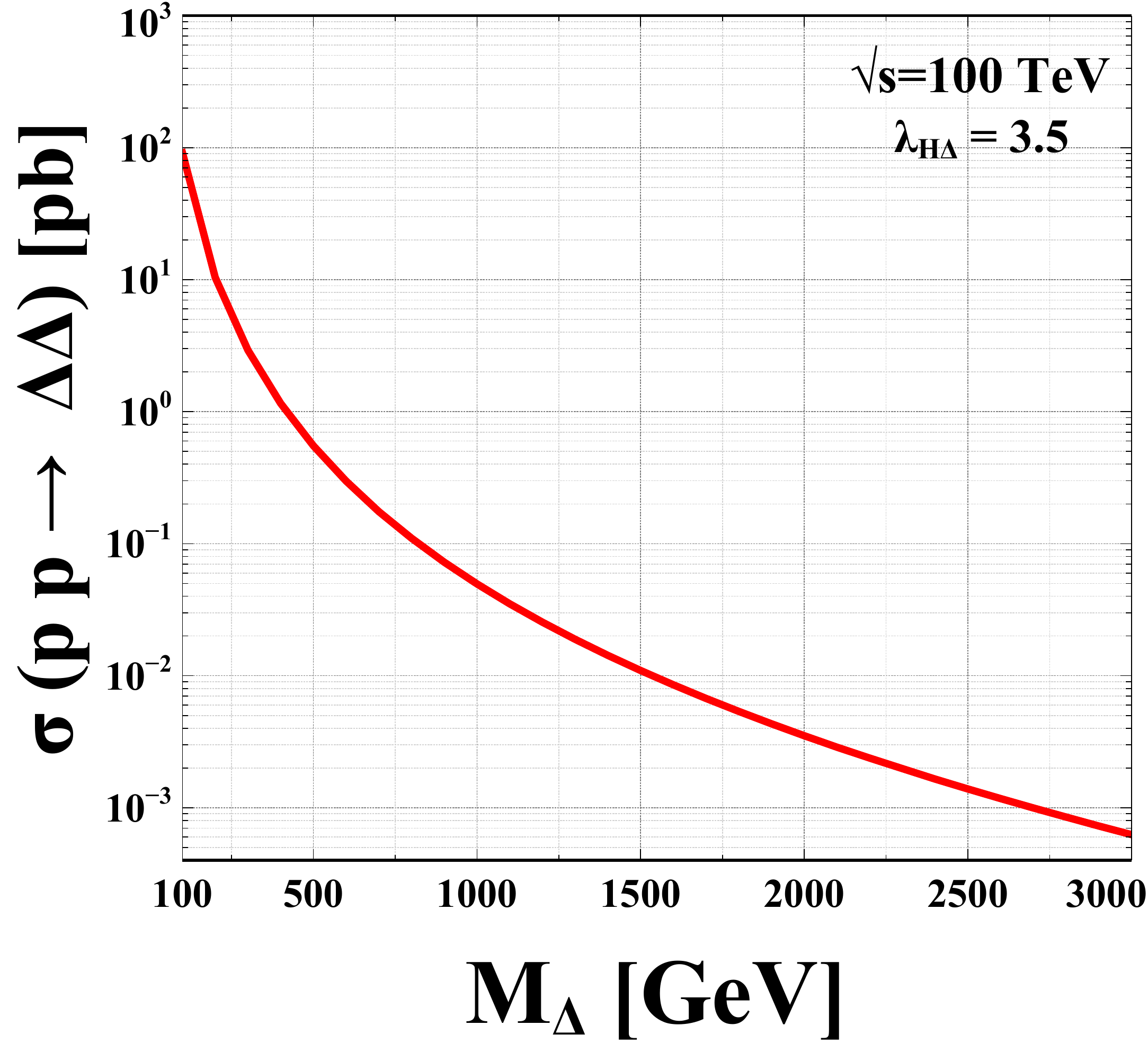}
			\caption{Production cross-section of triplet scalar $\Delta = \delta^{\pm,0}$ as a function of $M_\Delta$ at 13.6 TeV (left), and at 100 TeV (right).}
			\label{fig:pptodd}
		\end{figure}

		We have calculated the production cross-section of the triplet scalar in $p-p$ collisions, which is shown as a function of $M_\Delta$ at center of mass energies $\sqrt{s}=13.6$ TeV and $\sqrt{s}=100$ TeV in the left and right panels of Fig \ref{fig:pptodd} using CalcHEP-3.8.9 \cite{Belyaev:2012qa} with the NNPDF23\_lo\_as\_0120\_qed  parton distribution function. Clearly, with an increase in $M_{\Delta}$, the production cross-section decreases.  
		\begin{figure}[htb]
			\centering
			\includegraphics[width=7.5cm,height=7.5cm]{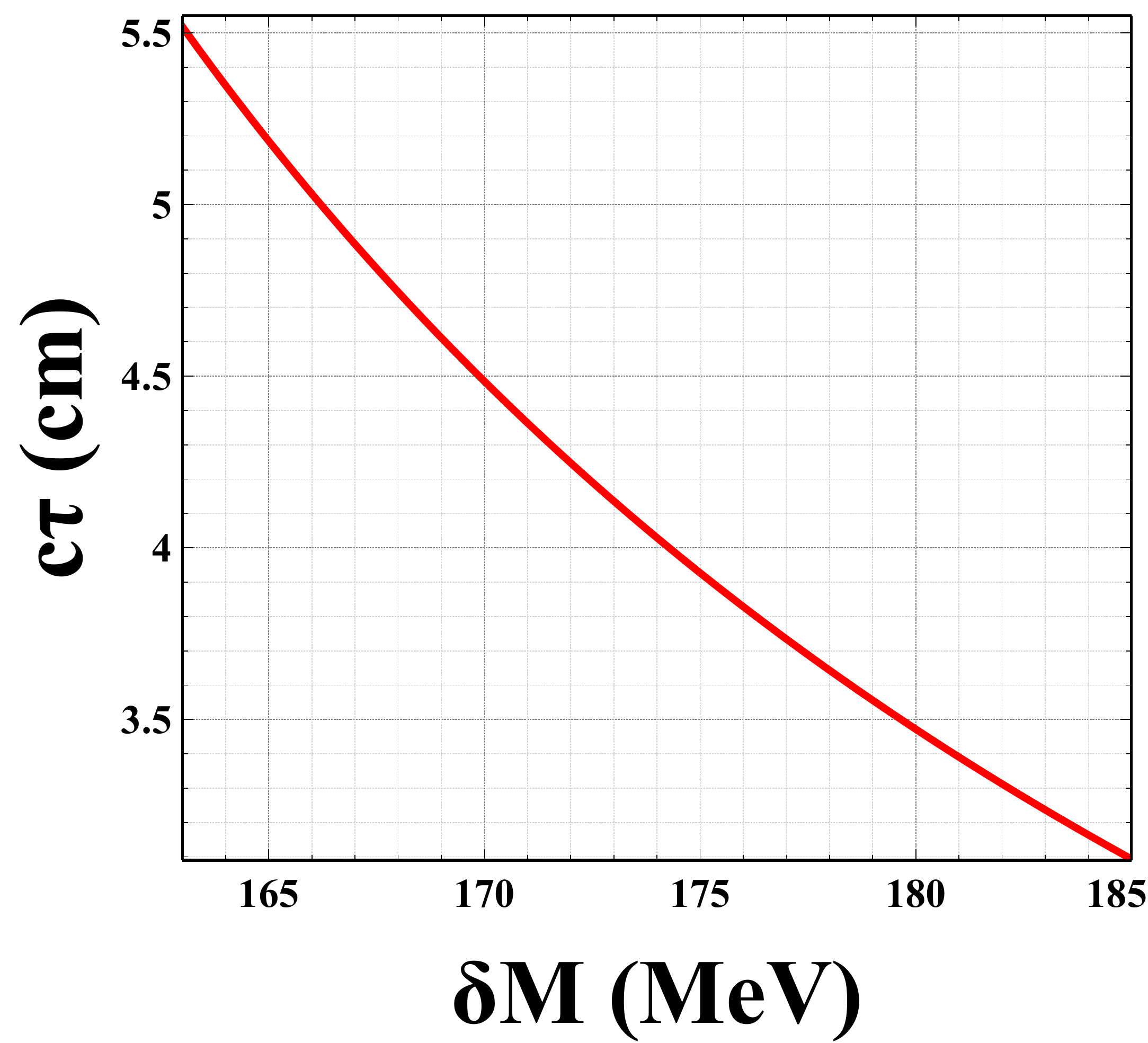}
			\caption{Variation of $c\tau$ with the mass splitting of the scalar triplet.}
			\label{fig:ctauvsdm}
		\end{figure}
		
		The mass splitting ($\delta m$) between $\delta^\pm$ and $\delta^0$ generated due to radiative corrections facilitates the ($\delta^\pm \rightarrow \delta^0 \pi^\pm$) decay channel, and its corresponding decay width is given as \cite{Cirelli:2005uq}:
		\begin{equation}
			\Gamma(\delta^\pm \rightarrow \delta^0 \pi^\pm)=\frac{2 G_F^2}{\pi}f_\pi^2V_{ud}^2(\delta M)^3\sqrt{1-\frac{m_\pi^2}{(\delta m)^2}},\
			\label{Eq:deltadecay}
		\end{equation}
		where $G_F$ is the Fermi constant, $f_\pi$($=131$ MeV) is the pion decay constant, $V_{ud}$ is the relevant CKM matrix element, and $m_\pi$ is the mass of pion. 
		Here it should be mentioned that, in addition to this dominant decay channel which accounts for 98\% branching fraction, there are two other decay modes of $\delta^\pm$, ( $\delta^\pm \rightarrow\delta^0\mu^\pm\nu_\mu(\bar{\nu_\mu})$ and $\delta^\pm \rightarrow\delta^0e^\pm\nu_e(\bar{\nu_e})$ ) accounting for 2\% of the branching fraction. We calculate the corresponding decay length $c\tau$, which is shown as a function of the mass splitting in Fig \ref{fig:ctauvsdm}.  Such displaced vertex signatures
		can be probed as a signature of the verifiability of this model
		at present and future colliders.	
		
		\section{Conclusion}
		\label{sec:conclude}
		In this paper, we studied an economical model to explain the neutrino mass, DM, baryon asymmetry of the universe within the type-III seesaw framework. We exploited the possibility of asymmetric DM in this scenario such that lepton and DM asymmetries are generated through similar mechanisms in both visible and dark sectors. We extended the type-III seesaw model with a vector-like singlet fermion ($\chi$) and a hypercharge zero scalar triplet ($\Delta$). A $Z_2$ symmetry is enforced, with $\chi$ and $\Delta$ being odd and all other particles being even. As a result, the lightest $Z_2$ odd particle $\chi$ acts as a DM candidate. Because of the assumed hierarchy among the triplet fermion masses ($M_{\Sigma_1}<M_{\Sigma_2}<M_{\Sigma_3}$), only the decay of the lightest hypercharge zero fermion triplet, $\Sigma_1$ to the SM sector and the dark sector produces lepton asymmetry and DM asymmetry simultaneously, and the symmetric component of the DM is annihilated away by the process $\bar{\chi}\chi\rightarrow\phi\phi$ ($\phi$ being a light $Z_2$ even scalar) to obtain the observed relic abundance. We note that, unlike leptogenesis in type-I seesaw, here in type-III seesaw scenario, because of the additional gauge interactions, the fermion triplets get thermalized in the early universe and for $M_{\Sigma_1}<10^{14}$ GeV, even if one starts with zero initial abundance, the situation resembles the case where thermal initial abundance is assumed. Thus, the final lepton asymmetry remains independent of the initial condition of $\Sigma_1$ number density in the early Universe. 
		
		After EWSB, $\Delta$ acquires a small induced vev and makes the DM unstable. To establish its viability as a DM candidate, it becomes imperative to guarantee its stability throughout the cosmological time scale. Our analysis reveals that a substantial range of DM masses remains permissible, provided one meticulously adjusts the vev of $\Delta$, thereby addressing the lifetime constraint. We find the relevant parameter space consistent with all the pertinent constraints from DM phenomenology and successful leptogenesis.
		In addition to the detection prospects of DM, we also explored the possibility of probing this scenario at colliders where because of the tiny mass splitting between $\delta^{\pm}$ and $\delta^0$, which is induced by the radiative corrections. This small mass splitting can give rise to displaced vertex signatures of $\delta^{\pm}$.
		
		\acknowledgments 
		The work of P.K.P., N.S., and P.S. is supported by the Department of Atomic Energy-Board of Research in Nuclear 
		Sciences, Government of India (Ref. Number: 58/14/15/2021- BRNS/37220). S.M. acknowledges the
		support from National Research Foundation of Korea
		grant 2022R1A2C1005050.
		
		\appendix
		\section{Lepton number to baryon number conversion}\label{App:LtoB}
		The asymmetry in the equilibrium number densities of
		particle $n_i$ over antiparticle $\bar{n}_i$ can be written in the limit $\mu_i/T\ll1$ as,\cite{Harvey:1990qw}
		\begin{eqnarray}
			n_i-\bar{n}_i&=&\frac{1}{6}g_i T^3 \bigg(\frac{\mu_i}{T} \bigg), ~~ \textrm{fermion}\nonumber\\&=&\frac{1}{3}g_i T^3 \bigg(\frac{\mu_i}{T} \bigg), ~~ \textrm{boson}
		\end{eqnarray}
		For a general derivation, we assume that the SM consists of $N$ generations of quarks and leptons, $m$ complex Higgs doublets. We extended the SM by $N^\prime$ generations of Fermion triplets ($\Sigma$), one scalar triplet ($\Delta$), one vector-like fermion ($\chi$), and one singlet scalar ($\phi$). The chemical potentials of the SM fields are assigned as follows: $\mu_W$ for $W^-$, $\mu_0$ for $m$ $\phi^0$ Higgs fields, $\mu_{-}$ for $m$ $\phi^-$ Higgs fields, $\mu_{uL}$ for all left-handed up-quark fields, $\mu_{dL}$ for all left-handed down-quark fields,  $\mu_{u_R}$ for all the right-handed
		up-quark fields, $\mu_{d_R}$ for all the right-handed
		down-quark fields, $\mu_i$, for the left-handed neutrino fields,
		$\mu_{iL}$ for the left-handed charged lepton fields, and $\mu_{iR}$ for
		the right-handed charged Lepton fields. The chemical potentials for the BSM fields are assigned as: $\mu_{\Sigma^0}$ for all $N^\prime$ $\Sigma^0$, $\mu_{\Sigma^+}$ for all $\Sigma^+$, $\mu_\delta$ for the $\delta^-$, $\mu_{\delta^0}$ for $\delta^0$, $\mu_\chi$ for $\chi$, and $\mu_\phi$ for $\phi$.  \\
		Now, rapid interactions in the early Universe enforce the following equilibrium relations among the chemical potentials:
		{
			\begin{eqnarray}
				W^-\leftrightarrow \phi^-+\phi^0 &\Rightarrow& \mu_W=\mu_{-}+\mu_0\\
				W^-\leftrightarrow \bar{u}_L+d_L &\Rightarrow& \mu_{d_L}=\mu_W+\mu_{u_L}\\
				W^-\leftrightarrow \bar{\nu}_{iL}+e_{iL} &\Rightarrow& \mu_{iL}=\mu_W+\mu_{i}\\
				\phi^0\leftrightarrow \bar{d}_R+d_L &\Rightarrow& \mu_{d_R}=-\mu_0+\mu_W+\mu_{u_L}~~~~~~\\
				\phi^0\leftrightarrow e_{iL}+\bar{e}_{iR} &\Rightarrow& \mu_{iR}=-\mu_0+\mu_W+\mu_{i}~~~~\\
				\phi^0\leftrightarrow \bar{u}_L+u_R &\Rightarrow& \mu_{u_R}=\mu_0+\mu_{u_L}\\
				W^-+W^-\leftrightarrow \delta^-+\delta^- &\Rightarrow& \mu_{W}+\mu_W=\mu_\delta+\mu_\delta\nonumber\\&\Rightarrow& \mu_\delta=\mu_W\\
				W^-+W^3\leftrightarrow \delta^-+\delta^0 &\Rightarrow& \mu_{W}+0=\mu_\delta+\mu_{\delta^0}\nonumber\\&\Rightarrow& \mu_{\delta^0}=0\\
				\phi\leftrightarrow \bar{\chi}+\chi&\Rightarrow& \mu_{\phi}=-\mu_\chi+\mu_\chi\nonumber\\&\Rightarrow&\mu_\phi=0\\
				\Sigma^0\leftrightarrow W^-+\Sigma^+&\Rightarrow& \mu_{\Sigma^0}=\mu_W+\mu_{\Sigma^+}\\
				\Sigma^0\leftrightarrow\delta^0+\chi&\Rightarrow&\mu_{\Sigma^0}=\mu_{\delta^0}+\mu_\chi\nonumber\\&\Rightarrow&\mu_\chi=\mu_{\Sigma^0}\\
				\nu_{iL}\leftrightarrow\Sigma^0+\phi^0&\Rightarrow&\mu_i=\mu_{\Sigma^0}+\mu_0\label{eq:A13}
		\end{eqnarray}}
		From Eq. \eqref{eq:A13},
		\begin{eqnarray}
			\sum_i \mu_i&=&\sum_i \mu_{\Sigma^0}+\sum_i\mu_0\nonumber\\&\Rightarrow&\mu=N\mu_{\Sigma^0}+N\mu_0\Rightarrow\mu=N\mu_{\chi}+N\mu_0\nonumber\\&\Rightarrow&\mu_\chi=\frac{\mu-N\mu_{0}}{N}.
		\end{eqnarray}
		Now, the EW$B+L$ anomaly implies the existence of processes that correspond to the creation of $u_Ld_Ld_L\nu_L$ from each generation out of the vacuum.	As long as these interactions are rapid, we have
		\begin{eqnarray}
			N(\mu_{u_L}+2\mu_{d_L})+\sum_i\mu_i=0\Rightarrow3N\mu_{u_L}+2N\mu_W+\mu=0.\nonumber\\
		\end{eqnarray}
		Let us now express the baryon ($B$), lepton ($L$), charge ($Q$), and third component of weak isospin ($Q_3$) number densities as
		\begin{eqnarray}
			B&=&3N(\frac{1}{3}\mu_{u_L}+\frac{1}{3}\mu_{u_R})+3N(\frac{1}{3}\mu_{d_L}+\frac{1}{3}\mu_{d_R})=4N\mu_{u_L}+2N\mu_W
		\end{eqnarray}
		\begin{eqnarray}
			L=\sum_i(\mu_i+\mu_{iL}+\mu_{iR})+\mu_\chi=3\mu+2N\mu_W-N{\mu_0}+\mu_\chi\nonumber\\
		\end{eqnarray}
		\begin{eqnarray}
			Q&=&3N(\frac{2}{3}\mu_{u_L}-\frac{1}{3}\mu_{d_L}+\frac{2}{3}\mu_{u_R}-\frac{1}{3}\mu_{d_R})-\big(\sum_i\mu_{iL}+\sum_i\mu_{iR}\big)-2*2\mu_W-2m\mu_--2*2\mu_\delta+2*2(-\mu_\delta)\nonumber\\&=& 2N\mu_{u_L}-2\mu-(4N+2m+12)\mu_W+(4N+2m)\mu_0\nonumber\\
		\end{eqnarray}
		\begin{eqnarray}
			Q_3&=&3N(\frac{1}{2}\mu_{u_L}-\frac{1}{2}\mu_{d_L})+\sum_i\big(\frac{1}{2}\mu_i-\frac{1}{2}\mu_{iL}\big)-2*2\mu_W+2m(\frac{1}{2}\mu_+-\frac{1}{2}\mu_0)+2*2(-1)\mu_\delta+2*2(+1)(-\mu_\delta)\nonumber\\&=&-(2N+m+12)\mu_W
		\end{eqnarray}
		Now above the critical temperature both $Q$, and $Q_3$ are zero, which will give the $B$, and $L$ in terms of $\mu_{u_L}$ as
		\begin{eqnarray}
			L=\mu_{uL} N \left(\frac{4 (N+1)}{m+2 N}-3 \left(\frac{1}{N}+3\right)\right)
		\end{eqnarray}
		\begin{eqnarray}
			B=4 N \mu_{u_L}
		\end{eqnarray}
		\begin{eqnarray}
			B=-\frac{4}{3 \left(3+\frac{1}{N}\right)-\frac{4 (N+1)}{m+2 N}}L
		\end{eqnarray}
		
		\section{Relevant cross section and decay widths}	
		\label{appen1}
		\subsection{Decay Rates}
		The total decay rate is given in Eq \ref{Eq:Decayrate}, and the thermal averaged decay width is given by
		{\small\begin{equation}
				\Gamma_D=<\Gamma_{\Sigma_1}>=\Gamma_{\Sigma_1}\frac{K_1(z)}{K_2(z)},\
		\end{equation}}
		where $z=\frac{M_{\Sigma_1}}{T}$, $K_1(z)$ and $K_2(z)$ are the modified Bessel functions of 1st and 2nd kind respectively.
		\subsection{Gauge Scatterings}
		The reduced gauge scattering cross-section is given as \cite{Hambye:2003rt,AristizabalSierra:2010mv}
		{\begin{eqnarray}
				\hat{\sigma}_A&=\frac{6g^4}{\pi}\big(1+\frac{2}{x} \big)r+\frac{2g^4}{\pi}\bigg[ -r\big(4+\frac{17}{x}\big)+3\big(1+\frac{4}{x}-\frac{4}{x^2}\big)\ln\big[\frac{1+r}{1-r}\big] \bigg]
				\label{Eq:gaugescatteringcross},\ \nonumber\\
		\end{eqnarray}}
		where $r=\sqrt{1-4/x}$, $x=s/M^2_{\Sigma_1}$, $s$ is the center of mass energy.
		
		The total cross-section is given as,
		\begin{eqnarray}
			\sigma_A=\frac{\hat{\sigma}_A}{2s^{-1}\lambda[s,M^2_{\Sigma_1},M^2_{\Sigma_1}]},
		\end{eqnarray}
		where	$	\lambda[a,b,c]=\sqrt{(a-b-c)^2-4bc}.
		$
		
		Thermal averaged cross-section for the annihilation of AA to BB is given by \cite{Gondolo:1990dk}
		\begin{eqnarray}
			<\sigma v>_{AA\leftrightarrow BB}=\frac{z}{2[K_1^2(z)+K_2^2(z)]}\int_2^\infty dy~\sigma_{(AA\leftrightarrow BB)}(y^2-4)y^2K_1(yz),\label{eq:thermalavg}
		\end{eqnarray}
		where $y=\frac{\sqrt{s}}{M_A}$.
		\subsection{$\Delta$L=2 Scatterings}
		
		\begin{eqnarray}
			\Gamma_{2W}=<\sigma_{\chi\Delta\rightarrow LH}v> \sqrt{n^{eq}_l* n^{eq}_h} +<\sigma_{\chi L\rightarrow \Delta H}v> \sqrt{n^{eq}_h* n^{eq}_\Delta}+<\sigma_{\chi H\rightarrow \Delta L}v> \sqrt{n^{eq}_l* n^{eq}_\Delta}\nonumber\\
		\end{eqnarray}
		
		\begin{eqnarray}
			\Gamma^W_{2L}=<\sigma_{LL\rightarrow HH(t-channel)}v> n^{eq}_h+<\sigma_{LH\rightarrow LH}v> \sqrt{n^{eq}_l*n^{eq}_h}+<\sigma_{ \Sigma_1\Sigma_1\leftrightarrow  LL}v>n^{eq}_l
		\end{eqnarray}
		
		\begin{eqnarray}
			\Gamma^{W}_{2\chi}=<\sigma_{ \chi \Delta\leftrightarrow  \Delta\chi}v>\sqrt{n^{eq}_\chi*n^{eq}_\Delta}+<\sigma_{ \chi \chi\leftrightarrow  \Delta\Delta}v>n^{eq}_\Delta+<\sigma_{ \Sigma_1 \Sigma_1\leftrightarrow  \chi\chi}v>n^{eq}_\chi+<\sigma_{ \chi \Sigma_1\leftrightarrow  \chi\Sigma_1}v>\sqrt{n^{eq}_\chi*n^{eq}_\Sigma}\nonumber\\
		\end{eqnarray}
		
		{\small\begin{eqnarray}
				\sigma_{LL\leftrightarrow HH (t-channel)}&=   \frac{y_{_\Sigma}^4 M_{\Sigma_1}^2}{64 \pi}\Bigg(\frac{\sqrt{1-\frac{4 M_{h_1}^2}{s}}}{(M_{h_1}^2-M_{\Sigma_1}^2)^2+M_{\Sigma_1}^2s}-\frac{2\log \Bigg[\frac{-2M_{h_1}^2+2M_{\Sigma_1}^2+s-\sqrt{1-\frac{4M_{h_1}^2}{s}}s}{-2M_{h_1}^2+2M_{\Sigma_1}^2+s+\sqrt{1-\frac{4M_{h_1}^2}{s}}s}\Bigg]}{s(-2M_{h_1}^2+2M_{\Sigma_1}^2+s)}\Bigg)
				\label{Eq:delL2lltohh}
		\end{eqnarray}}
		{\small\begin{eqnarray}
				\sigma_{LH\leftrightarrow \bar{L}H^\dagger (t-channel)}&=\frac{y_{_\Sigma}^4 M_{\Sigma_1}^2}{64 \pi}\Bigg(\frac{1}{-M_{h_1}^4+M_{\Sigma_1}^2s}-\frac{\log\Big[\frac{(2M_{h_1}^2-M_{\Sigma_1}^2-s)s}{M_{h_1}^4-M_{\Sigma_1}^2s}\Big]}{(M_{h_1}^2-s)^2}\Bigg)
				\label{Eq:delL2lhtolbarh}
		\end{eqnarray}}
		{\small\begin{eqnarray}
				\sigma_{\chi\Delta\leftrightarrow \bar{L} H^\dagger}&=-\frac{y_\chi^2y_{_\Sigma}^2(M_{h_1}^2-s)^3(4M_{\Sigma_1} M_\chi s+M_{\Sigma_1}^2(-M_{\Sigma_1}^2+M_\chi^2+s)+s(-M_{\Sigma_1}^2+M_\chi^2+s))}{128\pi(M_{\Sigma_1}^2-s)^2s^4\sqrt{\frac{(M_{h_1}^2-s)^2(M_{\Sigma_1}^4+(M_\chi^2-s)^2-sM_{\Sigma_1}^2(M_\chi^2+s))}{s^4}}}
				\label{Eq:delL2xdlh}
		\end{eqnarray}}
		{\small\begin{eqnarray}
				\sigma_{\chi L\leftrightarrow \Delta^\dagger H^\dagger}&=&\frac{y_{_\Sigma}^2 y_\chi^2}{64\pi (M_\chi^2-s)^2} \Bigg(-\Bigg(\Bigg(s\sqrt{\frac{(M_\chi^2-s)^2(M_{h_1}^4+(M_\Delta^2-s)^2-2M_{h_1}^2(M_\Delta^2+s))}{s^4}}(-M_\Delta^4(M_\chi^2-4s)\nonumber\\&+&M_{h_1}^2(M_\chi^2-2s)(-M_{\Sigma_1}^2+M_\Delta^2+M_\chi^2-s)+2M_{\Sigma_1}^4s+2M_{\Sigma_1}^3M_\chi s- 2M_{\Sigma_1} M_\Delta^2 M_\chi s\nonumber\\&+&M_\Delta^2(M_\chi^2-2s)s+M_{\Sigma_1}^2(M_{\Sigma_1}^2(M_\chi^2-6s)+s^2))\Bigg)\Big/(M_{h_1}^2(M_\chi^2-s)(-M_{\Sigma_1}^2+M_\Delta^2\nonumber\\&+&M_\chi^2-s)+(M_{\Sigma_1}-M_\Delta)(M_{\Sigma_1}+M_\Delta)(M_\Delta^2(M_\chi^2-2s)+s(N_\Sigma^2-M_\chi^2+s)))\Bigg)\nonumber\\&+&(M_{h_1}^2-3M_\Delta^2+2M_{\Sigma_1}(M_{\Sigma_1}+M_\chi)+s)\log\Bigg[\Bigg(M_\Delta^2(M_\chi^2-3s)+M_{h_1}^2(-M_\chi^2+s)\nonumber\\&+&s\Bigg(2M_{\Sigma_1}^2-M_\chi^2+s+s\sqrt{\frac{(M_\chi^2-s)^2(M_{h_1}^4+(M_\Delta^2-s)^2-2M_{h_1}^2(M_\Delta^2+s))}{s^4}} \Bigg)\Bigg)\Big/\Bigg(M_\Delta^2(M_\chi^2-3s)\nonumber\\&&M_{h_1}^2(-M_\chi^2+s)+s\Bigg(2M_{\Sigma_1}^2-M_\chi^2+s-s\sqrt{\frac{(M_\chi^2-s)^2(M_{h_1}^4+(M_\Delta^2-s)^2-2M_{h_1}^2(M_\Delta^2+s))}{s^4}}\Bigg)\Bigg)\Bigg]\Bigg)\nonumber\\
				\label{Eq:delL2xldh}
		\end{eqnarray}}
		{\small\begin{eqnarray}
				\sigma_{\chi H\leftrightarrow  \Delta^\dagger\bar{L}}&=&-\frac{y_{_\Sigma}^2y_\chi^2}{(32\pi(-4M_{h_1}^2M_\chi^2+(M_{h_1}^2+M_\chi^2-s)^2))}\Bigg(\Bigg((M_{h_1}-M_{\Sigma_1})(M_{h_1}+M_{\Sigma_1})(M_{\Sigma_1}-M_\Delta+M_\chi)\nonumber\\&&(M_{\Sigma_1}+M_\Delta+M_\chi)s^2\sqrt{\frac{(M_\Delta^2-s)^2(M_{h_1}^4+(M_\chi^2-s)^2-2M_{h_1}^2(M_\chi^2+s))}{s^4}}\Bigg)\Big/(2(M_{h_1}^4M_\Delta^2+M_{\Sigma_1}^2(M_\Delta^2-s)\nonumber\\&&(M_\chi^2-s)+M_{\Sigma_1}^4s+M_{h_1}^2((M_\Delta-M_\chi)(M_\Delta+M_\chi)(M_\Delta^2-s)-M_{\Sigma_1}^2(M_\Delta^2+s))))\nonumber\\&+&\frac{1}{2}((M_{\Sigma_1}+M_\chi)^2-s)\log\Bigg[\Bigg(M_{h_1}^2(M_\Delta^2+s)+M_\Delta^2(-M_\chi^2+s)-s\Bigg(2M_{\Sigma_1}^2\nonumber\\&-&M_\chi^2+s+s\sqrt{\frac{(M_\Delta^2-s)^2(M_{h_1}^4+(M_\chi^2-s)^2-2M_{h_1}^2(M_\chi^2+s))}{s^4}}\Bigg)\Bigg)\Big/\Bigg(M_{h_1}^2(M_\Delta^2+s)\nonumber\\&+&M_\Delta^2(-M_\chi^2+s)+s\Bigg(-2M_{\Sigma_1}^2+M_\chi^2+s\Bigg(-1+\nonumber\\&&\sqrt{\frac{(M_\Delta^2-s)^2(M_{h_1}^4+(M_\chi^2-s)^2-2M_{h_1}^2(M_\chi^2+s))}{s^4}}\Bigg)\Bigg)\Bigg)\Bigg]\Bigg)
				\label{Eq:delL2xhdl}\nonumber\\
		\end{eqnarray}}
		{\small\begin{eqnarray}
				\sigma_{\chi \chi\leftrightarrow  \Delta\Delta(t-channel)}&=&\frac{y_\chi^4}{64\pi s(s-4M_\chi^2)}\Bigg(-\Big(\Big( 2s\sqrt{\frac{(s-4M_\Delta^2)(s-4M_\chi^2)}{s^2}}(3M_\Delta^4+(M_{\Sigma_1}+M_\chi)^2\nonumber\\&&(3M_{\Sigma_1}^2-2M_{\Sigma_1}M_\chi+3M_\chi^2)-2M_\Delta^2(3M_{\Sigma_1}^2+2M_{\Sigma_1}M_\chi+3M_\chi^2)\nonumber\\&+&2M_{\Sigma_1}^2s)  \Big)\Big/(M_\Delta^4+(M_{\Sigma_1}^2-M_\chi^2)^2-2M_\Delta^2(M_{\Sigma_1}^2+M_\chi^2)+M_{\Sigma_1}^2s)\Big)\nonumber\\&+&\frac{2}{2(-M_{\Sigma_1}^2+M_\Delta^2+M_\chi^2)-s}2( 6M_\Delta^4+2(3M_{\Sigma_1}-5M_\chi)(M_{\Sigma_1}+M_\chi)^3+\nonumber\\&&8M_{\Sigma_1}(M_{\Sigma_1}+M_\chi)s+s^2-4M_\Delta^2((3M_{\Sigma_1}-M_\chi)(M_{\Sigma_1}+M_\chi)+s)) \nonumber\\&& \log\Big[\frac{-2M_{\Sigma_1}^2+2M_\Delta^2+2M_\chi^2+s\Big(-1+\sqrt{\frac{(s-4M_{\Delta}^2)(s-4M_\chi^2)}{s^2}}\Big)}{-2M_{\Sigma_1}^2+2M_\Delta^2+2M_\chi^2-s\Big(1+\sqrt{\frac{(s-4M_{\Delta}^2)(s-4M_\chi^2)}{s^2}}\Big)}\Big] \Bigg)
		\end{eqnarray}}
		{\small\begin{eqnarray}
				\sigma_{\chi \Delta\leftrightarrow  \Delta\chi(t-channel)}&=&\Bigg(y_\chi^4\Bigg( -\frac{(M_{\Sigma_1}-M_\Delta+M_\chi)^2(M_{\Sigma_1}+M_\Delta+M_\chi)^2s^2\sqrt{\frac{(M_\Delta^4+(M_\chi^2-s)^2-2M_\Delta^2(M_\chi^2+s))^2}{s^4}}} {(M_{\Sigma_1}^2-2M_\Delta^2-2M_\chi^2+s)((M_\Delta^2-M_\chi^2)^2-M_{\Sigma_1}^2s)} +\nonumber\\& & ((M_{\Sigma_1}+2M_\chi)^2-s) \log\Big[\Big( M_\Delta^4+M_\chi^4-2M_{\Sigma_1}^2s+2M_\chi^2s+2M_\Delta^2(-M_\chi^2+s) +\nonumber\\&& s^2\big(-1+ \sqrt{\frac{(M_\Delta^4+(M_\chi^2-s)^2-2M_\Delta^2(M_\chi^2+s))^2}{s^4}}  \big) \Big)\Big/  \Big( M_\Delta^4+M_\chi^4+ 2M_\chi^2s+\nonumber\\&& 2M_\Delta^2(-M_\chi^2+s)-s\big(2M_{\Sigma_1}^2+s+s\sqrt{\frac{(M_\Delta^4+(M_\chi^2-s)^2-2M_\Delta^2(M_\chi^2+s))^2}{s^4}}  \big) \Big) \Big]\Bigg)\Bigg) \Bigg/\nonumber\\&& (16\pi(-4M_\Delta^2M_\chi^2+(M_\Delta^2+M_\chi^2-s)^2))
		\end{eqnarray}}
		
		{\scriptsize 
			\begin{eqnarray}
				\sigma_{ \chi \Sigma_1 \rightarrow  \chi \Sigma_1(s-channel)}&=& \frac{ y_{\chi }^4 s \left(s-\left(M_{\Sigma }+M_{\chi }\right){}^2\right) \sqrt{\frac{\left(-2 M_{\Sigma }^2 \left(M_{\chi }^2+s\right)+\left(s-M_{\chi
							}^2\right){}^2+M_{\Sigma }^4\right){}^2}{s^4}}}{16 \pi  \left(s-M_{\Delta }^2\right){}^2 \left(s-\left(M_{\Sigma }-M_{\chi }\right){}^2\right)}
		\end{eqnarray}}
		{\scriptsize
			\begin{eqnarray}
				\sigma_{ \chi \Sigma_1 \rightarrow  \chi \Sigma_1(t-channel)}&=& \frac{y_{\chi }^4}{16 \pi  \left(\left(M_{\Sigma }^2+M_{\chi }^2-s\right){}^2-4 M_{\Sigma }^2 M_{\chi }^2\right)} \bigg( \frac{1}{\left(s M_{\Delta }^2-\left(M_{\Sigma }^2-M_{\chi }^2\right){}^2\right) \left(M_{\Delta }^2-2 \left(M_{\Sigma }^2+M_{\chi }^2\right)+s\right)}\bigg(\nonumber\\&&
				s \sqrt{\frac{\left(-2 M_{\Sigma }^2 \left(M_{\chi }^2+s\right)+\left(M_{\chi }^2-s\right){}^2+M_{\Sigma }^4\right){}^2}{s^4}} (-\left(M_{\Sigma }-M_{\chi }\right){}^2
				\left(M_{\Sigma }+M_{\chi }\right){}^2 \left(M_{\Delta }^2-2 \left(M_{\Sigma }^2+M_{\chi }^2\right)\right)+s^2 M_{\Delta }^2\nonumber\\&+&2 s \left(-2 M_{\Delta }^2 \left(M_{\Sigma }
				M_{\chi }+M_{\Sigma }^2+M_{\chi }^2\right)+M_{\Delta }^4+2 M_{\Sigma } M_{\chi } \left(M_{\Sigma }+M_{\chi }\right){}^2\right)) \bigg)+(M_{\Delta }^2-\left(M_{\Sigma }+M_{\chi }\right){}^2)\nonumber\\&&
				2 \log \left(\frac{s^2 \left(\sqrt{\frac{\left(-2 M_{\Sigma }^2 \left(M_{\chi }^2+s\right)+\left(M_{\chi }^2-s\right){}^2+M_{\Sigma }^4\right){}^2}{s^4}}-1\right)-2 s
					M_{\Delta }^2+2 M_{\Sigma }^2 \left(s-M_{\chi }^2\right)+2 s M_{\chi }^2+M_{\Sigma }^4+M_{\chi }^4}{-s \left(2 M_{\Delta }^2+s \sqrt{\frac{\left(-2 M_{\Sigma }^2
							\left(M_{\chi }^2+s\right)+\left(M_{\chi }^2-s\right){}^2+M_{\Sigma }^4\right){}^2}{s^4}}+s\right)+2 M_{\Sigma }^2 \left(s-M_{\chi }^2\right)+2 s M_{\chi }^2+M_{\Sigma
					}^4+M_{\chi }^4}\right) \bigg)\nonumber\\
			\end{eqnarray}
		}
		
		\subsection{$\Delta$L=1 Scatterings}
		We define,
		
		\begin{eqnarray}
			\Gamma_1&=&<\sigma_{L \Sigma_1\leftrightarrow  tt}v>*\sqrt{n^{eq}_\Sigma*n^{eq}_{top}}+<\sigma_{ \Sigma_1 t\leftrightarrow  Lt}v>*\sqrt{n^{eq}_\Sigma*n^{eq}_{top}}\nonumber\\&+&<\sigma_{\chi \Sigma_1\leftrightarrow  HH}v>\sqrt{n^{eq}_{\Sigma}*n^{eq}_h}+ <\sigma_{ \Sigma_1 H\leftrightarrow  \chi H}v>\sqrt{n^{eq}_{\Sigma}*n^{eq}_\chi}\nonumber\\&+&(< \sigma_{L\Sigma_1\rightarrow ZH}~v >+< \sigma_{L\Sigma_1\rightarrow WH}~v >+< \sigma_{L\Sigma_1\rightarrow HW}~v >)\sqrt{n^{eq}_l*n^{eq}_\Sigma}\nonumber\\&+&<\sigma_{\chi\Sigma_1\rightarrow W\Delta}~v>\sqrt{n^{eq}_{\Sigma_1}*n^{eq}_\chi}
		\end{eqnarray}
		
		\begin{eqnarray}
			\Gamma^W_{1L}&=&<\sigma_{L \Sigma_1\leftrightarrow  tt}v>*n^{eq}_{top}+<\sigma_{ \Sigma_1 t\leftrightarrow  Lt}v>*\sqrt{n^{eq}_{top}*n^{eq}_l}+(< \sigma_{L\Sigma_1\rightarrow ZH}~v >)\sqrt{n^{eq}_Z*n^{eq}_h}\nonumber\\&+&(< \sigma_{L\Sigma_1\rightarrow WH}~v >+< \sigma_{L\Sigma_1\rightarrow HW}~v >)\sqrt{n^{eq}_Z*n^{eq}_W}
		\end{eqnarray}

		\begin{eqnarray}
			\Gamma^{W}_{1\chi}=<\sigma_{\chi \Sigma_1\leftrightarrow  HH}v>n^{eq}_h+ <\sigma_{ \Sigma_1 H\leftrightarrow  \chi H}v>\sqrt{n^{eq}_{h}*n^{eq}_\chi}+<\sigma_{\chi\Sigma_1\rightarrow W\Delta}~v>\sqrt{n^{eq}_{\Delta}*n^{eq}_W}
		\end{eqnarray}

		{\small\begin{eqnarray}
				\sigma_{L \Sigma_1\leftrightarrow  tt}&=\frac{s M_{\text{top}}^2 y_{\Sigma }^2 \left(s-4 M_{\text{top}}^2\right) \sqrt{\frac{\left(M_{\Sigma _1}^2-s\right){}^2 \left(s-4 M_{\text{top}}^2\right)}{s^3}} \left(\sqrt{2} c_w M_Z s_w+246q_{e}\right){}^2}{5164032 \pi  M_W^2 s_w^2 \left(M_{h_1}^2-s\right){}^2 \left(s-M_{\Sigma _1}^2\right)},\
		\end{eqnarray}}
		where $q_{e}$ is electromagnetic coupling constant (0.3133).\\	
		{\small\begin{eqnarray}
				\sigma_{ \Sigma_1 t\leftrightarrow  Lt}&=&\frac{M_{\text{top}}^2 y_{\Sigma }^2 \left(\sqrt{2} c_w M_Z s_w+246 q_{e}\right){}^2}{30984192 \pi  M_W^2 s_w^2 \left(\left(M_{\Sigma _1}^2+M_{\text{top}}^2-s\right){}^2-4 M_{\Sigma _1}^2 M_{\text{top}}^2\right)} \nonumber\\&&\Bigg( \Bigg(s \sqrt{\frac{(M_{\text{top}}^2-s){}^2 (-2 M_{\Sigma _1}^2 M_{\text{top}}^2+s)+(M_{\text{top}}^2-s){}^2+M_{\Sigma _1}^4)}{s^4}} (M_{h_1}^2 (-M_{\Sigma _1}^2 (M_{\text{top}}^2+2 s)-6 s M_{\text{top}}^2 \nonumber\\&+&M_{\text{top}}^4+s^2) +2 s M_{h_1}^4+M_{\Sigma _1}^2 M_{\text{top}}^2
				(M_{\Sigma _1}^2+4 s)) \Bigg)\Big/ (M_{h_1}^2 ((M_{\text{top}}^2-s){}^2-M_{\Sigma _1}^2 (M_{\text{top}}^2+s)) \nonumber\\&+&s M_{h_1}^4+M_{\Sigma _1}^4 M_{\text{top}}^2)+(-2 M_{h_1}^2+M_{\Sigma _1}^2+4 M_{\text{top}}^2) \log[ (2 s M_{h_1}^2+s^2  \nonumber\\&& \sqrt{\frac{(M_{\text{top}}^2-s){}^2 (-2 M_{\Sigma _1}^2
						(M_{\text{top}}^2+s)+(M_{\text{top}}^2-s){}^2+M_{\Sigma _1}^4)}{s^4}}-M_{\Sigma _1}^2
				(M_{\text{top}}^2+s)+(M_{\text{top}}^2-s){}^2)\big/\nonumber\\&&(2 s M_{h_1}^2-s^2 \sqrt{\frac{(M_{\text{top}}^2-s){}^2 (-2 M_{\Sigma _1}^2
						(M_{\text{top}}^2+s)+(M_{\text{top}}^2-s){}^2+M_{\Sigma _1}^4)}{s^4}}-M_{\Sigma _1}^2
				(M_{\text{top}}^2+s)+(M_{\text{top}}^2-s){}^2)]\Bigg) \nonumber\\
		\end{eqnarray}}
		{\small\begin{eqnarray}
				\sigma_{\chi \Sigma_1\leftrightarrow  HH}&=&-\frac{s v_1^2 y_{\chi }^2 \sqrt{\frac{\left(s-4 M_{h_1}^2\right) \left(-2 M_{\Sigma _1}^2 \left(\text{M$\chi $}^2+s\right)+\left(M_{\chi }^2-s\right){}^2+M_{\Sigma
								_1}^4\right)}{s^3}} \left(q_{e}^2 M_{\Delta }^2+2 \lambda _{\text{H$\Delta $}} M_W^2 s_w^2\right){}^2}{16 \pi  q_{e}^4 v_0^4 \left(M_{\Delta }^2-s\right){}^2
					\left(\left(M_{\Sigma _1}-M_{\chi }\right){}^2-s\right)}\nonumber\\
		\end{eqnarray}}
		{\small\begin{eqnarray}
				\sigma_{ \Sigma_1 H\leftrightarrow  \chi H}&=& -\Bigg(\Bigg(v_1^2 y_{\chi }^2 (q_{e}^2 M_{\Delta }^2+2 \lambda _{\text{H$\Delta $}} M_W^2 s_w^2){}^2 \Bigg( -\bigg(s^2 (-M_{\Delta }+M_{\Sigma _1}+M_{\chi })(M_{\Delta }M_{\Sigma _1}+\nonumber\\&&M_{\chi })\sqrt{\frac{(-2 M_{h_1}^2(M_{\Sigma _1}^2+s)+M_{h_1}^4+(M_{\Sigma _1}^2-s){}^2) (-2 M_{h_1}^2 (M_{\chi
						}^2+s)+M_{h_1}^4+(M_{\chi }^2-s){}^2)}{s^4}}\bigg)\bigg/\nonumber\\&&\bigg( 2 (M_{\Delta }^2 (-M_{h_1}^2 (M_{\Sigma _1}^2+M_{\chi }^2+2 s)+M_{h_1}^4+(M_{\Sigma _1}^2-s) (M_{\chi }^2-s))+M_{h_1}^2
				(M_{\Sigma _1}^2-M_{\chi }^2){}^2\nonumber\\&+&s M_{\Delta }^4)	
				\bigg)-\frac{1}{2} \log\big[\big( M_{h_1}^4+(M_{\Sigma_1}^2-s)(M_\chi^2-s)+2M_\Delta^2s-M_{h_1}^2(M_{\Sigma_1}^2+M_\chi^2+2s)+\nonumber\\&&
				s^2\sqrt{\frac{(M_{h_1}^4+(M_{\Sigma_1}^2-s)^2-2M_{h_1}^2(M_{\Sigma_1}^2+s))(M_{h_1}^2+(M_\chi^2-s)^2-2M_{h_1}^2(M_\chi^2+s))}{s^4}}  \big)/\nonumber\\&&
				\big(M_{h_1}^4+(M_{\Sigma_1}^2-s)(M_\chi^2-s)+2M_\Delta^2s-M_{h_1}^2(M_{\Sigma_1}^2+M_\chi^2+2s)-\nonumber\\&&
				s^2\sqrt{\frac{(M_{h_1}^4+(M_{\Sigma_1}^2-s)^2-2M_{h_1}^2(M_{\Sigma_1}^2+s))(M_{h_1}^2+(M_\chi^2-s)^2-2M_{h_1}^2(M_\chi^2+s))}{s^4}} \big)\big]\Bigg)    \Bigg) \bigg/ \nonumber\\&&		 
				(2 \pi  q_{e}^4 v_0^4 ((M_{h_1}^2+M_{\Sigma _1}^2-s){}^2-4 M_{h_1}^2 M_{\Sigma _1}^2))\Bigg)\nonumber\\
		\end{eqnarray}}
		{\scriptsize
			\begin{eqnarray}
				\sigma_{ L\Sigma_1 \leftrightarrow  Z H}&=&-\frac{q_e^2 y_{\Sigma }^2 \sqrt{\frac{\left(s-M_{\Sigma }^2\right){}^2}{s^2}} \left(2 M_{\Sigma }^2+s\right)}{512 \pi  s_w^2 \left(s_w^2-1\right) \left(M_{\Sigma
					}^2-s\right){}^2}
		\end{eqnarray}}
		{\scriptsize
			\begin{eqnarray}
				\sigma_{ L\Sigma_1 \leftrightarrow W H}&=&-\frac{s q_e^2 y_{\Sigma }^2 \sqrt{\frac{\left(M_{\Sigma }^2-s\right){}^2}{s^2}}}{128 \pi  s_w^2 \left(M_{\Sigma }^2-s\right){}^2}
		\end{eqnarray}}
		{\scriptsize
			\begin{eqnarray}
				\sigma_{ L\Sigma_1 \leftrightarrow HW}&=&\frac{q_e^2 y_{\Sigma }^2 \left(2 \left(M_{\Sigma }+0.16\right) \left(0.16\, -M_{\Sigma }\right)+s\right) \left(\frac{s \sqrt{\frac{\left(M_{\Sigma }^2-s\right){}^2}{s^2}}}{-M_{\Sigma }^2+\left(M_{\Sigma
						}+0.16\right){}^2+s}+\log \left(\frac{s \left(\sqrt{\frac{\left(M_{\Sigma }^2-s\right){}^2}{s^2}}-1\right)+M_{\Sigma }^2-2 \left(M_{\Sigma }+0.16\right){}^2}{-s \left(\sqrt{\frac{\left(M_{\Sigma
								}^2-s\right){}^2}{s^2}}+1\right)+M_{\Sigma }^2-2 \left(M_{\Sigma }+0.16\right){}^2}\right)\right)}{64 \pi  s_w^2 \left(M_{\Sigma }^2-s\right){}^2}\nonumber\\
		\end{eqnarray}}

		{\scriptsize
			\begin{eqnarray}
				\sigma_{ \chi\Sigma_1 \leftrightarrow W \Delta}&=&\frac{q_e^2 y_{\chi }^2}{16 \pi  s_w^2 \left(\left(\left(M_{\Sigma }+0.16\right){}^2+M_{\chi }^2-s\right){}^2-4 \left(M_{\Sigma }+0.16\right){}^2 M_{\chi }^2\right)} \bigg( (M_{\Delta }^2-2 \left(M_{\Sigma }-2 \left(M_{\Sigma }+0.16\right)\right) \left(M_{\Sigma }+M_{\chi }\right)-s)\nonumber\\&&
				\log\big( (s (s \left(\sqrt{\frac{\left(M_{\Delta }^2-s\right){}^2 \left(-2 \left(M_{\Sigma }+0.16\right){}^2 \left(M_{\chi }^2+s\right)+\left(M_{\chi }^2-s\right){}^2+\left(M_{\Sigma
						}+0.16\right){}^4\right)}{s^4}}-1\right)-2 M_{\Sigma }^2+\left(M_{\Sigma }+0.16\right){}^2\nonumber\\&+&M_{\chi }^2)+M_{\Delta }^2 \left(\left(M_{\Sigma }+0.16\right){}^2-M_{\chi }^2+s\right))/( M_{\Delta }^2 \left(\left(M_{\Sigma }+0.16\right){}^2-M_{\chi }^2+s\right)+s (-2 M_{\Sigma }^2+\left(M_{\Sigma }+0.16\right){}^2+M_{\chi }^2 \nonumber\\&& -s \left(\sqrt{\frac{\left(M_{\Delta }^2-s\right){}^2 \left(-2 \left(M_{\Sigma }+0.16\right){}^2 \left(M_{\chi }^2+s\right)+\left(M_{\chi }^2-s\right){}^2+\left(M_{\Sigma
						}+0.16\right){}^4\right)}{s^4}}+1\right))  ) \big) \nonumber\\&-&
				s \sqrt{\frac{\left(M_{\Delta }^2-s\right){}^2 \left(-2 \left(M_{\Sigma }+0.16\right){}^2 \left(M_{\chi }^2+s\right)+\left(M_{\chi }^2-s\right){}^2+\left(M_{\Sigma }+0.16\right){}^4\right)}{s^4}} \big( M_{\Delta }^4 \left(M_{\Sigma }+0.16\right){}^2-0.16 M_{\Delta }^2\nonumber\\&& -0.16 M_{\Delta }^2 \left(0.32 s-\left(2 M_{\Sigma }+0.16\right) \left(M_{\Sigma }-M_{\chi }+0.16\right) \left(M_{\Sigma }+M_{\chi }+0.16\right)\right) + s (s M_{\Sigma }^2+2 \left(M_{\Sigma }+M_{\chi }\right)\nonumber\\&& (-2 \left(M_{\Sigma }+0.16\right) M_{\Sigma } M_{\chi }+\left(M_{\Sigma }+0.16\right){}^2 M_{\chi }+M_{\Sigma }^3-2 \left(M_{\Sigma
				}+0.16\right) M_{\Sigma }^2))   \big)/\big( M_{\Delta }^4 \left(M_{\Sigma }+0.16\right){}^2+s (s M_{\Sigma }^2\nonumber\\&-&0.16 \left(2 M_{\Sigma }+0.16\right) \left(M_{\Sigma }-M_{\chi }\right) \left(M_{\Sigma }+M_{\chi }\right))+M_{\Delta }^2 (0.16 \left(2 M_{\Sigma }+0.16\right) \left(M_{\Sigma }-M_{\chi }+0.16\right) \left(M_{\Sigma }+M_{\chi }+0.16\right)\nonumber\\&-&s \left(M_{\Sigma }^2+\left(M_{\Sigma
				}+0.16\right){}^2\right)) \big)   \bigg)
		\end{eqnarray}}
		
		\subsection{$\Delta$L=0 processes}
		
		{\scriptsize\begin{eqnarray}
				\sigma_{ \Sigma_1 \bar{\Sigma}_1 \rightarrow  HH^\dagger}&=&	\frac{y_{\Sigma }^4 \left(\frac{\left(4 M_{h_1}^2 \left(M_{\Sigma }^2-s\right)-2 s M_{\Sigma }^2-2
						M_{\Sigma }^4+6 M_{h_1}^4+s^2\right) \log \left(\frac{s \left(\sqrt{\frac{\left(s-4 M_{h_1}^2\right)
									\left(s-4 M_{\Sigma }^2\right)}{s^2}}-1\right)+2 M_{\Sigma }^2+2 M_{h_1}^2}{-s
							\left(\sqrt{\frac{\left(s-4 M_{h_1}^2\right) \left(s-4 M_{\Sigma }^2\right)}{s^2}}+1\right)+2 M_{\Sigma
							}^2+2 M_{h_1}^2}\right)}{2 \left(M_{\Sigma }^2+M_{h_1}^2\right)-s}-3 s \sqrt{\frac{\left(s-4
							M_{h_1}^2\right) \left(s-4 M_{\Sigma }^2\right)}{s^2}}\right)}{1024 \pi  s \left(s-4 M_{\Sigma
					}^2\right)}\nonumber\\
		\end{eqnarray}}
		
		{\scriptsize\begin{eqnarray}
				\sigma_{ \Sigma_1 \bar{\Sigma}_1 \rightarrow  \Delta \Delta^\dagger}&=&	\frac{y^4_{\chi}}{64 \pi  s \left(s-4 M_{\Sigma }^2\right)}\bigg(  \frac{2 \log \left(\frac{2 M_{\Delta }^2+s \left(\sqrt{\frac{\left(s-4 M_{\Delta }^2\right) \left(s-4
								M_{\Sigma }^2\right)}{s^2}}-1\right)+2 M_{\Sigma }^2-2 M_{\chi }^2}{2 M_{\Delta }^2-s
						\left(\sqrt{\frac{\left(s-4 M_{\Delta }^2\right) \left(s-4 M_{\Sigma }^2\right)}{s^2}}+1\right)+2
						M_{\Sigma }^2-2 M_{\chi }^2}\right)}{2 \left(M_{\Delta }^2+M_{\Sigma }^2-M_{\chi }^2\right)-s} \nonumber\\&&( 6 M_{\Delta }^4+4 M_{\Delta }^2 \left(\left(M_{\Sigma }-3 M_{\chi }\right) \left(M_{\Sigma }+M_{\chi
				}\right)-s\right)+8 s M_{\chi } \left(M_{\Sigma }+M_{\chi }\right)-2 \left(5 M_{\Sigma }-3 M_{\chi
				}\right) \left(M_{\Sigma }+M_{\chi }\right){}^3+s^2)\nonumber\\&-&\frac{2 s \sqrt{\frac{\left(s-4 M_{\Delta }^2\right) \left(s-4 M_{\Sigma }^2\right)}{s^2}} \left(-2
					M_{\Delta }^2 \left(2 M_{\Sigma } M_{\chi }+3 M_{\Sigma }^2+3 M_{\chi }^2\right)+3 M_{\Delta }^4+2 s
					M_{\chi }^2+\left(M_{\Sigma }+M_{\chi }\right){}^2 \left(-2 M_{\Sigma } M_{\chi }+3 M_{\Sigma }^2+3
					M_{\chi }^2\right)\right)}{-2 M_{\Delta }^2 \left(M_{\Sigma }^2+M_{\chi }^2\right)+M_{\Delta }^4+s M_{\chi
					}^2+\left(M_{\Sigma }^2-M_{\chi }^2\right){}^2}
				\bigg)\nonumber\\
		\end{eqnarray}}
		
		{\scriptsize \begin{eqnarray}
				\sigma_{ \Sigma_1 \bar{\Sigma}_1 \rightarrow  \chi\bar{\chi}}&=&\frac{y_{\chi }^4}{16 \pi  s (s-4 M_{\Sigma }^2)} \bigg( \frac{1}{2  (M_{\Delta }^4+M_{\Delta }^2  (s-2  (M_{\Sigma }^2+M_{\chi }^2 ) )+ (M_{\Sigma }^2-M_{\chi }^2 ){}^2 )}\bigg( s \sqrt{\frac{ (s-4 M_{\Sigma }^2 )  (s-4 M_{\chi }^2 )}{s^2}}  (5 M_{\Delta }^4\nonumber\\&-&M_{\Delta }^2  (8 M_{\Sigma } M_{\chi }+10 M_{\Sigma }^2+10 M_{\chi }^2-3 s )+ (M_{\Sigma }+M_{\chi } ){}^2  (-2 M_{\Sigma } M_{\chi }+5 M_{\Sigma }^2+5 M_{\chi
				}^2 ) ) \bigg) + \frac{1}{2 M_{\Delta }^2-2 \left(M_{\Sigma }^2+M_{\chi }^2\right)+s} \bigg(\nonumber\\ && \log \left(\frac{2 M_{\Delta }^2+s \sqrt{\frac{\left(s-4 M_{\Sigma }^2\right) \left(s-4 M_{\chi }^2\right)}{s^2}}-2 M_{\Sigma }^2-2 M_{\chi }^2+s}{2 M_{\Delta }^2-s \sqrt{\frac{\left(s-4 M_{\Sigma }^2\right) \left(s-4 M_{\chi }^2\right)}{s^2}}-2 M_{\Sigma }^2-2 M_{\chi }^2+s}\right) (-5 M_{\Delta }^4+M_{\Delta }^2 \left(8 M_{\Sigma } M_{\chi }+10 M_{\Sigma }^2+10 M_{\chi }^2-3 s\right)\nonumber\\&-&\left(M_{\Sigma }+M_{\chi }\right){}^2 \left(-2 M_{\Sigma } M_{\chi }+5 M_{\Sigma }^2+5 M_{\chi }^2-s\right)) \bigg) \bigg) 
		\end{eqnarray}}
		
		{\scriptsize \begin{eqnarray}
				\sigma_{ \Sigma_1 \bar{\Sigma}_1 \rightarrow  L\bar{L}}&=&\frac{y_{\Sigma}^4}{512 \pi  s \left(s-4 M_{\Sigma }^2\right) \left(2 M_{\Sigma }^2+M_{\chi }^2-s\right)}	\bigg( 5 s \sqrt{1-\frac{4 M_{\Sigma }^2}{s}} \left(2 M_{\Sigma }^2+M_{\chi }^2-s\right) + M_{\Sigma }^2 \left(-9 M_{\Sigma }^2-4 M_{\chi }^2+3 s\right)\nonumber\\&& \log \left(\frac{-s \sqrt{1-\frac{4 M_{\Sigma }^2}{s}}-2 M_{\Sigma }^2+s}{s \sqrt{1-\frac{4 M_{\Sigma }^2}{s}}-2 M_{\Sigma }^2+s}\right)-s \left(M_{\Sigma }^2-M_{\chi }^2\right)+\left(M_{\Sigma }^2+M_{\chi }^2\right){}^2 \log \left(\frac{s \left(\sqrt{1-\frac{4 M_{\Sigma }^2}{s}}-1\right)+2 M_{\Sigma }^2+2 M_{\chi }^2}{-s \left(\sqrt{1-\frac{4 M_{\Sigma }^2}{s}}+1\right)+2 M_{\Sigma }^2+2
					M_{\chi }^2}\right)  \bigg)\nonumber\\
		\end{eqnarray}}	
		\subsection{$\Delta$L=0 Processes (Transfer term)}
		Form \cite{Falkowski:2011xh},
		{\small\begin{equation}
				I_i=\frac{\hat{\Gamma}_{\Sigma_1}}{\pi}\int^\infty_0 t^2 K_1(t)f_i(t^2/z^2)dt
		\end{equation}}
		{\small\begin{equation}
				f_{T_+}(s)=\frac{s/2}{(s-1)^2+\hat{\Gamma}_{\Sigma_1}^2}+\frac{s}{s+1}+\frac{s-\log(s+1)}{s}
		\end{equation}}
		{\small\begin{equation}
				f_{T_-}(s)=\frac{s/2}{(s-1)^2+\hat{\Gamma}_{\Sigma_1}^2}+\frac{(s+1)\log(s+1)-s}{s+1}+\frac{(s+2)\log(s+1)-2s}{s}
		\end{equation}}
		where $\hat{\Gamma}_{\Sigma_1}=\frac{\Gamma_{\Sigma_1}}{M_{\Sigma_1}}$.
		
		\section{Fermion Mixings}
		\subsection{$\nu-\Sigma$ mixing}
		The Lagrangian responsible $\nu$ and $\Sigma$ mixing is
		\begin{equation}
			-\mathcal{L}_{\nu- \Sigma} \supset  m_\Sigma Tr[\overline{\Sigma^c}\Sigma]+ \sqrt{2} y_{_\Sigma} \overline{L}  \Tilde{H} \Sigma + h.c.,
		\end{equation}
		which can be written in the basis $((\nu_{L})^c ~~~\Sigma^0)^T$ as
		\begin{equation}
			-\mathcal{L}_{\nu -\Sigma} \supset\begin{pmatrix}
				\overline{\nu_{L}} & \overline{(\Sigma^{0})^c}
			\end{pmatrix} \begin{pmatrix}
				0 & \frac{y_{_\Sigma} v_0}{\sqrt{2}}\\
				\frac{y_{_\Sigma} v_0}{\sqrt{2}} & m_\Sigma
			\end{pmatrix} \begin{pmatrix}
				(\nu_{L})^c\\
				\Sigma^0
			\end{pmatrix}.\
		\end{equation}
		Now the mixing angle between $\nu-\Sigma$ is given as
		\begin{equation}
			\tan2\theta_{\nu-\Sigma}=\frac{\sqrt{2}y_{_\Sigma}v_0}{m_\Sigma}.
		\end{equation}
		\begin{equation}
			\theta_{\nu-\Sigma}\sim \sqrt{\frac{m_\nu}{m_\Sigma}}\sim \sqrt{\frac{m_\nu}{M_\Sigma}}.
		\end{equation}

		\subsection{$\chi-\Sigma$ mixing}
		The Lagrangian responsible $\chi$ and $\Sigma$ mixing is
		\begin{equation}
			-\mathcal{L}_{\chi- \Sigma} \supset M_\chi \overline{\chi}\chi + y_\chi Tr[\overline{\Sigma}\Delta\chi]+  h.c.,
		\end{equation}
		which can be written in the basis $(\chi ~~~\Sigma^0)^T$ as
		\begin{equation}
			-\mathcal{L}_{\chi -\Sigma} \supset\begin{pmatrix}
				\overline{\chi} & \overline{(\Sigma^{0})^c}
			\end{pmatrix} \begin{pmatrix}
				M_\chi & y_\chi v_1\\
				y_\chi v_1 & m_\Sigma
			\end{pmatrix} \begin{pmatrix}
				\chi\\
				\Sigma^0
			\end{pmatrix}.\
		\end{equation}
		Now the mixing angle between $\chi-\Sigma$ is given as
		\begin{equation}
			\tan2\theta_{\chi-\Sigma}=\frac{2y_{_\chi}v_1}{m_\Sigma-M_\chi}.
		\end{equation}
		\begin{equation}
			\theta_{\chi-\Sigma}\sim \frac{\sqrt{2}y_\chi v_1}{m_\Sigma}\sim \frac{\sqrt{2}y_\chi v_1}{M_\Sigma}.
		\end{equation}	
		
		\section{Pseudo Dirac nature of DM}\label{app:pseduodirac}
		The DM Lagrangian is,
		\begin{eqnarray}
			-\mathcal{L_{{\rm DM}}}=M_\chi\bar{\chi}\chi+\frac{1}{2}M_m\overline{\chi^c}\chi+h.c.,
		\end{eqnarray}
		where $M_m=\frac{y_\chi^2v^2_\Delta}{M_\Sigma}$, is the Majorana mass of the DM originating from the mixing with $\Sigma$, via $y_\chi \Sigma\Delta\chi$. For a typical choice of parameter, say $y_\chi=1$, $v_{\Delta}=1$ eV, and $M_{\Sigma}=10^9$ GeV, $M_m=10^{-18}$ eV.
		Now in the basis $(\chi~~\chi^c)$ we can express the above Lagrangian as
		\begin{eqnarray}
			\begin{pmatrix}
				\bar{\chi}&\overline{\chi^c}
			\end{pmatrix} \begin{pmatrix}
				M_\chi&M_m\\
				M_m&M_\chi
			\end{pmatrix}\begin{pmatrix}
				\chi\\\chi^c
			\end{pmatrix}
		\end{eqnarray}
		The matrix can be diagonalised by an orthogonal matrix as
		\begin{eqnarray}
			\begin{pmatrix}
				\bar{\chi}&\overline{\chi^c}
			\end{pmatrix}R^TR \begin{pmatrix}
				M_\chi&M_m\\
				M_m&M_\chi
			\end{pmatrix}R^TR\begin{pmatrix}
				\chi\\\chi^c
			\end{pmatrix}
		\end{eqnarray}
		where
		\begin{eqnarray}
			R= \begin{pmatrix}
				\cos\theta_{\chi\chi^c}&\sin\theta_{\chi\chi^c}\\
				-\sin\theta_{\chi\chi^c}&\cos\theta_{\chi\chi^c}
			\end{pmatrix}
		\end{eqnarray}
		We obtain two mass eigenstates $\chi_1$, and $\chi_2$, which can be written in terms of the flavor states as
		\begin{eqnarray}
			\begin{pmatrix}
				\chi_1\\\chi_2
			\end{pmatrix}= \begin{pmatrix}    \cos\theta_{\chi\chi^c}&\sin\theta_{\chi\chi^c}\\
				-\sin\theta_{\chi\chi^c}&\cos\theta_{\chi\chi^c}
			\end{pmatrix} \begin{pmatrix}
				\chi\\\chi^c
			\end{pmatrix} \label{eq:chichibarflavor}
		\end{eqnarray}
		with masses $M_1=M_\chi-M_m$, and $M_2=M_\chi+M_m$.
		
		\section{Dark matter/anti-dark matter oscillation}\label{app:dmantidmosc}
		Now from Eq \ref{eq:chichibarflavor}, the flavor state $|\chi>$ at $t=0$ can be written in terms of the mass eigenstates as
		\begin{eqnarray}
			|\chi>=|\chi(0)>=\cos\theta_{\chi\chi^c}|\chi_1(0)>-\sin\theta_{\chi\chi^c}|\chi_2(0)>
		\end{eqnarray}
		At any time $t$, the state can be written as,
		\begin{eqnarray}
			|\chi(t)>&=&\cos\theta_{\chi\chi^c}e^{-iE_1t}|\chi_1(t)>-\sin\theta_{\chi\chi^c}e^{-iE_2t}|\chi_2(t)>\nonumber\\&=&A_{\chi}(t)|\chi>+A_{\chi^c}(t)|\chi^c>,
		\end{eqnarray}
		where,
		\begin{eqnarray}
			A_\chi(t)=\cos^2\theta_{\chi\chi^c}e^{-iE_1t}+\sin^2\theta_{\chi\chi^c}e^{-iE_2t}\nonumber\\
			A_\chi^c(t)=\sin\theta_{\chi\chi^c}\cos\theta_{\chi\chi^c}e^{-iE_1t}-\sin\theta_{\chi\chi^c}\cos\theta_{\chi\chi^c}e^{-iE_2t}
		\end{eqnarray}
		Now the probability of oscillating to $\chi^c$ from the state $\chi$ can be written as
		\begin{eqnarray}
			P_{\chi\rightarrow\chi^c}(t)&=&1-P_{\chi\rightarrow\chi}(t)\nonumber\\&=&1-|A_\chi(t)|^2\nonumber\\P_{\chi\rightarrow\chi^c}(t)&=&\sin^2(2\theta_{\chi\chi^c})\sin^2\bigg[\frac{(E_2-E_1)t}{2} \bigg].
		\end{eqnarray}
		With the mixing angle $\theta_{\chi\chi^c}=\frac{\pi}{4}$, the probability of oscillation can be expressed as
		\begin{equation}
			P_{\chi\rightarrow\chi^c}(t)=\left\{
			\begin{array}{l}
				\sin^2\big[\frac{M_\chi M_m}{T}t \big] ~,~~T>M_{\chi}\\
				\sin^2[M_mt]~~~~~~~,~~T\leq M_{\chi}.\\
			\end{array}
			\right.
		\end{equation}
		Now we also normalize the time of evolution from the time of EW phase transition so that at $t=t_{EW}$, $P_{\chi\rightarrow\chi^c}(t_{EW})=0$. Thus the probability of oscillation can be expressed as
		\begin{equation}
			P_{\chi\rightarrow\chi^c}(t)=\left\{
			\begin{array}{l}
				\sin^2\big[\frac{M_\chi M_m}{T}(t-t_{EW}) \big] ~,~~T>M_{\chi}\\
				\sin^2[M_m(t-t_{EW})]~~~~~~~,~~T\leq M_{\chi}.\\
			\end{array}
			\right.
		\end{equation}
		For the BP2, $M_m=2.3\times10^{-39}$ GeV, $t\sim t_{EW}$, $M_\chi/T=0.1$, it gives, $P_{\chi\rightarrow\chi^c}=2.6\times10^{-55}$. This implies that the DM anti-DM oscillation is negligible.

		\section{Cross-section for dark matter symmetric component annihilation}
		\label{appen2}
		{\small\begin{eqnarray}
				\sigma_{(\chi\chi\rightarrow \phi\phi)}&=&\frac{\cos^4\gamma \lambda_{\rm DM}^4}{2\pi s(s-4M^2_\chi)}\bigg(\frac{(6M^2_{h_2}-32M^4_\chi+16M^2_\chi s+s^2-4M^2_{h_2}(4M^2_\chi+s))\log\big[\frac{2M^2_{h_2}+s(A-1)}{2M^2_{h_2}-s(A+1)}\big]}{2M^2_{h_2}-s}\nonumber\\& -&\frac{s A(3M^4_{h_2}-16M^2_{h_2}M^2_\chi +2M^2_\chi(8M^2_\chi+s))}{M^4_{h_2}-4M^2_{h_2}M^2_{\chi}+M^2_\chi s} \bigg)
		\end{eqnarray}}
		where 
		{\small\begin{equation}
				A=\sqrt{\frac{(s-4M^2_{h_2})(s-4M^2_\chi)}{s^2}}
		\end{equation}}
		and $M_{h_2}$ is the light mass eigen state after $H-\phi$ mixing, which is essentially $\phi$ mass.


\begin{thebibliography}{57}%
			\makeatletter
			\providecommand \@ifxundefined [1]{%
				\@ifx{#1\undefined}
			}%
			\providecommand \@ifnum [1]{%
				\ifnum #1\expandafter \@firstoftwo
				\else \expandafter \@secondoftwo
				\fi
			}%
			\providecommand \@ifx [1]{%
				\ifx #1\expandafter \@firstoftwo
				\else \expandafter \@secondoftwo
				\fi
			}%
			\providecommand \natexlab [1]{#1}%
			\providecommand \enquote  [1]{``#1''}%
			\providecommand \bibnamefont  [1]{#1}%
			\providecommand \bibfnamefont [1]{#1}%
			\providecommand \citenamefont [1]{#1}%
			\providecommand \href@noop [0]{\@secondoftwo}%
			\providecommand \href [0]{\begingroup \@sanitize@url \@href}%
			\providecommand \@href[1]{\@@startlink{#1}\@@href}%
			\providecommand \@@href[1]{\endgroup#1\@@endlink}%
			\providecommand \@sanitize@url [0]{\catcode `\\12\catcode `\$12\catcode
				`\&12\catcode `\#12\catcode `\^12\catcode `\_12\catcode `\%12\relax}%
			\providecommand \@@startlink[1]{}%
			\providecommand \@@endlink[0]{}%
			\providecommand \url  [0]{\begingroup\@sanitize@url \@url }%
			\providecommand \@url [1]{\endgroup\@href {#1}{\urlprefix }}%
			\providecommand \urlprefix  [0]{URL }%
			\providecommand \Eprint [0]{\href }%
			\providecommand \doibase [0]{https://doi.org/}%
			\providecommand \selectlanguage [0]{\@gobble}%
			\providecommand \bibinfo  [0]{\@secondoftwo}%
			\providecommand \bibfield  [0]{\@secondoftwo}%
			\providecommand \translation [1]{[#1]}%
			\providecommand \BibitemOpen [0]{}%
			\providecommand \bibitemStop [0]{}%
			\providecommand \bibitemNoStop [0]{.\EOS\space}%
			\providecommand \EOS [0]{\spacefactor3000\relax}%
			\providecommand \BibitemShut  [1]{\csname bibitem#1\endcsname}%
			\let\auto@bib@innerbib\@empty
			\bibitem [{\citenamefont {Aghanim}\ \emph {et~al.}(2018)\citenamefont {Aghanim}
				\emph {et~al.}}]{Aghanim:2018eyx}%
			\BibitemOpen
			\bibfield  {author} {\bibinfo {author} {\bibfnamefont {N.}~\bibnamefont
					{Aghanim}} \emph {et~al.} (\bibinfo {collaboration} {Planck}),\ }\bibfield
			{title} {\bibinfo {title} {{Planck 2018 results. VI. Cosmological
						parameters}},\ }\href@noop {} {\  (\bibinfo {year} {2018})},\ \Eprint
			{https://arxiv.org/abs/1807.06209} {arXiv:1807.06209 [astro-ph.CO]}
			\BibitemShut {NoStop}%
			\bibitem [{\citenamefont {de~Salas}\ \emph {et~al.}(2017)\citenamefont
				{de~Salas}, \citenamefont {Forero}, \citenamefont {Ternes}, \citenamefont
				{Tortola},\ and\ \citenamefont {Valle}}]{deSalas:2017kay}%
			\BibitemOpen
			\bibfield  {author} {\bibinfo {author} {\bibfnamefont {P.~F.}\ \bibnamefont
					{de~Salas}}, \bibinfo {author} {\bibfnamefont {D.~V.}\ \bibnamefont
					{Forero}}, \bibinfo {author} {\bibfnamefont {C.~A.}\ \bibnamefont {Ternes}},
				\bibinfo {author} {\bibfnamefont {M.}~\bibnamefont {Tortola}},\ and\ \bibinfo
				{author} {\bibfnamefont {J.~W.~F.}\ \bibnamefont {Valle}},\ }\bibfield
			{title} {\bibinfo {title} {{Status of neutrino oscillations 2018: first hint
						for normal mass ordering and improved CP sensitivity}},\ }\href@noop {} {\
				(\bibinfo {year} {2017})},\ \Eprint {https://arxiv.org/abs/1708.01186}
			{arXiv:1708.01186 [hep-ph]} \BibitemShut {NoStop}%
			\bibitem [{\citenamefont {Aartsen}\ \emph {et~al.}(2015)\citenamefont {Aartsen}
				\emph {et~al.}}]{Aartsen:2015zva}%
			\BibitemOpen
			\bibfield  {author} {\bibinfo {author} {\bibfnamefont {M.~G.}\ \bibnamefont
					{Aartsen}} \emph {et~al.} (\bibinfo {collaboration} {IceCube}),\ }\bibfield
			{title} {\bibinfo {title} {{The IceCube Neutrino Observatory - Contributions
						to ICRC 2015 Part II: Atmospheric and Astrophysical Diffuse Neutrino Searches
						of All Flavors}},\ }in\ \href
			{https://inspirehep.net/record/1398539/files/arXiv:1510.05223.pdf} {\emph
				{\bibinfo {booktitle} {{Proceedings, 34th International Cosmic Ray Conference
							(ICRC 2015): The Hague, The Netherlands, July 30-August 6, 2015}}}}\
			(\bibinfo {year} {2015})\ \Eprint {https://arxiv.org/abs/1510.05223}
			{arXiv:1510.05223 [astro-ph.HE]} \BibitemShut {NoStop}%
			\bibitem [{\citenamefont {Abe}\ \emph {et~al.}(2015)\citenamefont {Abe} \emph
				{et~al.}}]{Abe:2015awa}%
			\BibitemOpen
			\bibfield  {author} {\bibinfo {author} {\bibfnamefont {K.}~\bibnamefont
					{Abe}} \emph {et~al.} (\bibinfo {collaboration} {T2K}),\ }\bibfield  {title}
			{\bibinfo {title} {{Measurements of neutrino oscillation in appearance and
						disappearance channels by the T2K experiment with $6.6*10^{20}$ protons on
						target}},\ }\href {https://doi.org/10.1103/PhysRevD.91.072010} {\bibfield
				{journal} {\bibinfo  {journal} {Phys. Rev.}\ }\textbf {\bibinfo {volume}
					{D91}},\ \bibinfo {pages} {072010} (\bibinfo {year} {2015})},\ \Eprint
			{https://arxiv.org/abs/1502.01550} {arXiv:1502.01550 [hep-ex]} \BibitemShut
			{NoStop}%
			\bibitem [{\citenamefont {Lesgourgues}\ and\ \citenamefont
				{Pastor}(2006)}]{Lesgourgues:2006nd}%
			\BibitemOpen
			\bibfield  {author} {\bibinfo {author} {\bibfnamefont {J.}~\bibnamefont
					{Lesgourgues}}\ and\ \bibinfo {author} {\bibfnamefont {S.}~\bibnamefont
					{Pastor}},\ }\bibfield  {title} {\bibinfo {title} {{Massive neutrinos and
						cosmology}},\ }\href {https://doi.org/10.1016/j.physrep.2006.04.001}
			{\bibfield  {journal} {\bibinfo  {journal} {Phys. Rept.}\ }\textbf {\bibinfo
					{volume} {429}},\ \bibinfo {pages} {307} (\bibinfo {year} {2006})},\ \Eprint
			{https://arxiv.org/abs/astro-ph/0603494} {arXiv:astro-ph/0603494}
			\BibitemShut {NoStop}%
			\bibitem [{\citenamefont {Wong}(2011)}]{Wong:2011ip}%
			\BibitemOpen
			\bibfield  {author} {\bibinfo {author} {\bibfnamefont {Y.~Y.~Y.}\
					\bibnamefont {Wong}},\ }\bibfield  {title} {\bibinfo {title} {{Neutrino mass
						in cosmology: status and prospects}},\ }\href
			{https://doi.org/10.1146/annurev-nucl-102010-130252} {\bibfield  {journal}
				{\bibinfo  {journal} {Ann. Rev. Nucl. Part. Sci.}\ }\textbf {\bibinfo
					{volume} {61}},\ \bibinfo {pages} {69} (\bibinfo {year} {2011})},\ \Eprint
			{https://arxiv.org/abs/1111.1436} {arXiv:1111.1436 [astro-ph.CO]}
			\BibitemShut {NoStop}%
			\bibitem [{\citenamefont {Lattanzi}\ and\ \citenamefont
				{Gerbino}(2018)}]{Lattanzi:2017ubx}%
			\BibitemOpen
			\bibfield  {author} {\bibinfo {author} {\bibfnamefont {M.}~\bibnamefont
					{Lattanzi}}\ and\ \bibinfo {author} {\bibfnamefont {M.}~\bibnamefont
					{Gerbino}},\ }\bibfield  {title} {\bibinfo {title} {{Status of neutrino
						properties and future prospects - Cosmological and astrophysical
						constraints}},\ }\href {https://doi.org/10.3389/fphy.2017.00070} {\bibfield
				{journal} {\bibinfo  {journal} {Front. in Phys.}\ }\textbf {\bibinfo {volume}
					{5}},\ \bibinfo {pages} {70} (\bibinfo {year} {2018})},\ \Eprint
			{https://arxiv.org/abs/1712.07109} {arXiv:1712.07109 [astro-ph.CO]}
			\BibitemShut {NoStop}%
			\bibitem [{\citenamefont {Minkowski}(1977)}]{Minkowski:1977sc}%
			\BibitemOpen
			\bibfield  {author} {\bibinfo {author} {\bibfnamefont {P.}~\bibnamefont
					{Minkowski}},\ }\bibfield  {title} {\bibinfo {title} {{$\mu \to e\gamma$ at a
						Rate of One Out of $10^{9}$ Muon Decays?}},\ }\href
			{https://doi.org/10.1016/0370-2693(77)90435-X} {\bibfield  {journal}
				{\bibinfo  {journal} {Phys. Lett. B}\ }\textbf {\bibinfo {volume} {67}},\
				\bibinfo {pages} {421} (\bibinfo {year} {1977})}\BibitemShut {NoStop}%
			\bibitem [{\citenamefont {Mohapatra}\ and\ \citenamefont
				{Senjanovic}(1980)}]{Mohapatra:1979ia}%
			\BibitemOpen
			\bibfield  {author} {\bibinfo {author} {\bibfnamefont {R.~N.}\ \bibnamefont
					{Mohapatra}}\ and\ \bibinfo {author} {\bibfnamefont {G.}~\bibnamefont
					{Senjanovic}},\ }\bibfield  {title} {\bibinfo {title} {{Neutrino Mass and
						Spontaneous Parity Nonconservation}},\ }\href
			{https://doi.org/10.1103/PhysRevLett.44.912} {\bibfield  {journal} {\bibinfo
					{journal} {Phys. Rev. Lett.}\ }\textbf {\bibinfo {volume} {44}},\ \bibinfo
				{pages} {912} (\bibinfo {year} {1980})}\BibitemShut {NoStop}%
			\bibitem [{\citenamefont {Ma}\ and\ \citenamefont {Sarkar}(1998)}]{Ma:1998dx}%
			\BibitemOpen
			\bibfield  {author} {\bibinfo {author} {\bibfnamefont {E.}~\bibnamefont
					{Ma}}\ and\ \bibinfo {author} {\bibfnamefont {U.}~\bibnamefont {Sarkar}},\
			}\bibfield  {title} {\bibinfo {title} {{Neutrino masses and leptogenesis with
						heavy Higgs triplets}},\ }\href {https://doi.org/10.1103/PhysRevLett.80.5716}
			{\bibfield  {journal} {\bibinfo  {journal} {Phys. Rev. Lett.}\ }\textbf
				{\bibinfo {volume} {80}},\ \bibinfo {pages} {5716} (\bibinfo {year}
				{1998})},\ \Eprint {https://arxiv.org/abs/hep-ph/9802445}
			{arXiv:hep-ph/9802445} \BibitemShut {NoStop}%
			\bibitem [{\citenamefont {Wetterich}(1981)}]{Wetterich:1981bx}%
			\BibitemOpen
			\bibfield  {author} {\bibinfo {author} {\bibfnamefont {C.}~\bibnamefont
					{Wetterich}},\ }\bibfield  {title} {\bibinfo {title} {{Neutrino Masses and
						the Scale of B-L Violation}},\ }\href
			{https://doi.org/10.1016/0550-3213(81)90279-0} {\bibfield  {journal}
				{\bibinfo  {journal} {Nucl. Phys.}\ }\textbf {\bibinfo {volume} {B187}},\
				\bibinfo {pages} {343} (\bibinfo {year} {1981})}\BibitemShut {NoStop}%
			\bibitem [{\citenamefont {Foot}\ \emph {et~al.}(1989)\citenamefont {Foot},
				\citenamefont {Lew}, \citenamefont {He},\ and\ \citenamefont
				{Joshi}}]{Foot:1988aq}%
			\BibitemOpen
			\bibfield  {author} {\bibinfo {author} {\bibfnamefont {R.}~\bibnamefont
					{Foot}}, \bibinfo {author} {\bibfnamefont {H.}~\bibnamefont {Lew}}, \bibinfo
				{author} {\bibfnamefont {X.}~\bibnamefont {He}},\ and\ \bibinfo {author}
				{\bibfnamefont {G.~C.}\ \bibnamefont {Joshi}},\ }\bibfield  {title} {\bibinfo
				{title} {{Seesaw Neutrino Masses Induced by a Triplet of Leptons}},\ }\href
			{https://doi.org/10.1007/BF01415558} {\bibfield  {journal} {\bibinfo
					{journal} {Z. Phys. C}\ }\textbf {\bibinfo {volume} {44}},\ \bibinfo {pages}
				{441} (\bibinfo {year} {1989})}\BibitemShut {NoStop}%
			\bibitem [{\citenamefont {Weinberg}(1979)}]{Weinberg:1979sa}%
			\BibitemOpen
			\bibfield  {author} {\bibinfo {author} {\bibfnamefont {S.}~\bibnamefont
					{Weinberg}},\ }\bibfield  {title} {\bibinfo {title} {{Baryon and Lepton
						Nonconserving Processes}},\ }\href
			{https://doi.org/10.1103/PhysRevLett.43.1566} {\bibfield  {journal} {\bibinfo
					{journal} {Phys. Rev. Lett.}\ }\textbf {\bibinfo {volume} {43}},\ \bibinfo
				{pages} {1566} (\bibinfo {year} {1979})}\BibitemShut {NoStop}%
			\bibitem [{\citenamefont {Ma}(1998)}]{Ma:1998dn}%
			\BibitemOpen
			\bibfield  {author} {\bibinfo {author} {\bibfnamefont {E.}~\bibnamefont
					{Ma}},\ }\bibfield  {title} {\bibinfo {title} {{Pathways to naturally small
						neutrino masses}},\ }\href {https://doi.org/10.1103/PhysRevLett.81.1171}
			{\bibfield  {journal} {\bibinfo  {journal} {Phys. Rev. Lett.}\ }\textbf
				{\bibinfo {volume} {81}},\ \bibinfo {pages} {1171} (\bibinfo {year}
				{1998})},\ \Eprint {https://arxiv.org/abs/hep-ph/9805219}
			{arXiv:hep-ph/9805219 [hep-ph]} \BibitemShut {NoStop}%
			\bibitem [{\citenamefont {Fukugita}\ and\ \citenamefont
				{Yanagida}(1986)}]{Fukugita:1986hr}%
			\BibitemOpen
			\bibfield  {author} {\bibinfo {author} {\bibfnamefont {M.}~\bibnamefont
					{Fukugita}}\ and\ \bibinfo {author} {\bibfnamefont {T.}~\bibnamefont
					{Yanagida}},\ }\bibfield  {title} {\bibinfo {title} {{Baryogenesis Without
						Grand Unification}},\ }\href {https://doi.org/10.1016/0370-2693(86)91126-3}
			{\bibfield  {journal} {\bibinfo  {journal} {Phys. Lett.}\ }\textbf {\bibinfo
					{volume} {B174}},\ \bibinfo {pages} {45} (\bibinfo {year}
				{1986})}\BibitemShut {NoStop}%
			\bibitem [{\citenamefont {Luty}(1992)}]{Luty:1992un}%
			\BibitemOpen
			\bibfield  {author} {\bibinfo {author} {\bibfnamefont {M.~A.}\ \bibnamefont
					{Luty}},\ }\bibfield  {title} {\bibinfo {title} {{Baryogenesis via
						leptogenesis}},\ }\href {https://doi.org/10.1103/PhysRevD.45.455} {\bibfield
				{journal} {\bibinfo  {journal} {Phys. Rev. D}\ }\textbf {\bibinfo {volume}
					{45}},\ \bibinfo {pages} {455} (\bibinfo {year} {1992})}\BibitemShut
			{NoStop}%
			\bibitem [{\citenamefont {Mohapatra}\ and\ \citenamefont
				{Zhang}(1992)}]{Mohapatra:1992pk}%
			\BibitemOpen
			\bibfield  {author} {\bibinfo {author} {\bibfnamefont {R.~N.}\ \bibnamefont
					{Mohapatra}}\ and\ \bibinfo {author} {\bibfnamefont {X.}~\bibnamefont
					{Zhang}},\ }\bibfield  {title} {\bibinfo {title} {{Electroweak baryogenesis
						in left-right symmetric models}},\ }\href
			{https://doi.org/10.1103/PhysRevD.46.5331} {\bibfield  {journal} {\bibinfo
					{journal} {Phys. Rev. D}\ }\textbf {\bibinfo {volume} {46}},\ \bibinfo
				{pages} {5331} (\bibinfo {year} {1992})}\BibitemShut {NoStop}%
			\bibitem [{\citenamefont {Flanz}\ \emph {et~al.}(1995)\citenamefont {Flanz},
				\citenamefont {Paschos},\ and\ \citenamefont {Sarkar}}]{Flanz:1994yx}%
			\BibitemOpen
			\bibfield  {author} {\bibinfo {author} {\bibfnamefont {M.}~\bibnamefont
					{Flanz}}, \bibinfo {author} {\bibfnamefont {E.~A.}\ \bibnamefont {Paschos}},\
				and\ \bibinfo {author} {\bibfnamefont {U.}~\bibnamefont {Sarkar}},\
			}\bibfield  {title} {\bibinfo {title} {{Baryogenesis from a lepton asymmetric
						universe}},\ }\href {https://doi.org/10.1016/0370-2693(94)01555-Q} {\bibfield
				{journal} {\bibinfo  {journal} {Phys. Lett. B}\ }\textbf {\bibinfo {volume}
					{345}},\ \bibinfo {pages} {248} (\bibinfo {year} {1995})},\ \bibinfo {note}
			{[Erratum: Phys.Lett.B 384, 487--487 (1996), Erratum: Phys.Lett.B 382,
				447--447 (1996)]},\ \Eprint {https://arxiv.org/abs/hep-ph/9411366}
			{arXiv:hep-ph/9411366} \BibitemShut {NoStop}%
			\bibitem [{\citenamefont {Buchmuller}\ \emph {et~al.}(2005)\citenamefont
				{Buchmuller}, \citenamefont {Di~Bari},\ and\ \citenamefont
				{Plumacher}}]{Buchmuller:2004nz}%
			\BibitemOpen
			\bibfield  {author} {\bibinfo {author} {\bibfnamefont {W.}~\bibnamefont
					{Buchmuller}}, \bibinfo {author} {\bibfnamefont {P.}~\bibnamefont
					{Di~Bari}},\ and\ \bibinfo {author} {\bibfnamefont {M.}~\bibnamefont
					{Plumacher}},\ }\bibfield  {title} {\bibinfo {title} {{Leptogenesis for
						pedestrians}},\ }\href {https://doi.org/10.1016/j.aop.2004.02.003} {\bibfield
				{journal} {\bibinfo  {journal} {Annals Phys.}\ }\textbf {\bibinfo {volume}
					{315}},\ \bibinfo {pages} {305} (\bibinfo {year} {2005})},\ \Eprint
			{https://arxiv.org/abs/hep-ph/0401240} {arXiv:hep-ph/0401240 [hep-ph]}
			\BibitemShut {NoStop}%
			\bibitem [{\citenamefont {Davidson}\ \emph {et~al.}(2008)\citenamefont
				{Davidson}, \citenamefont {Nardi},\ and\ \citenamefont
				{Nir}}]{Davidson:2008bu}%
			\BibitemOpen
			\bibfield  {author} {\bibinfo {author} {\bibfnamefont {S.}~\bibnamefont
					{Davidson}}, \bibinfo {author} {\bibfnamefont {E.}~\bibnamefont {Nardi}},\
				and\ \bibinfo {author} {\bibfnamefont {Y.}~\bibnamefont {Nir}},\ }\bibfield
			{title} {\bibinfo {title} {{Leptogenesis}},\ }\href
			{https://doi.org/10.1016/j.physrep.2008.06.002} {\bibfield  {journal}
				{\bibinfo  {journal} {Phys. Rept.}\ }\textbf {\bibinfo {volume} {466}},\
				\bibinfo {pages} {105} (\bibinfo {year} {2008})},\ \Eprint
			{https://arxiv.org/abs/0802.2962} {arXiv:0802.2962 [hep-ph]} \BibitemShut
			{NoStop}%
			\bibitem [{\citenamefont {Falkowski}\ \emph {et~al.}(2011)\citenamefont
				{Falkowski}, \citenamefont {Ruderman},\ and\ \citenamefont
				{Volansky}}]{Falkowski:2011xh}%
			\BibitemOpen
			\bibfield  {author} {\bibinfo {author} {\bibfnamefont {A.}~\bibnamefont
					{Falkowski}}, \bibinfo {author} {\bibfnamefont {J.~T.}\ \bibnamefont
					{Ruderman}},\ and\ \bibinfo {author} {\bibfnamefont {T.}~\bibnamefont
					{Volansky}},\ }\bibfield  {title} {\bibinfo {title} {{Asymmetric Dark Matter
						from Leptogenesis}},\ }\href {https://doi.org/10.1007/JHEP05(2011)106}
			{\bibfield  {journal} {\bibinfo  {journal} {JHEP}\ }\textbf {\bibinfo
					{volume} {05}},\ \bibinfo {pages} {106}},\ \Eprint
			{https://arxiv.org/abs/1101.4936} {arXiv:1101.4936 [hep-ph]} \BibitemShut
			{NoStop}%
			\bibitem [{\citenamefont {Patel}\ \emph {et~al.}(2022)\citenamefont {Patel},
				\citenamefont {Malhotra}, \citenamefont {Patra},\ and\ \citenamefont
				{Yajnik}}]{Patel:2022xyv}%
			\BibitemOpen
			\bibfield  {author} {\bibinfo {author} {\bibfnamefont {U.}~\bibnamefont
					{Patel}}, \bibinfo {author} {\bibfnamefont {L.}~\bibnamefont {Malhotra}},
				\bibinfo {author} {\bibfnamefont {S.}~\bibnamefont {Patra}},\ and\ \bibinfo
				{author} {\bibfnamefont {U.~A.}\ \bibnamefont {Yajnik}},\ }\bibfield  {title}
			{\bibinfo {title} {{Cogenesis of visible and dark sector asymmetry in a
						minimal seesaw framework}},\ }\href@noop {} {\  (\bibinfo {year} {2022})},\
			\Eprint {https://arxiv.org/abs/2211.04722} {arXiv:2211.04722 [hep-ph]}
			\BibitemShut {NoStop}%
			\bibitem [{\citenamefont {Biswas}\ \emph {et~al.}(2019)\citenamefont {Biswas},
				\citenamefont {Choubey}, \citenamefont {Covi},\ and\ \citenamefont
				{Khan}}]{Biswas:2018sib}%
			\BibitemOpen
			\bibfield  {author} {\bibinfo {author} {\bibfnamefont {A.}~\bibnamefont
					{Biswas}}, \bibinfo {author} {\bibfnamefont {S.}~\bibnamefont {Choubey}},
				\bibinfo {author} {\bibfnamefont {L.}~\bibnamefont {Covi}},\ and\ \bibinfo
				{author} {\bibfnamefont {S.}~\bibnamefont {Khan}},\ }\bibfield  {title}
			{\bibinfo {title} {{Common origin of baryon asymmetry, dark matter and
						neutrino mass}},\ }\href {https://doi.org/10.1007/JHEP05(2019)193} {\bibfield
				{journal} {\bibinfo  {journal} {JHEP}\ }\textbf {\bibinfo {volume} {05}},\
				\bibinfo {pages} {193}},\ \Eprint {https://arxiv.org/abs/1812.06122}
			{arXiv:1812.06122 [hep-ph]} \BibitemShut {NoStop}%
			\bibitem [{\citenamefont {Narendra}\ \emph {et~al.}(2018)\citenamefont
				{Narendra}, \citenamefont {Patra}, \citenamefont {Sahu},\ and\ \citenamefont
				{Shil}}]{Narendra:2018vfw}%
			\BibitemOpen
			\bibfield  {author} {\bibinfo {author} {\bibfnamefont {N.}~\bibnamefont
					{Narendra}}, \bibinfo {author} {\bibfnamefont {S.}~\bibnamefont {Patra}},
				\bibinfo {author} {\bibfnamefont {N.}~\bibnamefont {Sahu}},\ and\ \bibinfo
				{author} {\bibfnamefont {S.}~\bibnamefont {Shil}},\ }\bibfield  {title}
			{\bibinfo {title} {{Baryogenesis via Leptogenesis from Asymmetric Dark Matter
						and radiatively generated Neutrino mass}},\ }\href
			{https://doi.org/10.1103/PhysRevD.98.095016} {\bibfield  {journal} {\bibinfo
					{journal} {Phys. Rev. D}\ }\textbf {\bibinfo {volume} {98}},\ \bibinfo
				{pages} {095016} (\bibinfo {year} {2018})},\ \Eprint
			{https://arxiv.org/abs/1805.04860} {arXiv:1805.04860 [hep-ph]} \BibitemShut
			{NoStop}%
			\bibitem [{\citenamefont {Nagata}\ \emph {et~al.}(2017)\citenamefont {Nagata},
				\citenamefont {Olive},\ and\ \citenamefont {Zheng}}]{Nagata:2016knk}%
			\BibitemOpen
			\bibfield  {author} {\bibinfo {author} {\bibfnamefont {N.}~\bibnamefont
					{Nagata}}, \bibinfo {author} {\bibfnamefont {K.~A.}\ \bibnamefont {Olive}},\
				and\ \bibinfo {author} {\bibfnamefont {J.}~\bibnamefont {Zheng}},\ }\bibfield
			{title} {\bibinfo {title} {{Asymmetric Dark Matter Models in SO(10)}},\
			}\href {https://doi.org/10.1088/1475-7516/2017/02/016} {\bibfield  {journal}
				{\bibinfo  {journal} {JCAP}\ }\textbf {\bibinfo {volume} {02}},\ \bibinfo
				{pages} {016}},\ \Eprint {https://arxiv.org/abs/1611.04693} {arXiv:1611.04693
				[hep-ph]} \BibitemShut {NoStop}%
			\bibitem [{\citenamefont {Arina}\ and\ \citenamefont
				{Sahu}(2012)}]{Arina:2011cu}%
			\BibitemOpen
			\bibfield  {author} {\bibinfo {author} {\bibfnamefont {C.}~\bibnamefont
					{Arina}}\ and\ \bibinfo {author} {\bibfnamefont {N.}~\bibnamefont {Sahu}},\
			}\bibfield  {title} {\bibinfo {title} {{Asymmetric Inelastic Inert Doublet
						Dark Matter from Triplet Scalar Leptogenesis}},\ }\href
			{https://doi.org/10.1016/j.nuclphysb.2011.09.014} {\bibfield  {journal}
				{\bibinfo  {journal} {Nucl. Phys. B}\ }\textbf {\bibinfo {volume} {854}},\
				\bibinfo {pages} {666} (\bibinfo {year} {2012})},\ \Eprint
			{https://arxiv.org/abs/1108.3967} {arXiv:1108.3967 [hep-ph]} \BibitemShut
			{NoStop}%
			\bibitem [{\citenamefont {Arina}\ \emph {et~al.}(2012)\citenamefont {Arina},
				\citenamefont {Gong},\ and\ \citenamefont {Sahu}}]{Arina:2012fb}%
			\BibitemOpen
			\bibfield  {author} {\bibinfo {author} {\bibfnamefont {C.}~\bibnamefont
					{Arina}}, \bibinfo {author} {\bibfnamefont {J.-O.}\ \bibnamefont {Gong}},\
				and\ \bibinfo {author} {\bibfnamefont {N.}~\bibnamefont {Sahu}},\ }\bibfield
			{title} {\bibinfo {title} {{Unifying darko-lepto-genesis with scalar triplet
						inflation}},\ }\href {https://doi.org/10.1016/j.nuclphysb.2012.07.029}
			{\bibfield  {journal} {\bibinfo  {journal} {Nucl. Phys. B}\ }\textbf
				{\bibinfo {volume} {865}},\ \bibinfo {pages} {430} (\bibinfo {year}
				{2012})},\ \Eprint {https://arxiv.org/abs/1206.0009} {arXiv:1206.0009
				[hep-ph]} \BibitemShut {NoStop}%
			\bibitem [{\citenamefont {Arina}\ \emph {et~al.}(2013)\citenamefont {Arina},
				\citenamefont {Mohapatra},\ and\ \citenamefont {Sahu}}]{Arina:2012aj}%
			\BibitemOpen
			\bibfield  {author} {\bibinfo {author} {\bibfnamefont {C.}~\bibnamefont
					{Arina}}, \bibinfo {author} {\bibfnamefont {R.~N.}\ \bibnamefont
					{Mohapatra}},\ and\ \bibinfo {author} {\bibfnamefont {N.}~\bibnamefont
					{Sahu}},\ }\bibfield  {title} {\bibinfo {title} {{Co-genesis of Matter and
						Dark Matter with Vector-like Fourth Generation Leptons}},\ }\href
			{https://doi.org/10.1016/j.physletb.2013.01.059} {\bibfield  {journal}
				{\bibinfo  {journal} {Phys. Lett. B}\ }\textbf {\bibinfo {volume} {720}},\
				\bibinfo {pages} {130} (\bibinfo {year} {2013})},\ \Eprint
			{https://arxiv.org/abs/1211.0435} {arXiv:1211.0435 [hep-ph]} \BibitemShut
			{NoStop}%
			\bibitem [{\citenamefont {Narendra}\ \emph {et~al.}(2021)\citenamefont
				{Narendra}, \citenamefont {Sahu},\ and\ \citenamefont
				{Shil}}]{Narendra:2019cyt}%
			\BibitemOpen
			\bibfield  {author} {\bibinfo {author} {\bibfnamefont {N.}~\bibnamefont
					{Narendra}}, \bibinfo {author} {\bibfnamefont {N.}~\bibnamefont {Sahu}},\
				and\ \bibinfo {author} {\bibfnamefont {S.}~\bibnamefont {Shil}},\ }\bibfield
			{title} {\bibinfo {title} {{Dark matter to baryon ratio from scalar triplets
						decay in type-II seesaw}},\ }\href
			{https://doi.org/10.1140/epjc/s10052-021-09882-3} {\bibfield  {journal}
				{\bibinfo  {journal} {Eur. Phys. J. C}\ }\textbf {\bibinfo {volume} {81}},\
				\bibinfo {pages} {1098} (\bibinfo {year} {2021})},\ \Eprint
			{https://arxiv.org/abs/1910.12762} {arXiv:1910.12762 [hep-ph]} \BibitemShut
			{NoStop}%
			\bibitem [{\citenamefont {Petraki}\ and\ \citenamefont
				{Volkas}(2013)}]{Petraki:2013wwa}%
			\BibitemOpen
			\bibfield  {author} {\bibinfo {author} {\bibfnamefont {K.}~\bibnamefont
					{Petraki}}\ and\ \bibinfo {author} {\bibfnamefont {R.~R.}\ \bibnamefont
					{Volkas}},\ }\bibfield  {title} {\bibinfo {title} {{Review of asymmetric dark
						matter}},\ }\href {https://doi.org/10.1142/S0217751X13300287} {\bibfield
				{journal} {\bibinfo  {journal} {Int. J. Mod. Phys.}\ }\textbf {\bibinfo
					{volume} {A28}},\ \bibinfo {pages} {1330028} (\bibinfo {year} {2013})},\
			\Eprint {https://arxiv.org/abs/1305.4939} {arXiv:1305.4939 [hep-ph]}
			\BibitemShut {NoStop}%
			\bibitem [{\citenamefont {Zurek}(2014)}]{Zurek:2013wia}%
			\BibitemOpen
			\bibfield  {author} {\bibinfo {author} {\bibfnamefont {K.~M.}\ \bibnamefont
					{Zurek}},\ }\bibfield  {title} {\bibinfo {title} {{Asymmetric Dark Matter:
						Theories, Signatures, and Constraints}},\ }\href
			{https://doi.org/10.1016/j.physrep.2013.12.001} {\bibfield  {journal}
				{\bibinfo  {journal} {Phys. Rept.}\ }\textbf {\bibinfo {volume} {537}},\
				\bibinfo {pages} {91} (\bibinfo {year} {2014})},\ \Eprint
			{https://arxiv.org/abs/1308.0338} {arXiv:1308.0338 [hep-ph]} \BibitemShut
			{NoStop}%
			\bibitem [{\citenamefont {Abdelhameed}\ \emph {et~al.}(2019)\citenamefont
				{Abdelhameed} \emph {et~al.}}]{CRESST:2019jnq}%
			\BibitemOpen
			\bibfield  {author} {\bibinfo {author} {\bibfnamefont {A.~H.}\ \bibnamefont
					{Abdelhameed}} \emph {et~al.} (\bibinfo {collaboration} {CRESST}),\
			}\bibfield  {title} {\bibinfo {title} {{First results from the CRESST-III
						low-mass dark matter program}},\ }\href
			{https://doi.org/10.1103/PhysRevD.100.102002} {\bibfield  {journal} {\bibinfo
					{journal} {Phys. Rev. D}\ }\textbf {\bibinfo {volume} {100}},\ \bibinfo
				{pages} {102002} (\bibinfo {year} {2019})},\ \Eprint
			{https://arxiv.org/abs/1904.00498} {arXiv:1904.00498 [astro-ph.CO]}
			\BibitemShut {NoStop}%
			\bibitem [{\citenamefont {Aalbers}\ \emph {et~al.}(2023)\citenamefont {Aalbers}
				\emph {et~al.}}]{LZ:2022lsv}%
			\BibitemOpen
			\bibfield  {author} {\bibinfo {author} {\bibfnamefont {J.}~\bibnamefont
					{Aalbers}} \emph {et~al.} (\bibinfo {collaboration} {LZ}),\ }\bibfield
			{title} {\bibinfo {title} {{First Dark Matter Search Results from the
						LUX-ZEPLIN (LZ) Experiment}},\ }\href
			{https://doi.org/10.1103/PhysRevLett.131.041002} {\bibfield  {journal}
				{\bibinfo  {journal} {Phys. Rev. Lett.}\ }\textbf {\bibinfo {volume} {131}},\
				\bibinfo {pages} {041002} (\bibinfo {year} {2023})},\ \Eprint
			{https://arxiv.org/abs/2207.03764} {arXiv:2207.03764 [hep-ex]} \BibitemShut
			{NoStop}%
			\bibitem [{\citenamefont {Aprile}\ \emph {et~al.}(2023)\citenamefont {Aprile}
				\emph {et~al.}}]{XENON:2023cxc}%
			\BibitemOpen
			\bibfield  {author} {\bibinfo {author} {\bibfnamefont {E.}~\bibnamefont
					{Aprile}} \emph {et~al.} (\bibinfo {collaboration} {XENON}),\ }\bibfield
			{title} {\bibinfo {title} {{First Dark Matter Search with Nuclear Recoils
						from the XENONnT Experiment}},\ }\href
			{https://doi.org/10.1103/PhysRevLett.131.041003} {\bibfield  {journal}
				{\bibinfo  {journal} {Phys. Rev. Lett.}\ }\textbf {\bibinfo {volume} {131}},\
				\bibinfo {pages} {041003} (\bibinfo {year} {2023})},\ \Eprint
			{https://arxiv.org/abs/2303.14729} {arXiv:2303.14729 [hep-ex]} \BibitemShut
			{NoStop}%
			\bibitem [{\citenamefont {Agnes}\ \emph {et~al.}(2023)\citenamefont {Agnes}
				\emph {et~al.}}]{GlobalArgonDarkMatter:2022ppc}%
			\BibitemOpen
			\bibfield  {author} {\bibinfo {author} {\bibfnamefont {P.}~\bibnamefont
					{Agnes}} \emph {et~al.} (\bibinfo {collaboration} {Global Argon Dark
					Matter}),\ }\bibfield  {title} {\bibinfo {title} {{Sensitivity projections
						for a dual-phase argon TPC optimized for light dark matter searches through
						the ionization channel}},\ }\href
			{https://doi.org/10.1103/PhysRevD.107.112006} {\bibfield  {journal} {\bibinfo
					{journal} {Phys. Rev. D}\ }\textbf {\bibinfo {volume} {107}},\ \bibinfo
				{pages} {112006} (\bibinfo {year} {2023})},\ \Eprint
			{https://arxiv.org/abs/2209.01177} {arXiv:2209.01177 [physics.ins-det]}
			\BibitemShut {NoStop}%
			\bibitem [{\citenamefont {Aalbers}\ \emph {et~al.}(2016)\citenamefont {Aalbers}
				\emph {et~al.}}]{DARWIN:2016hyl}%
			\BibitemOpen
			\bibfield  {author} {\bibinfo {author} {\bibfnamefont {J.}~\bibnamefont
					{Aalbers}} \emph {et~al.} (\bibinfo {collaboration} {DARWIN}),\ }\bibfield
			{title} {\bibinfo {title} {{DARWIN: towards the ultimate dark matter
						detector}},\ }\href {https://doi.org/10.1088/1475-7516/2016/11/017}
			{\bibfield  {journal} {\bibinfo  {journal} {JCAP}\ }\textbf {\bibinfo
					{volume} {11}},\ \bibinfo {pages} {017}},\ \Eprint
			{https://arxiv.org/abs/1606.07001} {arXiv:1606.07001 [astro-ph.IM]}
			\BibitemShut {NoStop}%
			\bibitem [{\citenamefont {Cirelli}\ \emph {et~al.}(2006)\citenamefont
				{Cirelli}, \citenamefont {Fornengo},\ and\ \citenamefont
				{Strumia}}]{Cirelli:2005uq}%
			\BibitemOpen
			\bibfield  {author} {\bibinfo {author} {\bibfnamefont {M.}~\bibnamefont
					{Cirelli}}, \bibinfo {author} {\bibfnamefont {N.}~\bibnamefont {Fornengo}},\
				and\ \bibinfo {author} {\bibfnamefont {A.}~\bibnamefont {Strumia}},\
			}\bibfield  {title} {\bibinfo {title} {{Minimal dark matter}},\ }\href
			{https://doi.org/10.1016/j.nuclphysb.2006.07.012} {\bibfield  {journal}
				{\bibinfo  {journal} {Nucl. Phys.}\ }\textbf {\bibinfo {volume} {B753}},\
				\bibinfo {pages} {178} (\bibinfo {year} {2006})},\ \Eprint
			{https://arxiv.org/abs/hep-ph/0512090} {arXiv:hep-ph/0512090 [hep-ph]}
			\BibitemShut {NoStop}%
			\bibitem [{\citenamefont {Casas}\ and\ \citenamefont
				{Ibarra}(2001)}]{Casas:2001sr}%
			\BibitemOpen
			\bibfield  {author} {\bibinfo {author} {\bibfnamefont {J.~A.}\ \bibnamefont
					{Casas}}\ and\ \bibinfo {author} {\bibfnamefont {A.}~\bibnamefont {Ibarra}},\
			}\bibfield  {title} {\bibinfo {title} {{Oscillating neutrinos and muon --->
						e, gamma}},\ }\href {https://doi.org/10.1016/S0550-3213(01)00475-8}
			{\bibfield  {journal} {\bibinfo  {journal} {Nucl. Phys.}\ }\textbf {\bibinfo
					{volume} {B618}},\ \bibinfo {pages} {171} (\bibinfo {year} {2001})},\ \Eprint
			{https://arxiv.org/abs/hep-ph/0103065} {arXiv:hep-ph/0103065 [hep-ph]}
			\BibitemShut {NoStop}%
			\bibitem [{\citenamefont {Hambye}(2012)}]{Hambye:2012fh}%
			\BibitemOpen
			\bibfield  {author} {\bibinfo {author} {\bibfnamefont {T.}~\bibnamefont
					{Hambye}},\ }\bibfield  {title} {\bibinfo {title} {{Leptogenesis: beyond the
						minimal type I seesaw scenario}},\ }\href
			{https://doi.org/10.1088/1367-2630/14/12/125014} {\bibfield  {journal}
				{\bibinfo  {journal} {New J. Phys.}\ }\textbf {\bibinfo {volume} {14}},\
				\bibinfo {pages} {125014} (\bibinfo {year} {2012})},\ \Eprint
			{https://arxiv.org/abs/1212.2888} {arXiv:1212.2888 [hep-ph]} \BibitemShut
			{NoStop}%
			\bibitem [{\citenamefont {Vatsyayan}\ and\ \citenamefont
				{Goswami}(2023)}]{Vatsyayan:2022rth}%
			\BibitemOpen
			\bibfield  {author} {\bibinfo {author} {\bibfnamefont {D.}~\bibnamefont
					{Vatsyayan}}\ and\ \bibinfo {author} {\bibfnamefont {S.}~\bibnamefont
					{Goswami}},\ }\bibfield  {title} {\bibinfo {title} {{Lowering the scale of
						fermion triplet leptogenesis with two Higgs doublets}},\ }\href
			{https://doi.org/10.1103/PhysRevD.107.035014} {\bibfield  {journal} {\bibinfo
					{journal} {Phys. Rev. D}\ }\textbf {\bibinfo {volume} {107}},\ \bibinfo
				{pages} {035014} (\bibinfo {year} {2023})},\ \Eprint
			{https://arxiv.org/abs/2208.12011} {arXiv:2208.12011 [hep-ph]} \BibitemShut
			{NoStop}%
			\bibitem [{\citenamefont {Fischler}\ and\ \citenamefont
				{Flauger}(2008)}]{Fischler:2008xm}%
			\BibitemOpen
			\bibfield  {author} {\bibinfo {author} {\bibfnamefont {W.}~\bibnamefont
					{Fischler}}\ and\ \bibinfo {author} {\bibfnamefont {R.}~\bibnamefont
					{Flauger}},\ }\bibfield  {title} {\bibinfo {title} {{Neutrino Masses,
						Leptogenesis, and Unification in the Absence of Low Energy Supersymmetry}},\
			}\href {https://doi.org/10.1088/1126-6708/2008/09/020} {\bibfield  {journal}
				{\bibinfo  {journal} {JHEP}\ }\textbf {\bibinfo {volume} {09}},\ \bibinfo
				{pages} {020}},\ \Eprint {https://arxiv.org/abs/0805.3000} {arXiv:0805.3000
				[hep-ph]} \BibitemShut {NoStop}%
			\bibitem [{\citenamefont {Hambye}\ \emph {et~al.}(2004)\citenamefont {Hambye},
				\citenamefont {Lin}, \citenamefont {Notari}, \citenamefont {Papucci},\ and\
				\citenamefont {Strumia}}]{Hambye:2003rt}%
			\BibitemOpen
			\bibfield  {author} {\bibinfo {author} {\bibfnamefont {T.}~\bibnamefont
					{Hambye}}, \bibinfo {author} {\bibfnamefont {Y.}~\bibnamefont {Lin}},
				\bibinfo {author} {\bibfnamefont {A.}~\bibnamefont {Notari}}, \bibinfo
				{author} {\bibfnamefont {M.}~\bibnamefont {Papucci}},\ and\ \bibinfo {author}
				{\bibfnamefont {A.}~\bibnamefont {Strumia}},\ }\bibfield  {title} {\bibinfo
				{title} {{Constraints on neutrino masses from leptogenesis models}},\ }\href
			{https://doi.org/10.1016/j.nuclphysb.2004.06.027} {\bibfield  {journal}
				{\bibinfo  {journal} {Nucl. Phys. B}\ }\textbf {\bibinfo {volume} {695}},\
				\bibinfo {pages} {169} (\bibinfo {year} {2004})},\ \Eprint
			{https://arxiv.org/abs/hep-ph/0312203} {arXiv:hep-ph/0312203} \BibitemShut
			{NoStop}%
			\bibitem [{\citenamefont {Borah}\ \emph {et~al.}(2024)\citenamefont {Borah},
				\citenamefont {Mahapatra}, \citenamefont {Paul}, \citenamefont {Sahu},\ and\
				\citenamefont {Shukla}}]{Borah:2024wos}%
			\BibitemOpen
			\bibfield  {author} {\bibinfo {author} {\bibfnamefont {D.}~\bibnamefont
					{Borah}}, \bibinfo {author} {\bibfnamefont {S.}~\bibnamefont {Mahapatra}},
				\bibinfo {author} {\bibfnamefont {P.~K.}\ \bibnamefont {Paul}}, \bibinfo
				{author} {\bibfnamefont {N.}~\bibnamefont {Sahu}},\ and\ \bibinfo {author}
				{\bibfnamefont {P.}~\bibnamefont {Shukla}},\ }\bibfield  {title} {\bibinfo
				{title} {{Asymmetric self-interacting dark matter with a canonical seesaw
						model}},\ }\href {https://doi.org/10.1103/PhysRevD.110.035033} {\bibfield
				{journal} {\bibinfo  {journal} {Phys. Rev. D}\ }\textbf {\bibinfo {volume}
					{110}},\ \bibinfo {pages} {035033} (\bibinfo {year} {2024})},\ \Eprint
			{https://arxiv.org/abs/2404.14912} {arXiv:2404.14912 [hep-ph]} \BibitemShut
			{NoStop}%
			\bibitem [{\citenamefont {Aristizabal~Sierra}\ \emph
				{et~al.}(2010)\citenamefont {Aristizabal~Sierra}, \citenamefont {Kamenik},\
				and\ \citenamefont {Nemevsek}}]{AristizabalSierra:2010mv}%
			\BibitemOpen
			\bibfield  {author} {\bibinfo {author} {\bibfnamefont {D.}~\bibnamefont
					{Aristizabal~Sierra}}, \bibinfo {author} {\bibfnamefont {J.~F.}\ \bibnamefont
					{Kamenik}},\ and\ \bibinfo {author} {\bibfnamefont {M.}~\bibnamefont
					{Nemevsek}},\ }\bibfield  {title} {\bibinfo {title} {{Implications of Flavor
						Dynamics for Fermion Triplet Leptogenesis}},\ }\href
			{https://doi.org/10.1007/JHEP10(2010)036} {\bibfield  {journal} {\bibinfo
					{journal} {JHEP}\ }\textbf {\bibinfo {volume} {10}},\ \bibinfo {pages}
				{036}},\ \Eprint {https://arxiv.org/abs/1007.1907} {arXiv:1007.1907 [hep-ph]}
			\BibitemShut {NoStop}%
			\bibitem [{\citenamefont {Gondolo}\ and\ \citenamefont
				{Gelmini}(1991)}]{Gondolo:1990dk}%
			\BibitemOpen
			\bibfield  {author} {\bibinfo {author} {\bibfnamefont {P.}~\bibnamefont
					{Gondolo}}\ and\ \bibinfo {author} {\bibfnamefont {G.}~\bibnamefont
					{Gelmini}},\ }\bibfield  {title} {\bibinfo {title} {{Cosmic abundances of
						stable particles: Improved analysis}},\ }\href
			{https://doi.org/10.1016/0550-3213(91)90438-4} {\bibfield  {journal}
				{\bibinfo  {journal} {Nucl. Phys.}\ }\textbf {\bibinfo {volume} {B360}},\
				\bibinfo {pages} {145} (\bibinfo {year} {1991})}\BibitemShut {NoStop}%
			\bibitem [{\citenamefont {Cirelli}\ \emph {et~al.}(2012)\citenamefont
				{Cirelli}, \citenamefont {Panci}, \citenamefont {Servant},\ and\
				\citenamefont {Zaharijas}}]{Cirelli:2011ac}%
			\BibitemOpen
			\bibfield  {author} {\bibinfo {author} {\bibfnamefont {M.}~\bibnamefont
					{Cirelli}}, \bibinfo {author} {\bibfnamefont {P.}~\bibnamefont {Panci}},
				\bibinfo {author} {\bibfnamefont {G.}~\bibnamefont {Servant}},\ and\ \bibinfo
				{author} {\bibfnamefont {G.}~\bibnamefont {Zaharijas}},\ }\bibfield  {title}
			{\bibinfo {title} {{Consequences of DM/antiDM Oscillations for Asymmetric
						WIMP Dark Matter}},\ }\href {https://doi.org/10.1088/1475-7516/2012/03/015}
			{\bibfield  {journal} {\bibinfo  {journal} {JCAP}\ }\textbf {\bibinfo
					{volume} {03}},\ \bibinfo {pages} {015}},\ \Eprint
			{https://arxiv.org/abs/1110.3809} {arXiv:1110.3809 [hep-ph]} \BibitemShut
			{NoStop}%
			\bibitem [{\citenamefont {Tulin}\ \emph {et~al.}(2012)\citenamefont {Tulin},
				\citenamefont {Yu},\ and\ \citenamefont {Zurek}}]{Tulin:2012re}%
			\BibitemOpen
			\bibfield  {author} {\bibinfo {author} {\bibfnamefont {S.}~\bibnamefont
					{Tulin}}, \bibinfo {author} {\bibfnamefont {H.-B.}\ \bibnamefont {Yu}},\ and\
				\bibinfo {author} {\bibfnamefont {K.~M.}\ \bibnamefont {Zurek}},\ }\bibfield
			{title} {\bibinfo {title} {{Oscillating Asymmetric Dark Matter}},\ }\href
			{https://doi.org/10.1088/1475-7516/2012/05/013} {\bibfield  {journal}
				{\bibinfo  {journal} {JCAP}\ }\textbf {\bibinfo {volume} {05}},\ \bibinfo
				{pages} {013}},\ \Eprint {https://arxiv.org/abs/1202.0283} {arXiv:1202.0283
				[hep-ph]} \BibitemShut {NoStop}%
			\bibitem [{\citenamefont {Davidson}\ and\ \citenamefont
				{Ibarra}(2002)}]{Davidson:2002qv}%
			\BibitemOpen
			\bibfield  {author} {\bibinfo {author} {\bibfnamefont {S.}~\bibnamefont
					{Davidson}}\ and\ \bibinfo {author} {\bibfnamefont {A.}~\bibnamefont
					{Ibarra}},\ }\bibfield  {title} {\bibinfo {title} {{A Lower bound on the
						right-handed neutrino mass from leptogenesis}},\ }\href
			{https://doi.org/10.1016/S0370-2693(02)01735-5} {\bibfield  {journal}
				{\bibinfo  {journal} {Phys. Lett.}\ }\textbf {\bibinfo {volume} {B535}},\
				\bibinfo {pages} {25} (\bibinfo {year} {2002})},\ \Eprint
			{https://arxiv.org/abs/hep-ph/0202239} {arXiv:hep-ph/0202239 [hep-ph]}
			\BibitemShut {NoStop}%
			\bibitem [{\citenamefont {Hambye}\ \emph {et~al.}(2006)\citenamefont {Hambye},
				\citenamefont {Raidal},\ and\ \citenamefont {Strumia}}]{Hambye:2005tk}%
			\BibitemOpen
			\bibfield  {author} {\bibinfo {author} {\bibfnamefont {T.}~\bibnamefont
					{Hambye}}, \bibinfo {author} {\bibfnamefont {M.}~\bibnamefont {Raidal}},\
				and\ \bibinfo {author} {\bibfnamefont {A.}~\bibnamefont {Strumia}},\
			}\bibfield  {title} {\bibinfo {title} {{Efficiency and maximal CP-asymmetry
						of scalar triplet leptogenesis}},\ }\href
			{https://doi.org/10.1016/j.physletb.2005.11.007} {\bibfield  {journal}
				{\bibinfo  {journal} {Phys. Lett. B}\ }\textbf {\bibinfo {volume} {632}},\
				\bibinfo {pages} {667} (\bibinfo {year} {2006})},\ \Eprint
			{https://arxiv.org/abs/hep-ph/0510008} {arXiv:hep-ph/0510008} \BibitemShut
			{NoStop}%
			\bibitem [{\citenamefont {Baring}\ \emph {et~al.}(2016)\citenamefont {Baring},
				\citenamefont {Ghosh}, \citenamefont {Queiroz},\ and\ \citenamefont
				{Sinha}}]{Baring:2015sza}%
			\BibitemOpen
			\bibfield  {author} {\bibinfo {author} {\bibfnamefont {M.~G.}\ \bibnamefont
					{Baring}}, \bibinfo {author} {\bibfnamefont {T.}~\bibnamefont {Ghosh}},
				\bibinfo {author} {\bibfnamefont {F.~S.}\ \bibnamefont {Queiroz}},\ and\
				\bibinfo {author} {\bibfnamefont {K.}~\bibnamefont {Sinha}},\ }\bibfield
			{title} {\bibinfo {title} {{New Limits on the Dark Matter Lifetime from Dwarf
						Spheroidal Galaxies using Fermi-LAT}},\ }\href
			{https://doi.org/10.1103/PhysRevD.93.103009} {\bibfield  {journal} {\bibinfo
					{journal} {Phys. Rev. D}\ }\textbf {\bibinfo {volume} {93}},\ \bibinfo
				{pages} {103009} (\bibinfo {year} {2016})},\ \Eprint
			{https://arxiv.org/abs/1510.00389} {arXiv:1510.00389 [hep-ph]} \BibitemShut
			{NoStop}%
			\bibitem [{\citenamefont {Mahapatra}\ \emph {et~al.}(2023)\citenamefont
				{Mahapatra}, \citenamefont {Mohapatra},\ and\ \citenamefont
				{Sahu}}]{Mahapatra:2023zhi}%
			\BibitemOpen
			\bibfield  {author} {\bibinfo {author} {\bibfnamefont {S.}~\bibnamefont
					{Mahapatra}}, \bibinfo {author} {\bibfnamefont {R.~N.}\ \bibnamefont
					{Mohapatra}},\ and\ \bibinfo {author} {\bibfnamefont {N.}~\bibnamefont
					{Sahu}},\ }\bibfield  {title} {\bibinfo {title} {{Gauged
						$Le-Lmu-Ltau$
						symmetry, fourth generation, neutrino mass and dark matter}},\ }\href
			{https://doi.org/10.1016/j.physletb.2023.138011} {\bibfield  {journal}
				{\bibinfo  {journal} {Phys. Lett. B}\ }\textbf {\bibinfo {volume} {843}},\
				\bibinfo {pages} {138011} (\bibinfo {year} {2023})},\ \Eprint
			{https://arxiv.org/abs/2302.01784} {arXiv:2302.01784 [hep-ph]} \BibitemShut
			{NoStop}%
			\bibitem [{\citenamefont {Ellis}\ \emph {et~al.}(2008)\citenamefont {Ellis},
				\citenamefont {Olive},\ and\ \citenamefont {Savage}}]{Ellis:2008hf}%
			\BibitemOpen
			\bibfield  {author} {\bibinfo {author} {\bibfnamefont {J.~R.}\ \bibnamefont
					{Ellis}}, \bibinfo {author} {\bibfnamefont {K.~A.}\ \bibnamefont {Olive}},\
				and\ \bibinfo {author} {\bibfnamefont {C.}~\bibnamefont {Savage}},\
			}\bibfield  {title} {\bibinfo {title} {{Hadronic Uncertainties in the Elastic
						Scattering of Supersymmetric Dark Matter}},\ }\href
			{https://doi.org/10.1103/PhysRevD.77.065026} {\bibfield  {journal} {\bibinfo
					{journal} {Phys. Rev. D}\ }\textbf {\bibinfo {volume} {77}},\ \bibinfo
				{pages} {065026} (\bibinfo {year} {2008})},\ \Eprint
			{https://arxiv.org/abs/0801.3656} {arXiv:0801.3656 [hep-ph]} \BibitemShut
			{NoStop}%
			\bibitem [{\citenamefont {Ellis}\ \emph {et~al.}(2000)\citenamefont {Ellis},
				\citenamefont {Ferstl},\ and\ \citenamefont {Olive}}]{Ellis:2000ds}%
			\BibitemOpen
			\bibfield  {author} {\bibinfo {author} {\bibfnamefont {J.~R.}\ \bibnamefont
					{Ellis}}, \bibinfo {author} {\bibfnamefont {A.}~\bibnamefont {Ferstl}},\ and\
				\bibinfo {author} {\bibfnamefont {K.~A.}\ \bibnamefont {Olive}},\ }\bibfield
			{title} {\bibinfo {title} {{Reevaluation of the elastic scattering of
						supersymmetric dark matter}},\ }\href
			{https://doi.org/10.1016/S0370-2693(00)00459-7} {\bibfield  {journal}
				{\bibinfo  {journal} {Phys. Lett.}\ }\textbf {\bibinfo {volume} {B481}},\
				\bibinfo {pages} {304} (\bibinfo {year} {2000})},\ \Eprint
			{https://arxiv.org/abs/hep-ph/0001005} {arXiv:hep-ph/0001005 [hep-ph]}
			\BibitemShut {NoStop}%
			\bibitem [{\citenamefont {Agnese}\ \emph {et~al.}(2017)\citenamefont {Agnese}
				\emph {et~al.}}]{SuperCDMS:2016wui}%
			\BibitemOpen
			\bibfield  {author} {\bibinfo {author} {\bibfnamefont {R.}~\bibnamefont
					{Agnese}} \emph {et~al.} (\bibinfo {collaboration} {SuperCDMS}),\ }\bibfield
			{title} {\bibinfo {title} {{Projected Sensitivity of the SuperCDMS SNOLAB
						experiment}},\ }\href {https://doi.org/10.1103/PhysRevD.95.082002} {\bibfield
				{journal} {\bibinfo  {journal} {Phys. Rev. D}\ }\textbf {\bibinfo {volume}
					{95}},\ \bibinfo {pages} {082002} (\bibinfo {year} {2017})},\ \Eprint
			{https://arxiv.org/abs/1610.00006} {arXiv:1610.00006 [physics.ins-det]}
			\BibitemShut {NoStop}%
			\bibitem [{\citenamefont {Agnes}\ \emph {et~al.}(2018)\citenamefont {Agnes}
				\emph {et~al.}}]{Agnes:2018oej}%
			\BibitemOpen
			\bibfield  {author} {\bibinfo {author} {\bibfnamefont {P.}~\bibnamefont
					{Agnes}} \emph {et~al.} (\bibinfo {collaboration} {DarkSide}),\ }\bibfield
			{title} {\bibinfo {title} {{Constraints on Sub-GeV Dark-Matter--Electron
						Scattering from the DarkSide-50 Experiment}},\ }\href
			{https://doi.org/10.1103/PhysRevLett.121.111303} {\bibfield  {journal}
				{\bibinfo  {journal} {Phys. Rev. Lett.}\ }\textbf {\bibinfo {volume} {121}},\
				\bibinfo {pages} {111303} (\bibinfo {year} {2018})},\ \Eprint
			{https://arxiv.org/abs/1802.06998} {arXiv:1802.06998 [astro-ph.CO]}
			\BibitemShut {NoStop}%
			\bibitem [{\citenamefont {Belyaev}\ \emph {et~al.}(2013)\citenamefont
				{Belyaev}, \citenamefont {Christensen},\ and\ \citenamefont
				{Pukhov}}]{Belyaev:2012qa}%
			\BibitemOpen
			\bibfield  {author} {\bibinfo {author} {\bibfnamefont {A.}~\bibnamefont
					{Belyaev}}, \bibinfo {author} {\bibfnamefont {N.~D.}\ \bibnamefont
					{Christensen}},\ and\ \bibinfo {author} {\bibfnamefont {A.}~\bibnamefont
					{Pukhov}},\ }\bibfield  {title} {\bibinfo {title} {{CalcHEP 3.4 for collider
						physics within and beyond the Standard Model}},\ }\href
			{https://doi.org/10.1016/j.cpc.2013.01.014} {\bibfield  {journal} {\bibinfo
					{journal} {Comput. Phys. Commun.}\ }\textbf {\bibinfo {volume} {184}},\
				\bibinfo {pages} {1729} (\bibinfo {year} {2013})},\ \Eprint
			{https://arxiv.org/abs/1207.6082} {arXiv:1207.6082 [hep-ph]} \BibitemShut
			{NoStop}%
			\bibitem [{\citenamefont {Harvey}\ and\ \citenamefont
				{Turner}(1990)}]{Harvey:1990qw}%
			\BibitemOpen
			\bibfield  {author} {\bibinfo {author} {\bibfnamefont {J.~A.}\ \bibnamefont
					{Harvey}}\ and\ \bibinfo {author} {\bibfnamefont {M.~S.}\ \bibnamefont
					{Turner}},\ }\bibfield  {title} {\bibinfo {title} {{Cosmological baryon and
						lepton number in the presence of electroweak fermion number violation}},\
			}\href {https://doi.org/10.1103/PhysRevD.42.3344} {\bibfield  {journal}
				{\bibinfo  {journal} {Phys. Rev. D}\ }\textbf {\bibinfo {volume} {42}},\
				\bibinfo {pages} {3344} (\bibinfo {year} {1990})}\BibitemShut {NoStop}%
		\end{thebibliography}
		%

	\end{document}